\documentclass[12pt,final,twoside]{book}
\usepackage[a4paper,			%
  left		= 2.5cm,		
  right		= 2.5cm,		
  bindingoffset	= 1.0cm,		
  top		= 3.0cm,		
  bottom	= 3.0cm,		
  includehead	= true,			
  includefoot	= false,		
  showframe	= false			%
]{geometry}				%
\usepackage{amsmath,amsfonts,amsthm}
\usepackage[OT4]{fontenc}		
\usepackage[utf8]{inputenc}		
\usepackage[english]{babel}		%
\usepackage{graphicx}			
\usepackage{subfigure}			
\begin{document}
\begin{titlepage}
  \begin{center}
    \vspace{5.0cm}
    {\Huge Hydrodynamic description of particle production in relativistic heavy-ion collisions}\\
    \vspace{3.0cm}
    {\Large \bf Miko\l aj Chojnacki}\\
    \vspace{2.0cm}
    {\large
      The Henryk Niewodnicza\'{n}ski\\
      Institute of Nuclear Physics\\
      Polish Academy of Sciences\\
      Krak\'{o}w, Poland\\
    }
    \vspace{1.0cm}
    \includegraphics[width=0.4 \textwidth]{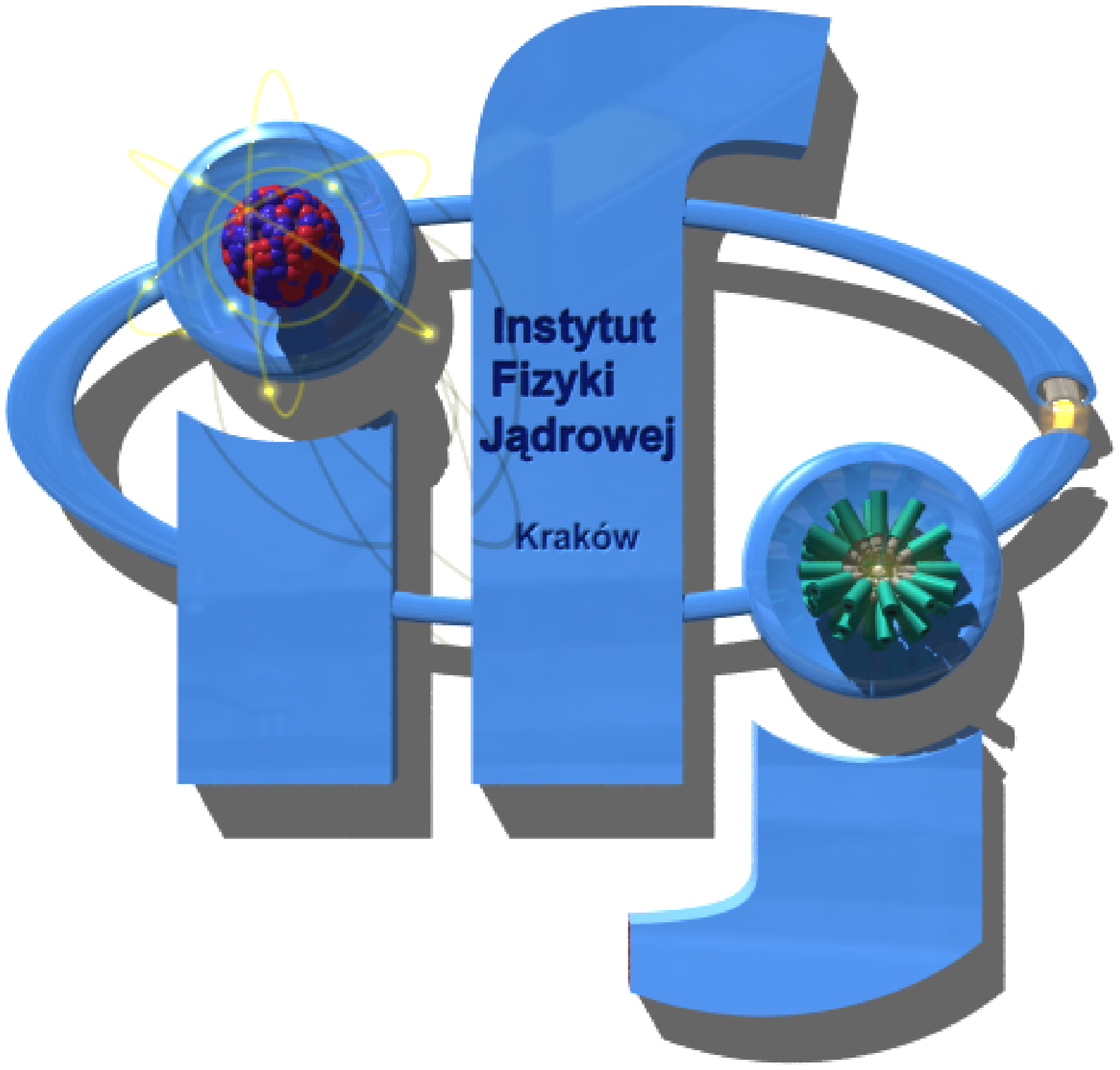}\\
    \vspace{2.0cm}
    {\it Thesis submitted for the Degree of Doctor of Philosophy in Physics}\\
    {\it Prepared under the supervision of Prof. Wojciech Florkowski}\\
    \vspace{2.0cm}
    {Krak\'{o}w, March 2009}
  \end{center}
\end{titlepage}
\pagebreak
\cleardoublepage
\begin{center}
  {\bf \large STRESZCZENIE}
  \end{center}
Niniejsza praca prezentuje nowo opracowany model ewolucji hydrodynamicznej, kt\'ory w po\l \k{a}czeniu z modelem  statystycznej hadronizacji {\tt THERMINATOR} s\l u\.zy nam do opisu zachowania silnie oddzia\l uj\k{a}cej materii  wyprodukowanej w relatywistycznych zderzeniach ci\k{e}\.{z}kich jon\'{o}w. Nasze oryginalne podej\'scie wykorzystano do wykonania dopasowa\'n dla danych pochodz\k{a}cych z eksperyment\'{o}w realizowanych na akceleratorze RHIC (Relativistic Heavy Ion Collider w Brookhaven National Laboratory) przy najwy\.{z}szej jego energii $\sqrt{s_{\rm NN}}$ = 200 GeV, oraz do sformu\l owania przewidywa\'{n} teoretycznych dla przysz\l ych eksperyment\'{o}w ci\k{e}\.{z}kojonowych przy wy\.{z}szych energiach ($\sqrt{s_{\rm NN}}$ = 5.5 TeV, dla akceleratora LHC, skr\'{o}t od Large Hadron Collider w CERN-ie). 
\par Nasze wyniki odnosz\k{a} si\k{e} do obserwabli jedno- i dwu-cz\k{a}stkowych w zakresie mi\k{e}kkiej fizyki ($p_{\rm T} \le$ 2 GeV). Opisujemy widma cz\k{a}stek w p\k{e}dzie poprzecznym, wsp\'{o}\l czynnik przep\l ywu eliptycznego $v_2$, oraz promienie HBT dla identycznych pio\-n\'ow (HBT jest skr\'otem od nazwisk Hanbury-Brown i Twiss). W ramach prac nad rozpraw\k{a} skonstruowali\'{s}my nowe r\'{o}wnanie stanu dla materii silnie oddzia\l uj\k{a}cej, kt\'{o}re \l \k{a}czy model gazu hadronowego z wynikami symulacji QCD na siatkach. Ca\l o\'{s}\'{c} program\'ow tworzy platform\k{e} obliczeniow\k{a} w sk\l ad kt\'{o}rej wchodzi kod hydrodynamiczny po\l \k{a}czony z modelem statystycznej hadronizacji {\tt THERMINATOR} (skr\'ot od THERMal heavy IoN generATOR). 
\par Stosuj\k{a}c standardowy model optyczny Glaubera jako warunek pocz\k{a}tkowy dla ewolucji hydrodynamicznej, osiagn\k{e}li\'{s}my bardzo dobry opis danych eksperymentalnych uzyskanych na akceleratorze RHIC. W szczeg\'olno\'sci osiagn\k{e}li\'{s}my zna\-cznie lepszy, od wcze\'sniej uzyskanych w modelach hydrodynamicznych, opis promieni korelacyjnych. Dla przysz\l ych eksperyment\'{o}w ci\k{e}\.{z}kojonowych na LHC uzyskali\'{s}my przewidywania teoretyczne dotycz\k{a}ce mi\k{e}kkich obserwabli. 
\par Zaproponowali\'{s}my r\'{o}wnie\.{z} spos\'ob rozwi\k{a}zania tzw. zagadki HBT na RHIC-u. Sugerujemy zmodyfikowanie warunk\'{o}w pocz\k{a}tkowych i wprowadznie Gaussowskiego profilu g\k{e}sto\'{s}ci materii jako warunku pocz\k{a}tkowego dla hydrodynamiki. Taka modyfikacja prowadzi do szybszej formacji poprzecznego przep\l ywu kolektywnego, co warunkuje uzyskanie wyj\k{a}tkowo dobrej zgodno\'sci naszego modelu z danymi eksperymentalnymi. 
\par Jako ostatni punkt, wprowadzili\'{s}my do naszego modelu proces swobodnego strumieniowania cz\k{a}stek, kt\'{o}ry w po\l \k{a}czeniu z mechanizmem nag\l ej, chocia\.z op\'o\'znionej w czasie termalizacji, tworzy nowe warunki pocz\k{a}tkowe dla kodu hydrodynami\-cznego. Wprowadzenie przedr\'{o}wnowagowej ewolucji pozwoli\l o nam na op\'{o}\'{z}nienie startu fazy hydrodynamicznej. Jest to po\.{z}\k{a}dany efekt, kt\'{o}ry ma na celu unikni\k{e}cie za\l o\.zenia o bardzo  wczesnej termalizacji uk\l adu, kt\'ore wydaje si\k{e} bardzo trudne do uzasadnienia na gruncie mikroskopowym. Podkre\'slmy, i\.z w\l \k{a}czenie swobodnego strumieniowania cz\k{a}stek nie zmienia wysokiej zgodno\'sci uzyskanych wynik\'ow mode\-lowych z danymi do\'swiadczalnymi.
%
\cleardoublepage
\begin{flushright}
 {\it I know that this defies the law of gravity,\\
 but, you see, I never studied law.}\\
 -- Bugs Bunny
\end{flushright}
\vspace{2cm}
\begin{flushright}
  {\it Physics isn't a religion.\\ 
  If it were, we'd have a much easier time raising money.}\\
  -- Leon Lederman
\end{flushright}
\vspace{15cm}
\begin{flushright}
  {\it \Large --- to my Parents ---}
\end{flushright}
\cleardoublepage
\begin{center}
  \bf \Large Acknowledgments
\end{center}
I would like to express my deepest thanks to my supervisor Prof. Wojciech Florkowski for his invaluable help, guidance and patience during the course of this Thesis. Furthermore I wish to thank Wojciech Broniowski, Adam Kisiel and Piotr Bo\.{z}ek for a chance of working with them and being part of the team and to everybody in the Department of Theory of Structure of Matter (NZ41).\\
\par I am very grateful to all of my family and friends for their support throughout my journey into obtaining this Degree. I thank you all.\\
\par Research was supported by the Polish Ministry of Science and Higher Education grant N202 153 32/4247 (2007-2009).

\tableofcontents
\chapter{Introduction}
\label{chaper:intro}
%
Nowadays, the relativistic hydrodynamics is regarded as the best theoretical framework for description of the spacetime evolution of strongly interacting matter produced in ultra-relativistic heavy-ion collisions 
\cite{Kolb:1999it,Teaney:1999gr,Kolb:2000sd,Teaney:2000cw,Kolb:2000fha,Huovinen:2001cy,Kolb:2001qz,Teaney:2001gc,Teaney:2001av,Heinz:2001xi,Heinz:2002sq,Kolb:2002cq,Kolb:2002ve,Kolb:2003gq,Kolb:2003dz,Nonaka:2000ek,Hirano:2001eu,Hirano:2001yi,Morita:2002av,Hirano:2002ds,Hirano:2002ds,Nonaka:2004pg,Hirano:2005xf,Nonaka:2006yn,Bozek:2009ty}
, for a recent review see \cite{Heinz:2009xj}. In particular, the soft hadronic one-particle data describing the transverse-momentum spectra and the elliptic flow coefficient $v_2$, collected in the RHIC experiments (Relativistic Heavy-Ion Collider at the Brookhaven National Laboratory), have been successfully explained in various approaches based on the perfect-fluid hydrodynamics. In fact, the explanation of the large value of the elliptic flow by perfect hydrodynamics is regarded as the evidence of early thermalization and suggests that the quark-gluon plasma created at RHIC is a strongly interacting system \cite{Shuryak:2004kh}. 
\par On the other hand, the approaches based on the hydrodynamics cannot reproduce the two-particle observables such as the pion correlation radii. The latter are commonly called the HBT radii -- after Hanbury-Brown and Twiss who in 1950s showed that it was possible to determine the angular sizes of astronomical radio sources and stars from the correlations of signal intensities, rather than amplitudes. The difficulty of the consistent description of the one- and two-particle observables picked up the name "HBT puzzle". The HBT puzzle and the problem of the microscopic explanation of very fast thermalization of the matter produced at RHIC  represent two issues that challenge the hydrodynamic picture. 
\par In this Thesis we present our recently developed hydrodynamic model and use it to describe the RHIC data. We address  both the one- and two-particle observables: the transverse momentum spectra, the elliptic flow coefficient $v_2$, and the pion HBT radii. We suggest how the HBT puzzle as well as the early thermalization problem may be solved. Similarly to other approaches, our model is based on the perfect fluid hydrodynamics and includes the symmetry against Lorentz boosts along the beam axis, the so called boost-invariance. This restriction means that our results may be applied only to the central regions of relativistic heavy-ion collisions. On the other hand, our framework differs from other approaches in several important aspects, in particular, in the use of a different equation of state, modification of the initial conditions, two-body method of the calculation of the correlation functions, and different treatment of the final hadronic stage.  
\bigskip
The main achievements of the Thesis are the following:
\begin{itemize}
  \item[1.] {\it The construction of the realistic equation of state for strongly interacting matter which interpolates between the hadron gas model and the results of the QCD lattice simulations.} This equation of state describes the crossover phase transition, i.e., the transition where thermodynamic variables such as energy density or entropy density change very rapidly in the narrow range of the temperature, however, no real discontinuities in the behavior of the thermodynamic variables are present. Noticeably, our equation of state has no pronounced soft point where the sound velocity is very small and possibly drops to zero. It is known that the presence of such a soft point leads to the ratio of the HBT radii $R_{\rm out}/R_{\rm side}$ that is larger than unity -- an effect which has not been confirmed by the experimental data.
  \item[2.] {\it The boost-invariant hydrodynamic equations for baryon free matter have been rewritten in the very concise form which reduces the number of the independent equations to two.} This may be done in the formal way if the range of the variable $r$ (the distance from the collision axis) is extended to the negative values. The applied procedure is a direct generalization of the formalism introduced earlier by Baym et al. in the studies of cylindrically symmetric systems with constant sound velocity. The new form of the hydrodynamic equations allows for the simple and natural inclusion of the boundary conditions at the origin of the system.
  \item[3.] {\it The computational platform has been constructed which combines the hydrodynamic code with the statistical hadronization model} {\tt THERMINATOR}\cite{Kisiel:2005hn}. This is arranged in such a way that the information about the freeze-out hypersurface obtained from the hydrodynamic code is exported and treated as an input for {\tt THERMINATOR}. We emphasize that our equation of state (in the region below the critical temperature $T_{\rm c} \sim 170$ MeV) describes the hadron gas with the same set of the hadronic species as that included in {\tt THERMINATOR}. Hence, there is a smooth change between the hydrodynamic and statistical description of the produced matter.  {\tt THERMINATOR} is a Monte Carlo program simulating the decays of resonances. The Monte-Carlo method allows for the direct comparison of our model results with the experimental data. In particular, one can easily include various experimental cuts.
  \item[4.] {\it The successful description of the soft hadronic RHIC data has been achieved with the standard initial conditions obtained from the optical limit of the Glauber model.}  By this we mean here that the use of the new equation of state helped to reduce the discrepancy between the theoretical and experimental ratio of the HBT radii $R_{\rm out}$ and $R_{\rm side}$. In this version of our calculations we find $R_{\rm out}/R_{\rm side} \sim 1.25$, significantly closer to the experimental values than in earlier hydrodynamic studies.  
  \item[5.] {\it Predictions for the future heavy-ion collisions at LHC have been formulated.} In the studied by us central region we expect that the transition from the RHIC energy, $\sqrt{s_{NN}}=$ 200 GeV, to the LHC energy, $\sqrt{s_{NN}}=$ 5.5 TeV, results essentially in a higher initial central temperature $T_{\rm i}$ used as the input for the hydrodynamic calculation. Thus, one can make predictions for the collisions at the LHC energies using a set of values for $T_{\rm i}$ which are higher than those used at RHIC. At RHIC we found $T_{\rm i}$ = 320 MeV, hence for LHC we studied the cases \mbox{$T_{\rm i}$ = 400, 450, and 500 MeV}. Our results for LHC indicate a moderate increase of the HBT radii and saturation of the pion elliptic flow (as compared to the RHIC experiments). 
  \item[6.] {\it The solution of the RHIC HBT puzzle has been proposed which suggests the use of the modified Gaussian-type initial conditions.} We find that the choice of the initial condition in the form of a two-dimensional Gaussian profile for the transverse energy leads to a complete and consistent description of soft observables measured at RHIC. The transverse-momentum spectra, the elliptic-flow, and the HBT correlation radii, including the ratio $R_{\rm out}/R_{\rm side}$ are very well described. 
  \item[7.] {\it The processes of the free streaming of partons followed by the sudden equilibration were incorporated in the model.} Those two processes deliver modified initial conditions for the hydrodynamics. In particular, the inclusion of the free-streaming stage allows for the delayed start of the hydrodynamic evolution, which is a desirable effect in the context of the early thermalization problem. 
\end{itemize}
\par The Thesis is organized as follows. In Chapter 2 we describe the construction of our equation of state. In Chapter 3 we present the hydrodynamic equations and transform them to the form used in the numerical calculations. The initial conditions and the freeze-out prescription are introduced in Chapters 4 and 5, respectively. The fits to the RHIC data obtained with the standard initial conditions are presented in Chapter 6. The predictions for the LHC are given in Chapter 7. The solution of the RHIC HBT puzzle with modified Gaussian initial conditions is presented and discussed in Chapter 8. In that Section we also discuss the inclusion of the parton free-streaming as the pre-hydrodynamics stage. The Summary and four Appendices close the Thesis.
\par We use everywhere the natural units with $c = \hbar = k_{\rm B} = 1$. The signature of the metric tensor is $(+---)$.
\newpage
The results discussed in this Thesis were published in the following articles:
{\small
\begin{itemize}
  \item[1.] M. Chojnacki, W. Florkowski, and T. Cs\"org\H{o},\\
  {\it Formation of Hubble - like flow in little bangs},\\
  Phys. Rev. {\bf C71} (2005) 044902, (nucl-th/0410036).
  \item[2.] M. Chojnacki and W. Florkowski,\\
  {\it Characteristic form of boost-invariant and cylindrically asymmetric hydrodynamic equations},\\
  Phys. Rev. {\bf C74} (2006) 034905, (nucl-th/0603065).
  \item[3.] M. Chojnacki and W. Florkowski,\\
  {\it Temperature dependence of sound velocity and hydrodynamics of ultra - relativistic heavy-ion collisions},\\
  Acta Phys. Pol. {\bf B38} (2007) 3249, (nucl-th/0702030).
  \item[4.] M. Chojnacki, W. Florkowski, W. Broniowski, and A. Kisiel,\\
  {\it Soft heavy-ion physics from hydrodynamics with statistical hadronization: Predictions for collisions at $\sqrt{s_{\rm NN}}$ = 5.5 TeV},\\
  Phys. Rev. {\bf C78} (2008) 014905, arXiv:0712.0947 [nucl-th].
  \item[5.] W. Broniowski, M. Chojnacki, W. Florkowski,  and A. Kisiel,\\
  {\it Uniform Description of Soft Observables in Heavy-Ion Collisions at\\ $\sqrt{s_{\rm NN}}$ = 200 GeV,}\\
  Phys. Rev. Lett. {\bf 101} (2008) 022301, arXiv:0801.4361 [nucl-th].
  \item[6.] A. Kisiel, W. Broniowski, M. Chojnacki, and W. Florkowski,\\
  {\it Azimuthally sensitive femtoscopy in hydrodynamics with statistical hadronization from the BNL Relativistic Heavy Ion Collider to the CERN Large Hadron Collider},\\
  Phys. Rev. {\bf C79} (2009) 014902, arXiv:0808.3363 [nucl-th].
  \item[7.] W. Broniowski, W. Florkowski, M. Chojnacki, A. Kisiel,\\
  {\it Free-streaming approximation in early dynamics of relativistic heavy-ion collisions},\\
  {\it submitted to Phys. Rev. {\bf C}}, arXiv:0812.3393 [nucl-th].
\end{itemize}
}
They were also presented during various international conferences including:
{\small
\begin{itemize}
  \item[1.] M. Chojnacki,\\
  {\it Hubble-like Flows in Relativistic Heavy-Ion Collisions},\\
  Acta Phys. Hung. {\bf A27} (2006) 331, (nucl-th/0510092).\\
  18th International Conference On Ultrarelativistic Nucleus-Nucleus Collisions: Quark Matter 2005 (QM 2005).
  \item[2.] M. Chojnacki,\\
  {\it Cylindrically asymmetric hydrodynamic equations},\\
  Acta Phys. Polon. {\bf B37} (2006) 3391, (nucl-th/0609060).\\
  Cracow School Of Theoretical Physics: 46th Course 2006.
  \item[3.] M. Chojnacki,\\
  {\it Temperature-dependent sound velocity in hydrodynamic equations for relativistic heavy-ion collisions},\\
  J. Phys. {\bf G35} (2008) 044074, arXiv:0709.1594 [nucl-th].\\
  International Conference On Strangeness In Quark Matter (SQM 2007).
  \item[4.] W. Florkowski, M. Chojnacki, W. Broniowski, A. Kisiel,\\
  {\it Soft-hadronic observables for relativistic heavy-ion collisions at RHIC and LHC},\\
  Acta Phys. Polon. {\bf B39} (2008) 1555, arXiv:0804.0974 [nucl-th].\\
  Cracow Epiphany Conference On LHC Physics.
  \item[5.] W. Florkowski, W. Broniowski, M. Chojnacki, A. Kisiel,\\
  {\it Hydrodynamics and perfect fluids: Uniform description of soft observables in Au+Au collisions at RHIC},\\
  arXiv:0811.3761 [nucl-th] and arXiv:0902.0377 [hep-ph].\\
  38th International Symposium On Multiparticle Dynamics ISMD08.
  \item[6.] W. Florkowski, W. Broniowski, M. Chojnacki, A. Kisiel,\\
  {\it Solution of the RHIC HBT puzzle with Gaussian initial conditions},\\
  arXiv:0812.4125 [nucl-th].\\
  International Conference On Strangeness In Quark Matter (SQM 2008).
  \item[7.] W. Broniowski, W. Florkowski, M. Chojnacki, A. Kisiel,\\
  {\it Initial conditions for hydrodynamics: implications for phenomenology},\\
  arXiv:0812.4935 [nucl-th].\\
  IV Workshop on Particle Correlations and Femtoscopy.
  \item[8.] W. Florkowski, W. Broniowski, M. Chojnacki, A. Kisiel,\\
  {\it Consistent hydrodynamic description of one- and two-particle observables in relativistic heavy-ion collisions at RHIC},\\
  arXiv:0901.1251 [nucl-th].\\
  IV Workshop on Particle Correlations and Femtoscopy.
\end{itemize}
}
\chapter[Relativistic thermodynamics]{Thermodynamics of relativistic baryon-free matter}
\label{chapter:thermo}
%
In our approach we concentrate on the description of the mid-rapidity region of ultra-relativistic heavy-ion collisions. Statistical analysis applied to the highest-energy RHIC data indicates that the baryon chemical potential \mbox{$\mu_{\rm B}$} at the chemical freeze-out is of about 25 MeV in this region \mbox{\cite{Florkowski:2001fp,BraunMunzinger:2001ip,Baran:2003nm,Cleymans:2004pp,Biedron:2006vf}}. The predictions of the statistical models for LHC give even smaller values, \mbox{$\mu_{\rm B} \approx$ 0.8 MeV} \cite{Andronic:2005yp}. On the other hand, the expected temperature is of about 150 - 170 MeV, hence the ratio $\mu_{\rm B}/T$ is small and in the hydrodynamic equations we can approximately assume that the baryon chemical potential vanishes. In this situation, as discussed in Ref. \cite{Chojnacki:2004ec}, the whole information about the equation of state is encoded in the temperature-dependent sound velocity $c_{\rm s}(T)$. We assume that at low temperatures the sound velocity is given by the hadron-gas model with a complete set of hadronic resonances. In this case the function $c_{\rm s}^2(T)$ approaches zero as $T/m_\pi$, which is the characteristics of  the pion gas ($m_\pi$ is the pion mass). On the other hand, at high temperatures our equation of state coincides with the recent lattice simulations of QCD \cite{Aoki:2005vt}. The thermodynamic properties of the hadron gas and the quark-gluon plasma are discussed below in more detail in Sects. \ref{section:thermo_HG} and \ref{section:thermo_QGP}, respectively. 
\par
\begin{figure}[!t]
  \begin{center}
    \includegraphics[width=0.55 \textwidth]{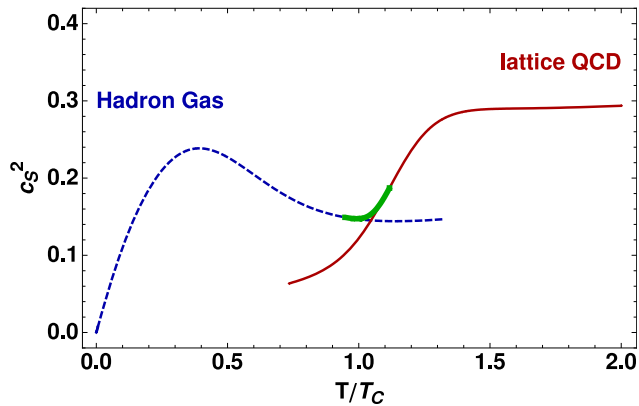}
  \end{center}
  \vspace{-3mm}
  \caption{\small Temperature dependence of the square of the sound velocity at zero baryon density. The plot shows the result of the lattice simulations of QCD \cite{Aoki:2005vt} (solid line) and the result obtained with the ideal hadron-gas model discussed in Sect. \ref{section:thermo_HG} (dashed line). A piece of the thick solid line describes the simplest interpolation between the two calculations. The critical temperature $T_{\rm c}$ equals 170 MeV. It is defined as the place where the sudden change of the thermodynamic variables occurs, see Fig. \ref{fig:thermo_thermo} }
  \label{fig:thermo_cs2}
\end{figure}

 In the transition region between the hadron gas and the plasma, whose position is characterized by the critical temperature $T_{\rm c}$,  different interpolations between the hadron gas result and the lattice result may be considered. In Ref. \cite{Chojnacki:2007jc} we showed, however, that the most promising equation of state is based on the simplest interpolation between the hadron-gas model and the lattice data, see Fig. \ref{fig:thermo_cs2}. This is so because the sound velocity function which does not exhibit a distinct minimum at the critical temperature leads to the relatively short evolution time and this effect helps to describe correctly the HBT data. The effects of different forms of the sound velocity are discussed in Sect. \ref{section:thermo_PhaseTr}. Since the simplest interpolation is the best, most of the results presented in this Thesis are obtained with the sound velocity function shown in Fig \ref{fig:thermo_cs2}. 
\par
\begin{figure}[!t]
  \begin{center}
    \includegraphics[width=0.8 \textwidth]{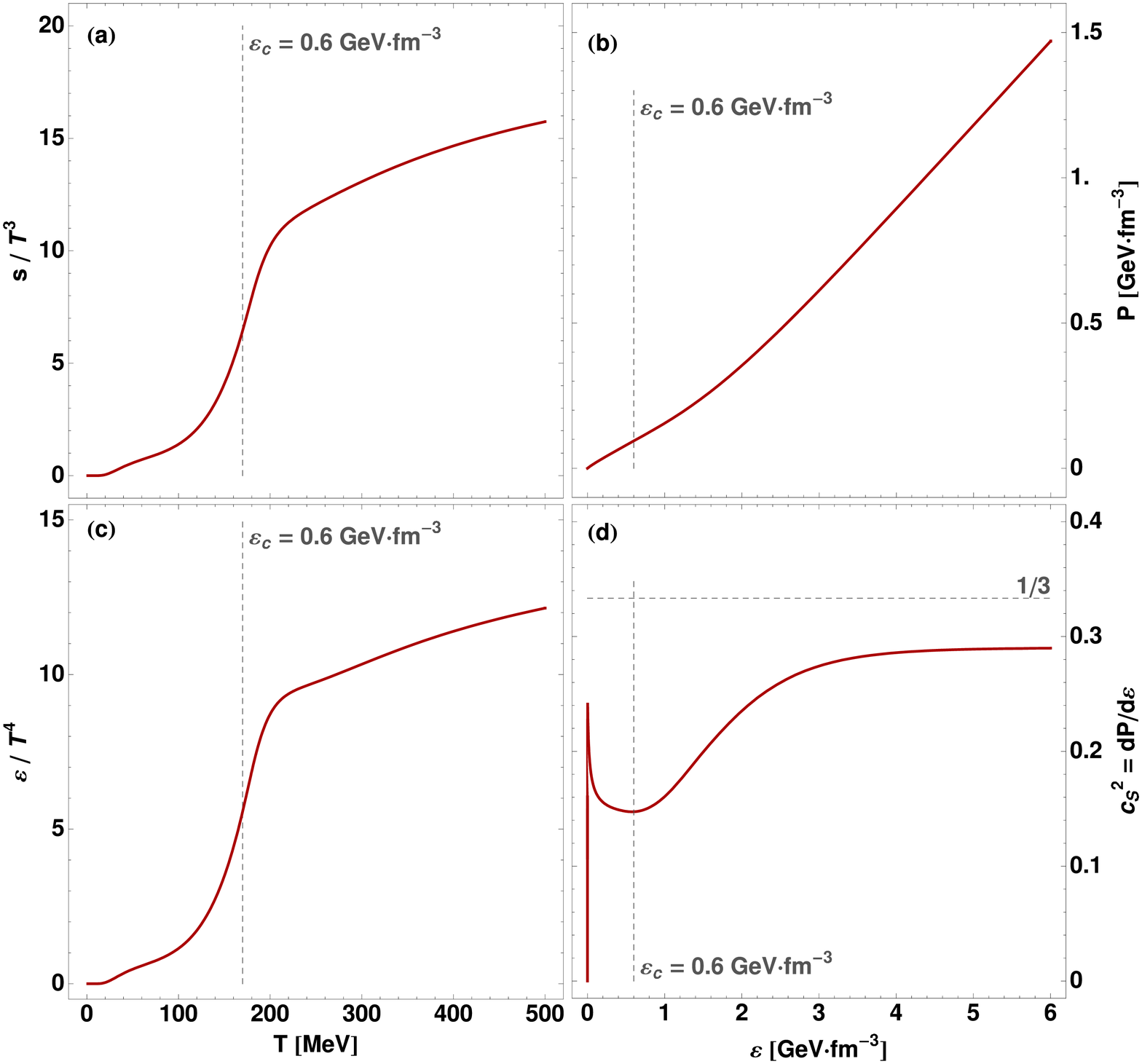}
  \end{center}
  \vspace{-3mm}
  \caption{\small The two left panels: the entropy and energy densities, scaled by $T^3$ and $T^4$, respectively, shown as functions of the temperature. The two right panels: the pressure and sound velocity shown as functions of the energy density. The presented thermodynamic functions follow directly from the temperature-dependent sound velocity shown in Fig.~\ref{fig:thermo_cs2}. We observe a sudden but smooth change of $s/T^3$ and $\varepsilon/T^4$ at $T \sim T_{\rm c}$. The vertical line indicates in all cases the critical energy density corresponding to $T_{\rm c}$ = 170 MeV. With our equation of state one finds $\varepsilon_{\rm c} = 0.6\,\mbox{GeV/fm}^3$.}
  \label{fig:thermo_thermo}
\end{figure}

 The knowledge of the function $c_{\rm s}(T)$ allows us to determine all other thermodynamic properties of our system. This is achieved with the help of the following thermodynamic identities
\begin{equation}
  \varepsilon + P = T s, \quad d\varepsilon = T ds, \quad dP = s dT, \quad c_{\rm s}^2 = \frac{dP}{d\varepsilon},
  \label{eqn:thermo_ident}
\end{equation}
where $\varepsilon$ is the energy density, $P$ is the pressure, $T$ is the temperature, and $s$ is the entropy density. In Fig.~\ref{fig:thermo_thermo} we display the entropy and energy densities as functions of $T$, and the pressure and sound velocity as functions of the energy density. These quantities follow directly from the assumed form of the function $c_{\rm s}(T)$, shown in Fig. \ref{fig:thermo_cs2}. 
\par We note that other equations of state and their impact on the physical observables were recently studied in Ref.~\cite{Huovinen:2005gy}. The result of that work was that the transverse-momentum spectra are quite insensitive to the assumed form of the equation of state. On the other hand, a noticeable dependence of the elliptic flow on the equation of state was observed. This dependence favored the strong first order phase transition. In Ref. \cite{Huovinen:2005gy} the effects concerning the HBT were not studied. Our recent work indicates that the constraints from the particle interferometry data exclude the strong first order phase transition since it leads to the unrealistically long evolution times. 
\par We stress that by using the lattice results we take into account the non - perturbative aspects of the plasma behavior, which may be regarded as the effective inclusion of the strongly-interacting quark-gluon plasma; large deviations from the ideal-gas behavior directly indicate the non-negligible interactions present in the plasma. In particular, $c_{\rm s}^2$ is significantly below the ideal-gas value of 1/3 also at temperatures way above $T_{\rm c}$. Note that in agreement with the present knowledge, no real phase transition is present in the system, but a {\em smooth cross-over}, therefore $c_{\rm s}$ does not drop to zero at $T_{\rm c}$ but remains a smooth function. 
%
\section{Hadron gas}
\label{section:thermo_HG}
%
The hadron-gas model is based on the assumption that all known hadrons (including hadronic resonances) form a multicomponent perfect gas. In this case, well known formulas for the thermodynamic variables of the relativistic perfect gases \cite{LandauSPh,Csernai:1994xw} may be applied. Of course, the hadron-gas model is applicable in the temperature region below the critical temperature $T_{\rm c}$ = 170 MeV. With increasing temperature the density of hadrons becomes so large that they start overlapping and the very idea of hadrons breaks down. We note that only in the case of pions the quantum statistics is necessary. For other particles, which are much heavier and less abundant, the Boltzmann classical limit is sufficient \cite{Michalec:2001um}.
%
\subsection{Pion gas}
\label{subsection:thermo_HG_piongas}
%
Pions are the most abundant particles produced in heavy-ion collisions. Because of their large multiplicity the statistical quantum effects cannot be neglected and one should use the Bose-Einstein distributions to describe the pion spectra. Additionally, since pions do not carry the baryon number and strangeness, their chemical potential may be assumed to be zero (possible values of the isospin chemical potential are very small, $\mu_{\rm I_3} < 1$ MeV, and are usually neglected). Therefore, in describing the thermodynamic properties of the pion gas ($\pi G$) we use the relations for a massive boson gas with zero chemical potential, which are worked out in Appendix \ref{chapter:Athermo}, namely
\begin{eqnarray}
 s          ^{\pi G}_i(T)	&=& \frac{1}{2\pi^2}\,       m_i^3\, \sum_{\kappa=1}^\infty \frac{1}{\kappa  } K_3\left( \frac{m_i}{T}\kappa \right), \label{eqn:thermo_hg_pg_s} \\
 P          ^{\pi G}_i(T)	&=& \frac{1}{2\pi^2}\, T^2\, m_i^2\, \sum_{\kappa=1}^\infty \frac{1}{\kappa^2} K_2\left( \frac{m_i}{T}\kappa \right), \label{eqn:thermo_hg_pg_P} \\
 \varepsilon^{\pi G}_i(T)	&=& \frac{1}{2\pi^2}\, T\,   m_i^2\, \sum_{\kappa=1}^\infty \frac{1}{\kappa^2} \left[ 3\, T\, K_2\left( \frac{m_i}{T} \kappa \right) + m_i\,\kappa\, K_1\left( \frac{m_i}{T} \kappa \right) \right], \label{eqn:thermo_hg_pg_e}
\end{eqnarray}
where the index $i$ specifies the isospin (distinguishes between $\pi^-$, $\pi^0$, $\pi^+$) and $m_i$ is the mass of the appropriate pion. The infinite sum over $\kappa$ is the technical way to include the Bose-Einstein statistics. In the numerical calculations it is sufficient to include only the first four terms -- the inverse powers of $\kappa$ reduce the higher-order terms fast enough.
\par Knowing the expression for the entropy density of the pion gas, we calculate the sound velocity from the last equation in (\ref{section:thermo_HG}). In the approximation that the pion masses are equal ($m_i=m_\pi$) we obtain
\begin{equation}
  \left(c_{\rm s}^2\right)^{\pi G}(T) = \frac{2\,T}{m_\pi} \frac{
    {\displaystyle \sum_{\kappa=1}^\infty } \frac{1}{\kappa}\, K_3\left( \frac{m_\pi}{T}\kappa \right)
  }{
    {\displaystyle \sum_{\kappa=1}^\infty } \left[ K_2\left( \frac{m_\pi}{T}\kappa \right) + K_4\left( \frac{m_\pi}{T}\kappa \right)\right]
  }.
  \label{eqn:thermo_hg_pg_cs2}
\end{equation}
It is interesting to observe the behavior of the sound velocity in the limit when temperature approaches zero. The arguments of the Bessel functions $K_\nu(z)$ tend to infinity and we can use the power series expansion to find that
\begin{equation}
  \left(c_{\rm s}^2\right)^{\pi G}(T \rightarrow 0) = \frac{T}{m_\pi} \frac{
    \sum_{\kappa=1}^\infty \left[ \kappa^{-\frac{3}{2}} e^{-\frac{m_\pi}{T}\kappa} + \cdots \right]
  }{
    \sum_{\kappa=1}^\infty \left[ \kappa^{-\frac{1}{2}} e^{-\frac{m_\pi}{T}\kappa} + \cdots \right]
  } \rightarrow \frac{T}{m_\pi},
  \label{eqn:thermo_hg_pg_cs2to0T}
\end{equation}
see Eq. (\ref{eqn:Zmath_BesselK_powser2}) from Appendix \ref{chapter:Zmath}. On the other hand, when temperature is large ($T \gg m_\pi$) we expand the modified Bessel functions according to (\ref{eqn:Zmath_BesselK_powser}) and get the following expression
\begin{equation}
  \left(c_{\rm s}^2\right)^{\pi G}(T \rightarrow \infty) = \frac{1}{3} \frac{
    \sum_{\kappa=1}^\infty \left[ \kappa^{-4} -\frac{1}{8 } \left(\frac{m_\pi}{T}\kappa\right)^2 + \frac{1}{8 } \left(\frac{m_\pi}{T}\kappa\right)^4 + \cdots \right]
  }{
    \sum_{\kappa=1}^\infty \left[ \kappa^{-4} -\frac{1}{24} \left(\frac{m_\pi}{T}\kappa\right)^2 + \frac{1}{96} \left(\frac{m_\pi}{T}\kappa\right)^4 + \cdots \right]
  } \rightarrow \frac{1}{3}.
  \label{eqn:thermo_hg_pg_cs2toinfT}
\end{equation}
The last limit illustrates the expected result -- at very high temperatures the massive pion gas behaves effectively like a massless gas with $P = \frac{1}{3} \varepsilon$ and its sound velocity squared equals $\frac{1}{3}$. (See Tables \ref{tab:Athermo_P} and \ref{tab:Athermo_E}, where the  complete formulas for pressure and energy density in various limits are given).
%
\subsection{Classical gas}
\label{subsection:thermo_HG_classgas}
%
Strictly speaking, all particles produced in heavy-ion collisions obey the quantum statistics, however for all particles other than pions the effects of quantum statistics are numerically negligible. Thus, to calculate the thermodynamic functions of the hadrons other than pions we may use the following formulas
\begin{eqnarray}
 s          ^{CG}_i(T)	&=& \frac{g_i}{2\pi^2}\,       m_i^3\, K_3\left( \frac{m_i}{T} \right), \label{eqn:thermo_hg_cg_s} \\
 P          ^{CG}_i(T)	&=& \frac{g_i}{2\pi^2}\, T^2\, m_i^2\, K_2\left( \frac{m_i}{T} \right), \label{eqn:thermo_hg_cg_P} \\
 \varepsilon^{CG}_i(T)	&=& \frac{g_i}{2\pi^2}\, T\,   m_i^2\, \left[ 3\, T\, K_2\left( \frac{m_i}{T} \right) + m_i\, K_1\left( \frac{m_i}{T} \right) \right], \label{eqn:thermo_hg_cg_e}
\end{eqnarray}
where $g_i = 2\,s_i + 1$ is the degeneration factor which holds the information about the spin degeneration of $i$-th particle. The sound velocity in the classical massive case is defined as follows
\begin{equation}
  \left(c_{\rm s}^2\right)^{CG}_i(T) = \frac{2\,T}{m_i} \frac{ K_3\left( \frac{m_i}{T} \right) }{ K_2\left( \frac{m_i}{T} \right) + K_4\left( \frac{m_i}{T} \right) },
  \label{eqn:thermo_hg_cg_cs2}
\end{equation}
and has the same asymptotic features as the pion gas, namely it tends to the value $\frac{1}{3}$ for very high temperatures and becomes proportional to $T$ if the temperature tends to zero.
%
\subsection{Massive hadron gas}
\label{subsection:thermo_HG_hadgas}
%
The massive hadron gas (HG) model is the sum of both massive pion gas and massive classical gas of all other hadrons. In our study, the information on the mass and spin of individual particles comes from the input file to the {\tt SHARE} program \cite{Torrieri:2004zz}. The table {\it particles.data} holds parameters for 371 particles consisting from {\it u}, {\it d} and {\it s} quarks. Thus, for all hadrons we may write
\begin{eqnarray}
 s^{HG}(T) &=& \sum_{i=1}^{3}\left(s\right)^{\pi G}_i(T) +\sum_{i=4}^{371} \left(s\right)^{CG}_i(T) \label{eqn:thermo_hg_hg_s}, \\
 \left(c_{\rm s}^2\right)^{HG}(T) &=& \frac{s^{HG}}{T} \frac{dT}{d(s^{HG})} \label{eqn:thermo_hg_hg_cs2},
\end{eqnarray}
and all other thermodynamic quantities ($s$, $P$ and $\varepsilon$) are expressed in the analogous way.
%
\section{Quark-Gluon Plasma}
\label{section:thermo_QGP}
\begin{figure}[!hbt]
  \begin{center}
    \includegraphics[width=0.55 \textwidth]{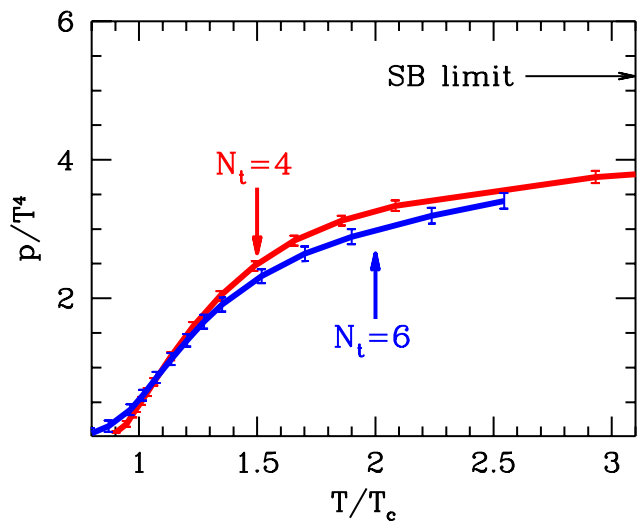}
  \end{center}
  \caption{\small QCD pressure as a function of temperature normalized by $T^4$ \cite{Aoki:2005vt}. Results are obtained for two lattice spacings $N_{\rm t} = 4$ (red) and $N_{\rm t} = 6$ (blue).}
  \label{fig:thermo_qqp_p}
\end{figure}

\begin{figure}[!hbt]
  \begin{center}
    \includegraphics[angle=-90,width=0.55 \textwidth]{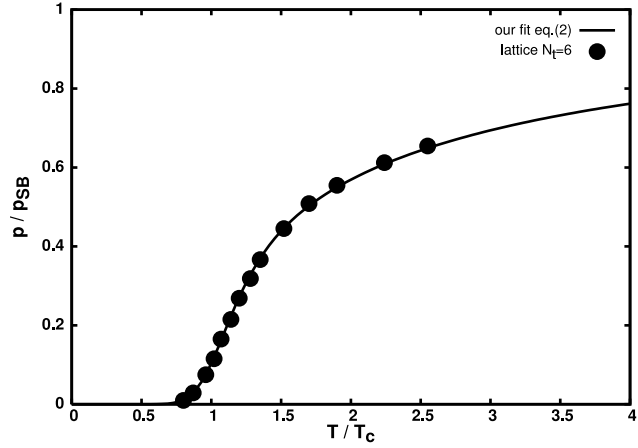}
  \end{center}
  \caption{\small Pressure as a function of temperature normalized to the pressure corresponding to the Stefan-Boltzmann limit. Points represent results from the QCD lattice simulations \cite{Aoki:2005vt}, whereas the line is the data fit from Ref. \cite{Biro:2006sv}.}
  \label{fig:thermo_qqp_pfit}
\end{figure}

%
In our approach we use the results of the lattice simulations of QCD at the finite temperature presented in Ref. \cite{Aoki:2005vt}, see Fig. \ref{fig:thermo_qqp_p}. They were obtained for physical masses of the light quarks and the strange quark.
The pressure data obtained from the QCD lattice calculation in Ref. \cite{Aoki:2005vt} have been recently parameterized for the case of $N_{\rm t}=6$ in Ref. \cite{Biro:2006sv}, see Fig. \ref{fig:thermo_qqp_pfit}. The parameterization has the form
\begin{equation}
  P = c\, T^4 \sigma\left( T_{\rm c} / T\right), \quad \sigma(g) = \frac{1+e^{-\frac{a}{b}}}{1+e^\frac{g-a}{b}} e^{-\lambda g},
  \label{eqn:thermo_qgp_pfit}
\end{equation}
where the dimensionless fit parameters equal: $a = 0.91$, $b = 0.11$, $c = 5.21$ and $\lambda = 1.08$.
Taking the lattice result at the face value, one expects that the sound velocity significantly drops down in the region $T \approx T_{\rm c}$. Similar behavior, with $c_{\rm s}(T_{\rm c})$ reaching zero,  is expected in the case of the first order phase transition where the changes of the energy happen at constant pressure. However, the lattice simulations suggest that for three massive quarks with realistic masses we deal with the cross-over rather than with the first order phase transition, hence the sound velocity remains finite, as is consistently shown in Fig.  \ref{fig:thermo_cs2}. 
%
\section{Modeling the crossover phase transition}
\label{section:thermo_PhaseTr}
%
\begin{figure}[!hb]
\begin{center}
\includegraphics[width=0.55 \textwidth]{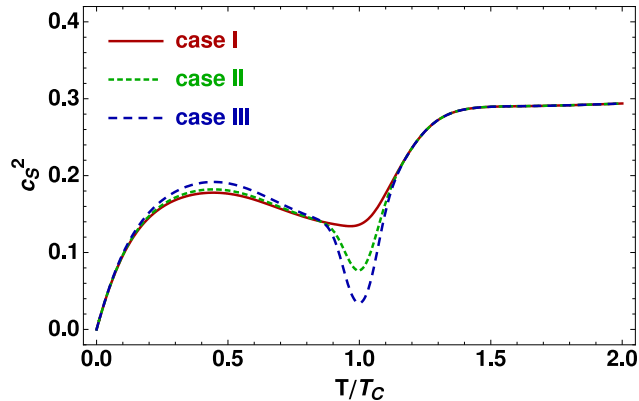}
\end{center}
\caption{\small Three different forms of the sound-velocity function analyzed in Ref. \cite{Chojnacki:2007jc}. The solid line describes the interpolation between the lattice and the hadron-gas results \cite{Chojnacki:2004ec} with a shallow minimum where $c_{\rm s}^2 = 0.14$  (case I), the dashed line describes the interpolation with a dip where $c_{\rm s}^2 = 0.08$ (case II), finally the long-dashed line describes the interpolation with a deep minimum where $c_{\rm s}^2 = 0.03$ (case III). Note that the case I is the {\it approximation} of the result shown previously in Fig. \ref{fig:thermo_cs2} -- see discussion in the text.}
\label{fig:thermo_cs2dip}
\end{figure}
\begin{figure}[!hbt]
\begin{center}
\includegraphics[width=0.8 \textwidth]{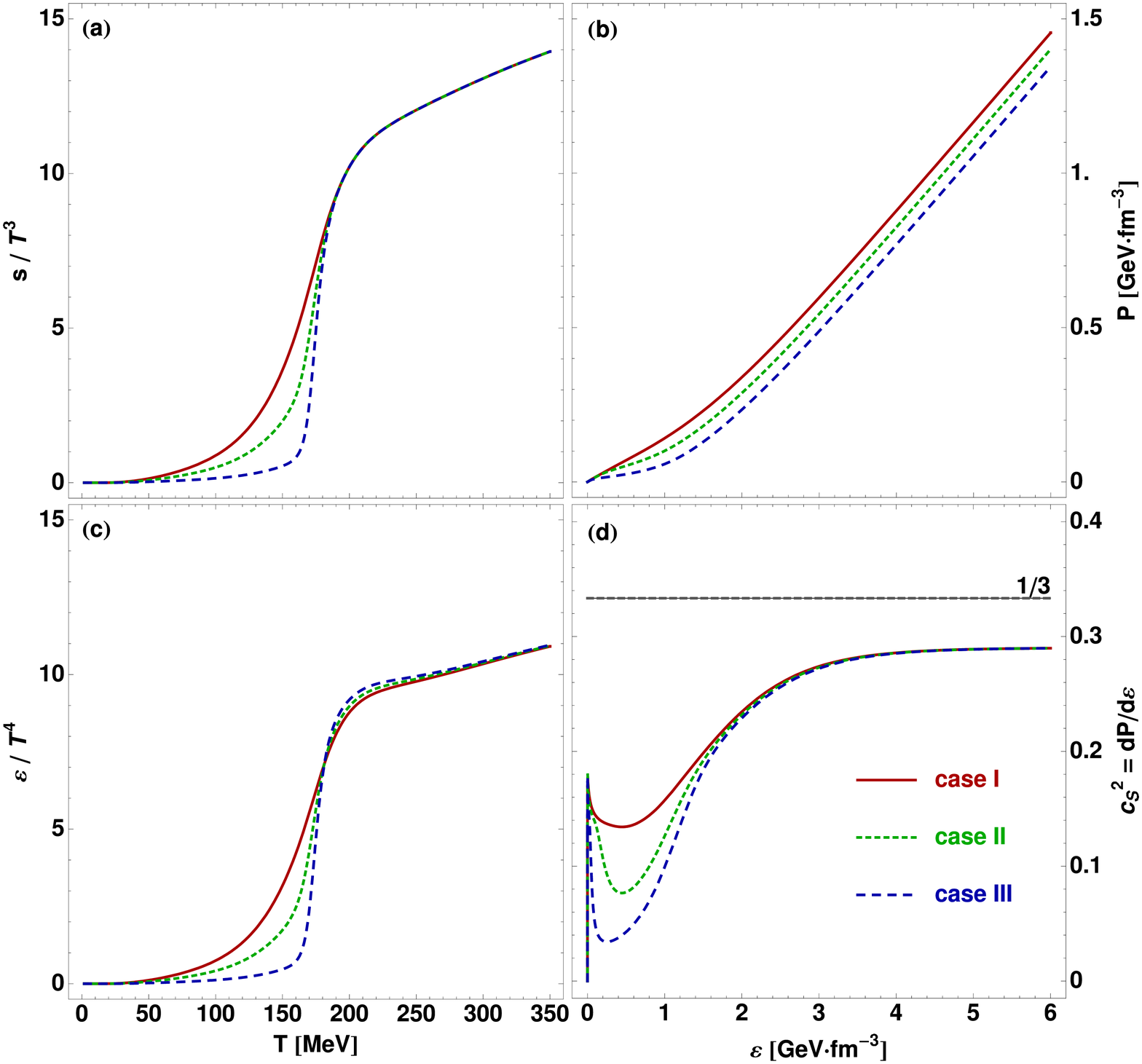}
\end{center}
\caption{\small The temperature dependence of the entropy density and energy density, panels ({\bf a}) and ({\bf c}), as well as the energy density dependence of the pressure and sound velocity, panels ({\bf b}) and ({\bf d}). One can observe that the deeper is the minimum of the sound velocity function, the steeper is the increase of the entropy density and the energy density.}
\label{fig:thermo_huo}
\end{figure}
The exact values of the sound velocity in the region $T \approx T_{\rm c}$ are  poorly known. The lattice calculations are not very much reliable for $T < T_{\rm c}$ and, at the same time,  the use of the hadron gas model with vacuum parameters becomes unrealistic for large densities (temperatures). The authors of Ref. \cite{Aoki:2005vt} state that in the hadronic phase the lattice spacing is larger than 0.3 fm and the lattice artifacts cannot be controlled in this region. In this situation, it is practical to consider different interpolations between the lattice and hadron-gas results and to analyze the physical effects of a particular choice of the interpolating function. This type of the study was performed in Ref. \cite{Chojnacki:2007jc}. Here we shortly discuss the main conclusions of this analysis.
\par In Ref. \cite{Chojnacki:2007jc} we considered three different sound-velocity functions $c_{\rm s}(T)$. Below, we refer to these three options as to the cases I, II and III, see Fig. \ref{fig:thermo_cs2dip}. In the case I,  we use the sound-velocity function which agrees with the ideal hadron gas model of Ref. \cite{Chojnacki:2004ec} in the temperature range $0 < T < 0.85 \,T_{\rm c}$ and with the lattice result in the temperature range  $T > 1.15 \,T_{\rm c}$\footnote{Ref. \cite{Chojnacki:2004ec} uses the approximation where the sum over hadronic states is replaced by the integral with the mass-density functions worked out in \cite{Broniowski:2000hd}. This leads to small differences between $c_{\rm s}(T)$ used in \cite{Chojnacki:2004ec,Chojnacki:2007jc} and \cite{Chojnacki:2007rq} }. In the region close to the critical temperature, $0.85 \,T_{\rm c} < T < 1.15 \,T_{\rm c}$, a simple interpolation between the two results is used. We have checked that such a simple interpolation yields directly the entropy density consistent with the lattice result. Namely, the use of the thermodynamic relation 
\begin{equation}
s(T) = s(T_{\rm min}) \exp\left[ \,\,\,\int\limits_{T_{\rm min}}^T \,\frac{dT^\prime}{T^\prime c_{\rm s}^2(T^\prime)}
\right],
\label{eqn:thermo_sumrule}
\end{equation}
relating the entropy density with the sound velocity for zero baryon chemical potential, gives the function $s(T)$ which agrees with the lattice result at high temperatures, $s(T)/T^3 \approx 12$ at $T = 1.5 \, T_{\rm c}$ \cite{Aoki:2005vt}. 
\par In the cases II and III, the sound-velocity interpolating functions have a distinct minimum at \mbox{$T = T_{\rm c}$}. Comparing to the case I [with \mbox{$c_{\rm s}(T_{\rm c})=0.37$} and \mbox{$c_{\rm s}^2(T_{\rm c})=0.14$}], the value of the sound velocity at \mbox{$T = T_{\rm c}$} is reduced by \mbox{25 \%} in the case II [where \mbox{$c_{\rm s}(T_{\rm c})=0.28$} and \mbox{$c_{\rm s}^2(T_{\rm c})=0.08$}], and by 50\% in the case III [where \mbox{$c_{\rm s}(T_{\rm c})=0.19$} and \mbox{$c_{\rm s}^2(T_{\rm c})=0.03$}].  From Eq. (\ref{eqn:thermo_sumrule}) one concludes that the decrease of the sound velocity at $T_{\rm c}$ leads to the increase of entropy density for high temperatures. Hence, in order to have the same value of the entropy density at high temperatures, a decrease of the sound velocity function in the region $T \approx  T_{\rm c}$ should be compensated by its increase in a different temperature range. For our interpolating functions in the cases II and III we assume that the values of $c_{\rm s}(T)$ in the range $0.15 \,T_{\rm c} < T < 0.85 \,T_{\rm c}$ are slightly higher than in the case I, see Fig. \ref{fig:thermo_cs2dip}. Such modifications may be regarded as the parameterization of the repulsive van der Waals forces in the hadron gas. The values of the maxima are chosen in such a way that the entropy densities for three considered cases are consistent with the lattice result, see the upper left panel of Fig. \ref{fig:thermo_huo} where the functions $s(T)/T^3$ are shown. We stress that in the three considered cases the values of $c_{\rm s}(T)$ in the temperature range $T_{\rm c} < T < 1.25 \, T_{\rm c}$ remain significantly below the massless limit $1/\sqrt{3}$. Such a limiting value is implicitly  used in many hydrodynamic codes assuming the equation of state of non-interacting massless quarks and gluons for $T > T_{\rm c}$, see for example the extended 3+1 hydrodynamic model of Ref. \cite{Nonaka:2006yn}.
\par The studies of the hydrodynamic spacetime evolution of matter described by the equations of state I, II and III were performed in Ref. \cite{Chojnacki:2007jc}. As expected, we found that the drop of $c_{\rm s}$ at $T = T_{\rm c}$ leads to the prolonged time of the evolution, hence leads to the increase of the $R_{\rm out} / R_{\rm side}$ ratio of the HBT radii. This behavior is in contradiction with the observed data which indicate $R_{\rm out} / R_{\rm side} \sim 1$. Thus, we have decided to exclude the cases II and III from further analysis and to restrict our consideration to the case I.
%

\chapter{Relativistic hydrodynamics of perfect fluid}
\label{chapter:hydro}
%
In this Chapter we present the main ingredients of our hydrodynamic model. We start with the general formulation of the relativistic hydrodynamics in the case of vanishing baryon number. Next, we implement the idea of boost-invariance that allows us to restrict our considerations to the plane $z=0$. In the subsequent Sections of this Chapter we show that the hydrodynamic equations may be rewritten in the form where only two equations are independent (at the expense of the formal extension of the variable $r = \sqrt{x^2+y^2}$ to negative values) and the boundary conditions at the origin are automatically fulfilled. Such a form, being the direct generalization of the approach introduced in Ref. \cite{Baym:1983sr},  turned out to be very convenient in the numerical analyses. 
%
\section{Hydrodynamic equations \\ for baryon free matter}
\label{section:hydro_eqn}
As the system reaches local thermodynamic equilibrium its further evolution is governed by the conservation laws for energy and momentum, which can be expressed by the formula
\begin{equation}
  \partial_\mu T^{\mu\nu}=0,
  \label{eqn:hydro_con_law}
\end{equation}
where the energy-momentum tensor has the form
\begin{equation}
  T^{\mu\nu}=(\varepsilon + P)\, u^\mu u^\nu - P g^{\mu\nu}.
  \label{eqn:hydro_Tmunu}
\end{equation}
Here $\varepsilon$ is the energy density, $P$ is the pressure, $g^{\mu\nu}$ is the metric tensor (we use the convention where $g^{00}=+1$) and $u^\mu$ is the four-velocity,  
\begin{equation}
  u^\mu=\gamma\,(1,{\bf v}). 
  \label{eqn:hydro_4veloc}
\end{equation}
In Eq. (\ref{eqn:hydro_4veloc}) ${\bf v}$ is the local three-velocity of the fluid and $\gamma$ is the Lorentz factor
\begin{equation}
  \gamma=\left(1-v^2\right)^{-\frac{1}{2}}.
  \label{eqn:hydro_gamma_def}
\end{equation}
In general situations, in the relativistic systems the energy density and pressure depend on the temperature and baryon number density. This requires that the baryon number conservation law,
\begin{equation}
  \partial_\mu\, j^\mu_{\rm B}=0,
  \label{eqn:hydro_con_bar}
\end{equation}
should be considered together with (\ref{eqn:hydro_con_law}). However, in the case of the central rapidity region of ultra-relativistic heavy-ion collisions the dominating degrees of freedom are mesons (initially gluons), hence the net baryon number is very close to zero. In such a case the local value of baryon chemical potential is negligible, $\mu_{\rm B} \approx 0$, and we can express the entropy density $s$, pressure $P$, and energy density $\varepsilon$ by temperature alone. This allows us to use the thermodynamic identities (\ref{eqn:thermo_ident}) and write the conservation laws (\ref{eqn:hydro_con_law}) in the form.
\begin{equation}
  \partial_\mu \left(T\, s\, u^\mu u^\nu \right) = \partial^\nu P.
  \label{eqn:hydro_con_law_2}
\end{equation}
After performing simple transformations shown explicitly in Appendix \ref{section:Bhydro_cov} Eq. (\ref{eqn:hydro_con_law_2}) leads to the two formulas
\begin{eqnarray}
  u^\mu \partial_\mu (T\, u^\nu) &=& \partial^\nu T,	\label{eqn:hydro_con_law_T} \\
  \partial_\mu (s\, u^\mu)       &=& 0.			\label{eqn:hydro_con_law_s}
\end{eqnarray}
Eq. (\ref{eqn:hydro_con_law_T}) is the acceleration equation and represents the relativistic analog of the Euler equation known from the classical hydrodynamics. Eq. (\ref{eqn:hydro_con_law_s}) states that the evolution is adiabatic (entropy is conserved).
In the non-covariant notation the hydrodynamic equations (\ref{eqn:hydro_con_law_T}) and (\ref{eqn:hydro_con_law_s}) are expressed by the following expressions \cite{Baym:1983sr}
\begin{eqnarray}
   \frac{\partial}{\partial t}(T \gamma {\bf v}) + {\bf \nabla} (T \gamma)  &=&  {\bf v} \times \left[ {\bf \nabla} \times (T \gamma {\bf v})\right],							\label{eqn:hydro_noncon_T} \\
  \frac{\partial}{\partial t}(s \gamma) + {\bf \nabla} (s \gamma {\bf v}) &=& 0.	\label{eqn:hydro_noncon_s}
\end{eqnarray}
The four equations above can be written in the Cartesian coordinates in the equivalent form as
\begin{eqnarray}
  (v^2 - 1) \frac{\partial \ln T}{\partial t} + \frac{d \ln T}{dt} + \frac{1}{1 - v^2} v \frac{d v}{dt} &=& 0,
  \label{eqn:hydro_T1_Car} \\
  (1 - v^2) \left( v_y \frac{\partial \ln T}{\partial x} - v_x \frac{\partial \ln T}{\partial y} \right) + v_y \frac{d v_x}{dt} - v_x \frac{d v_y}{dt} &=& 0,
  \label{eqn:hydro_T2_Car} \\
   (1-v^2) \frac{\partial \ln T}{\partial z} + v_z \frac{d \ln T}{dt} + \frac{d v_z}{d t} + \frac{v_z}{1-v^2}\, v \frac{d v}{dt} &=& 0, 
  \label{eqn:hydro_T3_Car} \\
  \frac{d \ln s}{dt} + \frac{v}{1-v^2} \frac{d v}{dt} + \frac{\partial v_x}{\partial x} + \frac{\partial v_y}{\partial y} + \frac{\partial v_z}{\partial z} &=& 0,
  \label{eqn:hydro_s_Car}
\end{eqnarray}
where the total time derivative is defined by the equation
\begin{equation}
\frac{d}{dt} = \frac{\partial}{\partial t} + {\bf v} \cdot {\bf \nabla}.
\label{eqn:hydro_dpodt}
\end{equation}
The details of the transformations leading from (\ref{eqn:hydro_con_law_2}) to (\ref{eqn:hydro_T1_Car}) - (\ref{eqn:hydro_s_Car}) can be found in Appendix \ref{section:Bhydro_noncov}. Equations (\ref{eqn:hydro_T1_Car}) - (\ref{eqn:hydro_s_Car}) do not form a closed system of equations since they contain five independent variables $T$, $s$, $v_x$, $v_y$ and $v_z$. An additional equation is needed to close them, i.e.,  the equation of state is required which introduces the relation between $T$ and $s$. Alternatively, the equation of state may be included by the use of the temperature dependent sound velocity 
\begin{equation}
  c_{\rm s}^2(T) = \frac{\partial P}{\partial \epsilon} = 
  \frac{s}{T}\frac{\partial T}{\partial s},
  \label{eqn:hydro_cs2}
\end{equation}
The temperature dependent sound velocity for strongly interacting matter that is used in our work was discussed thoroughly in Chapt. \ref{chapter:thermo} and is plotted in Fig. \ref{fig:thermo_cs2}.

\section{Implementation of boost-invariance}
\label{section:hydro_eqn_boost}
%
\begin{figure}[!t]
  \begin{center}
    \includegraphics[width=0.49\textwidth]{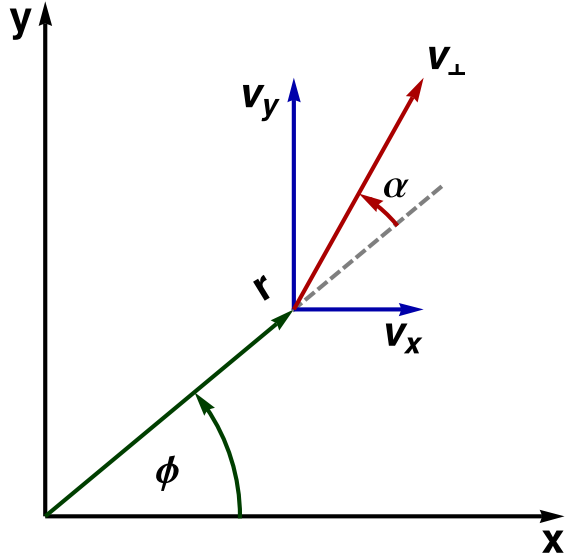}
  \end{center}
  \caption{\small Decomposition of the flow velocity vector in the plane $z=0$. In our approach we use the magnitude of the transverse flow $v_\perp = \sqrt{v_x^2 + v_y^2}$ and the angle $\alpha$ as two independent quantities, isted of $v_x$ and $v_y$. The longitudinal component of the flow \hbox{$v_z = z/t$}, a consequence of boost-invariance.}
  \label{fig:hydro_alpha}
\end{figure}
The experimental data collected at RHIC by the BRAHMS Collaboration \cite{Bearden:2004yx} suggests that in the midrapidity region the particle yields do not vary much with rapidity. Thus, we can assume that number of particles per unit rapidity in the range of $|y| \le 1$ is essentially constant and the midrapidity region (central region) is boost-invariant. This symmetry demands that the longitudinal component of the velocity has the form of the Bjorken flow \cite{Bjorken:1983qr}, 
\begin{equation}
  v_z = \frac{z}{t},
  \label{eqn:hydro_vz_BI}
\end{equation}
and the thermodynamic scalar variables like temperature or entropy density are functions of the longitudinal proper time $\tau=\sqrt{t^2-z^2}$ and the transverse coordinates $x$ and $y$. {\it In practice, these properties mean that  we can solve the hydrodynamic equations for $z=0$ and by using the appropriate Lorentz transformations we obtain solutions for $z \neq 0$}. 

Adopting the procedure outlined above and restricting our considerations to the plane $z=0$ we observe that Eq. (\ref{eqn:hydro_T3_Car}) is automatically fulfilled and we are left with only three independent equations 
\begin{eqnarray}
  (v_\perp^2 - 1) \frac{\partial \ln T}{\partial t} + \frac{d \ln T}{dt} + \frac{1}{1 - v_\perp^2} v_\perp \frac{d v_\perp}{dt} &=& 0,
  \label{eqn:hydro_T1_Car_BI} \\
  (1 - v_\perp^2) \left( v_y \frac{\partial \ln T}{\partial x} - v_x \frac{\partial \ln T}{\partial y} \right) + v_y \frac{d v_x}{dt} - v_x \frac{d v_y}{dt} &=& 0,
  \label{eqn:hydro_T2_Car_BI} \\
  \frac{d \ln s}{dt} + \frac{v_\perp}{1-v_\perp^2} \frac{d v_\perp}{dt} + \frac{\partial v_x}{\partial x} + \frac{\partial v_y}{\partial y} + \frac{1}{t} &=& 0,
  \label{eqn:hydro_s_Car_BI}
\end{eqnarray}
where $v_\perp = \sqrt{v_x^2 + v_y^2}$ is the transverse velocity. The equations describing transverse evolution can be rewritten in the cylindrical coordinates which are convenient for further analysis \cite{Dyrek:1984xz} (details can be found in Appendix \ref{subsection:Bhydro_Teq_BI} and \ref{subsection:Bhydro_seq_BI})
\begin{eqnarray}
  (v_\perp^2 - 1) \,\frac{\partial \ln T}{\partial t} + \frac{d \ln T}{dt} + \frac{1}{1-v_\perp^2}\, v_\perp \frac{d v_\perp}{dt} = 0,
  \nonumber \\ \label{eqn:hydro_T1_Cyl_BI} \\
  (1 - v_\perp^2) \left( v_\perp \sin \alpha \,\frac{\partial \ln T}{\partial r} - \frac{v_\perp \cos \alpha}{r} \,\frac{\partial \ln T}{\partial \phi}\right) - v_\perp^2 \left( \frac{d \alpha}{dt} + \frac{v_\perp \sin \alpha}{r} \right) = 0,
  \nonumber \\ \label{eqn:hydro_T2_Cyl_BI} \\
   v_\perp \frac{d \ln s}{dt} + \frac{1}{1-v_\perp^2} \frac{d v_\perp}{dt} - \frac{\partial v_\perp}{\partial t} - v_\perp^2 \sin \alpha \frac{\partial \alpha}{\partial r} + \frac{v_\perp^2 \cos \alpha}{r} \left( \frac{\partial \alpha}{\partial \phi} + 1 \right) + \frac{v_\perp}{t} = 0.
  \nonumber \\ \label{eqn:hydro_s_Cyl_BI}
\end{eqnarray}
Here $r$ is the distance from the beam axis, $r = \sqrt{x^2+y^2}$, and $\phi$ is the azimuthal angle, $\phi = \tan^{-1} (y/x)$. These two coordinates parameterize the plane $z=0$. The angle $\alpha$ is the function describing direction of the flow, $\alpha=~\tan^{-1}(v_y/v_x) - \phi$, see Fig.~\ref{fig:hydro_alpha}. The differential operator $d/dt$ used in Eqs. (\ref{eqn:hydro_T1_Cyl_BI}) - (\ref{eqn:hydro_s_Cyl_BI}) is defined by the formula
\begin{equation}
  \frac{d}{dt} = \frac{\partial}{\partial t} 
  + v_\perp \cos \alpha \frac{\partial}{\partial r} 
  + \frac{v_\perp \sin \alpha}{r} \frac{\partial}{\partial \phi}.
\end{equation} 
%
\section{Characteristic form of hydrodynamic\\ equations}
\label{subsection:hydro_numform}
%
In order to obtain the form of the hydrodynamic equations which is convenient for numerical studies we introduce  new independent variables. In this respect we follow the method originally proposed by Baym et al. in Ref. \cite{Baym:1983sr} and the new form of the equations will be called the characteristic form.  We introduce the potential $\Phi$ defined as
\begin{equation}
  d\Phi = \frac{1}{c_{\rm s}}\, d\ln T = c_{\rm s}\, d\ln s,
  \label{eqn:hydro_dPhi}
\end{equation}
and the transverse rapidity $\eta_\perp = \tanh^{-1} v_\perp$.  The new form of the hydrodynamic equations is expressed by the dimensionless auxiliary functions $A_+$ and $A_-$ defined by the equations
\begin{equation}
  A_\pm = \Phi \pm \eta_\perp.
  \label{eqn:hydro_Apm_def}
\end{equation}
The sum and difference of Eq. (\ref{eqn:hydro_T1_Cyl_BI}) and (\ref{eqn:hydro_s_Cyl_BI}) together with Eq. (\ref{eqn:hydro_T2_Cyl_BI}) take the form 
\begin{eqnarray}
  \frac{\partial A_\pm}{\partial t}
  &+& \frac{v_\perp \pm c_{\rm s}}{1 \pm c_{\rm s} v_\perp} \left[
    \cos \alpha			\,\frac{\partial A_\pm}{\partial r}
    + \frac{\sin \alpha}{r}	\,\frac{\partial A_\pm}{\partial \phi} 
  \right] \nonumber \\
  &-& \frac{c_{\rm s}}{1 \pm c_{\rm s} v_\perp} \left[
    v_\perp \sin \alpha		\,\frac{\partial \alpha}{\partial r} 
    - \frac{v_\perp \cos \alpha}{r}	\left( \frac{\partial \alpha }{\partial \phi } + 1 \right)
    - \frac{1}{t} \right] = 0,
  \label{eqn:hydro_Apm} \\
  \nonumber \\
  \frac{\partial \alpha }{\partial t}
  &+&v_\perp \cos \alpha				\,\frac{\partial \alpha }{\partial r}
   + \frac{v_\perp \sin \alpha }{r} \left(	\,\frac{\partial \alpha }{\partial \phi } + 1 \right)
  \nonumber \\
  &-&\frac{c_{\rm s} (1-v_\perp^2)}{v_\perp} \left[
    \sin \alpha			\,\frac{\partial \Phi }{\partial r}
    - \frac{\cos \alpha }{r}	\,\frac{\partial \Phi }{\partial \phi }
  \right] = 0,
  \label{eqn:hydro_alpha}
\end{eqnarray}
where the transverse velocity and the $\Phi$ potential are expressed through the auxiliary functions $A_\pm$ as follows
\begin{eqnarray}
  v_\perp = \tanh \left( \frac{A_+ - A_-}{2}\right), \quad \Phi = \frac{A_+ + A_-}{2}.
  \label{eqn:hydro_vPhiApm}
\end{eqnarray}
Temperature and all other temperature dependent variables, e.g., the sound velocity, can be also calculated from the functions $A_\pm$,
\begin{eqnarray}
  T = T_\Phi(\Phi) = T_\Phi \left( \frac{A_+ + A_-}{2}\right),
  \label{eqn:hydro_TApm} \\
  c_{\rm s}(T) = c_{\rm s} \left[ T_\Phi \left( \frac{A_+ + A_-}{2}\right)\right].
  \label{eqn:hydro_csApm}
\end{eqnarray}
{\it Here we have introduced the subscripts to make clear what kind of the argument is expected for a given function.} For example, the temperature may be considered as a function of entropy density or $\Phi$. In those two cases one should use the functions $T_S$ or $T_\Phi$, respectively. If the equation of state is known all such functions can be easily calculated and Eqs. (\ref{eqn:hydro_Apm}) and (\ref{eqn:hydro_alpha}) may be used to determine three unknown functions $A_+$, $A_-$, and $\alpha$.

%
\section{Boundary conditions}
\label{subsection:hydro_bond}
%
\begin{figure}[t]
  \begin{center}
    \includegraphics[width=0.6\textwidth]{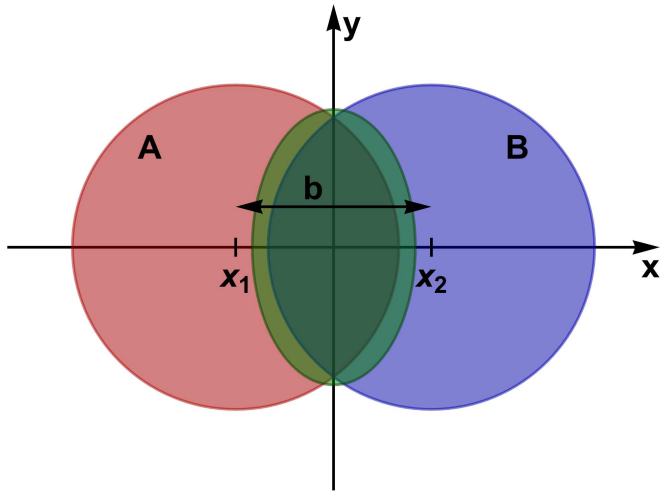}
  \end{center}
  \caption{\small Non-central collision of two identical nuclei at the impact vector {\bf b}, viewed in the transverse plane. The nucleus $A$ is located at $(x_1,y_1) = (-b/2,0)$ and the nucleus $B$ at $(x_2,y_2) = (b/2,0)$. The overlapping region has the shape of an almond elongated in the direction of the $y$-axis.}
  \label{fig:hydro_cent}
\end{figure}
\begin{figure}[!hbt]
  \begin{center}
    \includegraphics[width=0.7\textwidth]{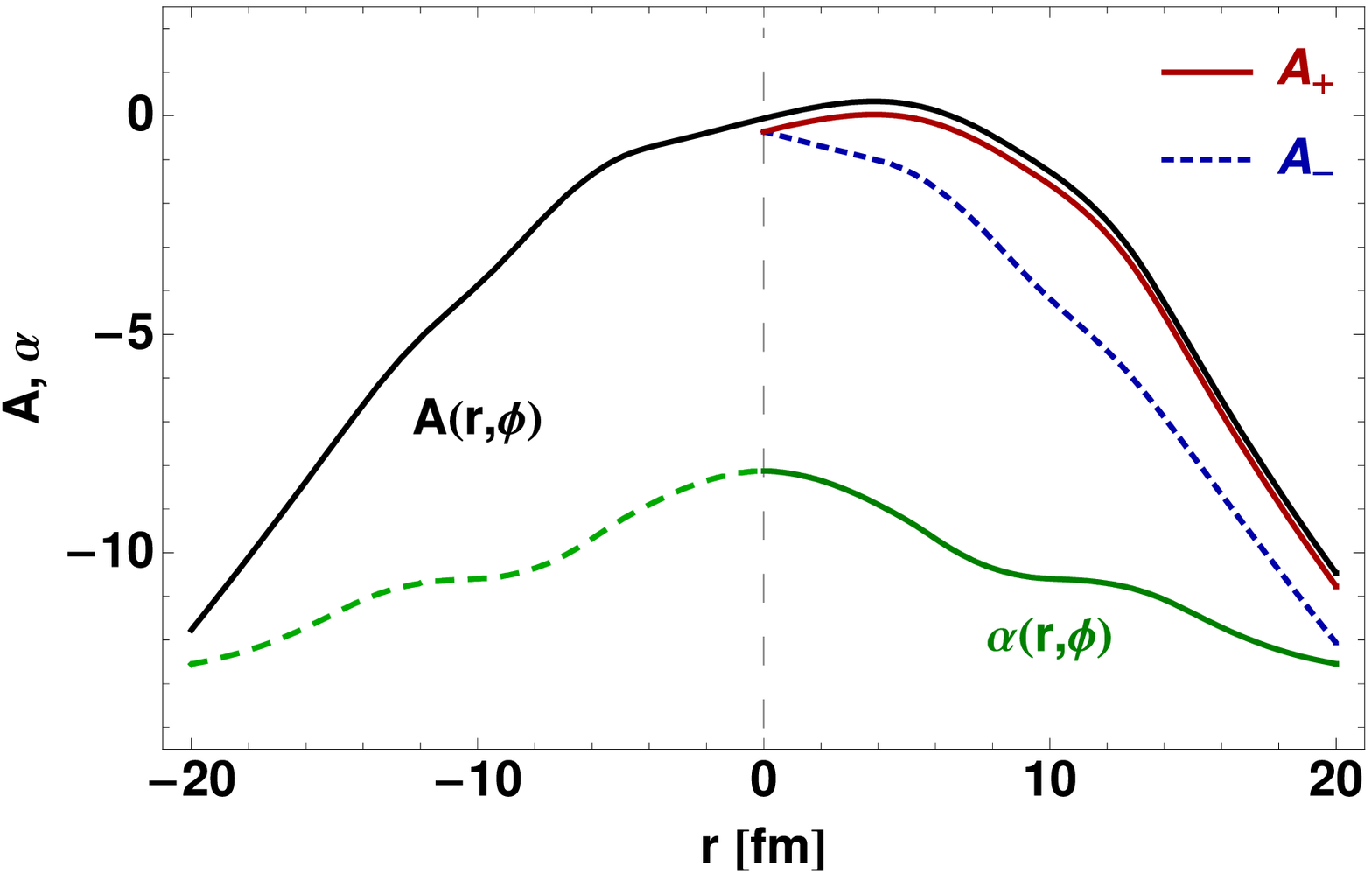}
  \end{center}
  \caption{\small Construction of the functions $A_\pm(t,r,\phi)$ in terms of a single function $A(t,r,\phi)$, see Eq. (\ref{eqn:hydro_Apm2A}). The function $\alpha(t,r,\phi)$ for negative values of $r$ is obtained from the symmetry condition
$\alpha(t,-r,\phi)=\alpha(t,r,\phi)$.}
  \label{fig:hydro_bound}
\end{figure}

Having in mind the heavy-ion collisions at RHIC and at LHC we consider the collisions of identical nuclei, Au+Au or Pb+Pb, which collide moving initially along the $z$-axis. The positions of the centers of nuclei depend on the impact parameter ${\bf b}$ and for non-central collision they may be located in the transverse plane at \mbox{${\bf x}_1 = (x_1,y_1)=(-b/2,0)$} and at ${\bf x}_2 = (x_2,y_2)=(b/2,0)$, see Fig. \ref{fig:hydro_cent}. The distribution of matter created just after the collision is not cylindrically symmetric and has an ellipsoidal shape (one commonly speaks of an almond shape). With the use of the coordinate system explained above the transverse velocity must vanish  at the origin of the system namely
\begin{equation}
  v_\perp(t,r=0,\phi) = 0.
  \label{eqn:hydro_vT_bound}
\end{equation}
Also the gradients with respect to the distance $r$ of temperature and entropy density must converge to zero at $r=0$
\begin{equation}
  \frac{\partial T(t,r,\phi)}{\partial r} \stackrel{r \rightarrow 0}{\longrightarrow} 0, \quad
  \frac{\partial s(t,r,\phi)}{\partial r} \stackrel{r \rightarrow 0}{\longrightarrow} 0.
  \label{eqn:hydro_dr_bound}
\end{equation}
The boundary conditions (\ref{eqn:hydro_vT_bound}) and (\ref{eqn:hydro_dr_bound}) can be naturally fulfilled by the following Ansatz
\begin{eqnarray}
 \left\{ \begin{array}{llll} 
   A_+(t,r,\phi) &=& A(t,r,\phi),		& \\
   A_-(t,r,\phi) &=& A(t,-r,\phi),		& \\
   \alpha(t,-r,\phi) &=& \alpha(t,r,\phi).	& \\
 \end{array} \right. r > 0
\label{eqn:hydro_Apm2A}
\end{eqnarray}
The domain of the transverse distance $r$ has been extended to the negative values of $r$. The two functions $A_+$ and $A_-$ are replaced in this way by a single function $A$. At the same time the function $\alpha$ has been generalized to have negative arguments by the condition that it is a symmetric function of $r$, see Fig. \ref{fig:hydro_bound}. 
With the help of the definitions (\ref{eqn:hydro_Apm2A}), Eqs. (\ref{eqn:hydro_Apm}) may be reduced to a single equation for the function $A(t,r,\phi)$,
\begin{eqnarray}
  \frac{\partial A}{\partial t}
  &+& \frac{v_\perp + c_{\rm s}}{1 + c_{\rm s} v_\perp} \left[
    \cos \alpha			\,\frac{\partial A}{\partial r}
    + \frac{\sin \alpha}{r}	\,\frac{\partial A}{\partial \phi} 
  \right] \nonumber \\
  &-& \frac{c_{\rm s}}{1 + c_{\rm s} v_\perp} \left[
    v_\perp \sin \alpha		\,\frac{\partial \alpha}{\partial r} 
    - \frac{v_\perp \cos \alpha}{r}	\left( \frac{\partial \alpha }{\partial \phi } + 1 \right)
    - \frac{1}{t} \right]  = 0.
  \label{eqn:hydro_A}
\end{eqnarray}
The transverse velocity and $\Phi$ potential from Eq. (\ref{eqn:hydro_vPhiApm}) are now expressed as
\begin{eqnarray}
  v_\perp(t,r,\phi) &=& \tanh \left( \frac{A(t,r,\phi) - A(t,-r,\phi)}{2}\right), \label{eqn:hydro_vTA}\\
 \Phi(t,r,\phi) &=& \frac{A(t,r,\phi) + A(t,-r,\phi)}{2}. \label{eqn:hydro_PhiA}
\end{eqnarray}
In addition, the use of cylindrical coordinates implies the periodicity of all functions in angle $\phi$ thus creating another set of periodic boundary conditions
\begin{eqnarray}
 \left\{ \begin{array}{lll} 
   A(t,r,\phi = 0) &=& A(t,r,\phi = 2\pi)\\
   \alpha(t,r,\phi = 0) &=& \alpha(t,r,\phi = 2\pi)
 \end{array} \right. .
\label{eqn:hydro_Aperiod}
\end{eqnarray}

\chapter{Initial conditions}
\label{chapter:initial}
%
Along with the formulation of the hydrodynamic equations that govern the evolution of matter we must also specify a set of initial conditions that are required to unambiguously solve such equations -- the hydrodynamic equations are first-order partial differential equations. The physical systems studied here are formed in the collisions of two identical gold nuclei (RHIC experiment) or two identical lead nuclei (the future heavy-ion program at LHC). In this Chapter we discuss in more detail the initial conditions used by us to analyze the collisions at RHIC and LHC. At first we discuss the most common form of the initial conditions that is based on the Glauber model. Next, we present the modified initial conditions where the initial energy density has a form of the two-dimensional Gaussian in the transverse plane. Finally, we present our method of including the parton free-streaming as a process which precedes the  hydrodynamic evolution. In particular, we show how to match the free-streaming stage with the hydrodynamic stage. 

\section{Standard initial conditions}
\label{section:initial_Glauber}
%
In the calculations presented in Chapter \ref{chaper:resrhic} and \ref{chaper:reslhc} we assume that the initial entropy density of the particles produced at the transverse position point ${\bf x}_{\bot }$ is proportional to a source profile obtained from the Glauber approach. This is done in the similar way to other hydrodynamic calculations. Specifically, for the particle source profile we use a mixed model \cite{Back:2001xy,Back:2004dy}, with a linear combination of the wounded-nucleon density $\rho_{\rm WN}\left( {\bf x}_{\bot } \right)$ and the density of binary collisions $\rho_{\rm BC}\left( {\bf x}_{\bot } \right)$, namely 
\begin{equation}
  s ({\bf x}_\bot) \propto \rho ({\bf x}_\bot) = 
  \frac{1-\kappa}{2}\, \rho_{\rm WN} ({\bf x}_\bot) + \kappa\, \rho_{\rm BC} ({\bf x}_\bot). 
  \label{eqn:initial_eps2}
\end{equation}
The case $\kappa=0$ corresponds to the standard wounded-nucleon model \cite{Bialas:1976ed}, while $\kappa=1$ would include  the binary collisions only. The PHOBOS analysis \cite{Back:2004dy} of the particle multiplicities yields $\kappa=0.12$ at $\sqrt{s_{\rm NN}}=17~{\rm GeV}$ and $\kappa=0.14$ at $\sqrt{s_{\rm NN}}=200~{\rm GeV}$. Since the density profile from the binary collisions is steeper than from the wounded nucleons, increased values of $\kappa$ yield steeper density profiles, which in turn result in steeper temperature profiles. 
\par In our hydrodynamic code, the initial conditions are specified for the temperature profile which takes the form
\begin{equation}
  T (\tau_{\rm i},{\bf x}_\bot) = T_S \left[ s_{\rm i} \frac{ \rho({\bf x}_\bot) }{ \rho(0) } \right],
  \label{eqn:initial_Tt0}
\end{equation}
where $T_S(s)$ is the inverse function to the function $s(T)$, and $s_{\rm i}$ is the initial entropy at the center of the system. The initial central temperature $T_{\rm i}$ equals $T_S(s_{\rm i})$. Throughout this work, the initial time for the start of the hydrodynamic evolution is denoted by $\tau_{\rm i}$.
\par The wounded-nucleon and the binary-collisions densities in Eq.~(\ref{eqn:initial_Tt0}) are obtained from the optical limit of the Glauber model, which is a very good approximation for not too peripheral collisions \cite{Miller:2007ri}. The standard formulas are \cite{Bialas:1976ed}
\begin{eqnarray}
  \rho_{\rm WN} \left( {\bf x}_\bot \right) &=& 
    T_A\left(  \frac{{\bf b}}{2} + {\bf x}_\bot \right) \left\{
      1 - \left[1 - \frac{\sigma_{\rm in}}{A}\, T_A\left( -\frac{{\bf b}}{2} + {\bf x}_\bot \right) \right]^A 
    \right\} \nonumber \\
    &+& T_A\left( -\frac{{\bf b}}{2} + {\bf x}_\bot \right) \left\{
      1 - \left[1 - \frac{\sigma_{\rm in}}{A}\, T_A\left(  \frac{{\bf b}}{2} + {\bf x}_\bot \right) \right]^A 
    \right\}
  \label{eqn:initial_rhoWN}
\end{eqnarray}
and
\begin{equation}
  \rho_{\rm BC} \left( {\bf x}_\bot \right) = \sigma_{\rm in}\,
    T_A\left( \frac{{\bf b}}{2} + {\bf x}_\bot \right)
    T_A\left(-\frac{{\bf b}}{2} + {\bf x}_\bot \right).
  \label{eqn:initial_rhoBC}
\end{equation}
In Eqs. (\ref{eqn:initial_rhoWN}) and  (\ref{eqn:initial_rhoBC}) ${\bf b}$ is the impact vector, $\sigma_{\rm in}$ is the nucleon-nucleon total inelastic cross section, and $T_A\left(x,y\right)$ is the nucleus thickness function 
\begin{equation}
  T_A(x,y) = \int dz\, \rho\left(x,y,z\right).
  \label{eqn:initial_TA}
\end{equation}
For RHIC energies we use the value $\sigma_{\rm in} = 42~{\rm mb}$, while for LHC we take $\sigma_{\rm in} = 63~{\rm mb}$. The function $\rho(r)$ in Eq. (\ref{eqn:initial_TA}) is the nuclear density profile given by the Woods-Saxon function with the conventional choice of the parameters: 
\begin{eqnarray}
  \rho_0 &=& 0.17\, \hbox{fm} ^{-3},			\nonumber \\
     r_0 &=& (1.12 A^{1/3} -0.86 A^{-1/3})\, \hbox{fm},	\nonumber \\
       a &=& 0.54\, \hbox{fm}.
  \label{eqn:initial_woodssaxon}
\end{eqnarray}
The values of the atomic mass $A$ are: 197 for RHIC (gold nuclei) and 208 for LHC (lead nuclei). The value of the impact parameter in Eqs. (\ref{eqn:initial_rhoWN}) and  (\ref{eqn:initial_rhoBC}) depends on the considered centrality class.
\par Besides the initial temperature profile (\ref{eqn:initial_Tt0}) we also specify the initial transverse flow profile,
\begin{eqnarray}
  v (\tau_{\rm i},r,\phi) &=& \frac{H_0\, r}{\sqrt{1 + H_0^2\, r^2}}, \nonumber \\
  \alpha(\tau_{\rm i},r,\phi) &=& 0.
  \label{eqn:initial_vt0}
\end{eqnarray} 
The results presented below are obtained with \mbox{$H_0 = 0.001$ fm$^{-1}$}. The very small value of the parameter $H_0$ means that the system is practically at rest at the moment when the hydrodynamic evolution starts. However, nonzero $H_0$ improves the stability of the numerical method.
\par In our numerical calculations we use the adaptive method of lines in the way as it is implemented in the MATHEMATICA package. The $\phi$ and $r$ directions are discretized, and the integration in time is treated as solving of the system of ordinary differential equations. Typically, we use grids with $\Delta r$ = 0.25 fm in the $r$ direction and $\Delta \phi = $ 6 degrees in the $\phi$ direction. This method fails in the case where the shock waves are formed, however with our regular equation of state and the regular initial conditions such shocks are not present. 
\par We stress that the shape of the initial condition (\ref{eqn:initial_Tt0}) is important, as it determines the development of the radial and elliptic flow, thus affecting such observables as the \mbox{$p_T$-spectra}, $v_2$, and the femtoscopic features. On qualitative grounds, sharper profiles lead to more rapid expansion. Several effects should be considered here. Firstly, as discussed in Ref.~\cite{Kolb:2000sd}, hydrodynamics may start a bit later, when the profile is less eccentric than originally due to initial free-streaming of partons in the pre-hydro phase. On the other hand, statistical fluctuations in the distribution of the Glauber sources (wounded nucleons, binary collisions) \cite{Aguiar:2000hw,Aguiar:2001ac,Voloshin:2006gz,Broniowski:2007ft,Broniowski:2007nz,Voloshin:2007pc,Alver:2008zza} lead to a significant enhancement of the eccentricity, especially at low values of the impact parameter. Thus the initial eccentricity may in fact be smaller or larger than what follows from the application of the Glauber model. This contributes to the systematic model uncertainty at the level of about 10-20\%. This uncertainty could only be reduced by the employment of a realistic model of the pre-hydrodynamic evolution. With this uncertainty in place, one should not expect or demand a better agreement with the physical observables than at the corresponding level of 10-20\%.\\
\medskip\\
{\bf The results obtained with the introduced here standard initial conditions are presented and discussed in Chapters \protect\ref{chaper:resrhic} and \protect\ref{chaper:reslhc}.}
%
\section{Gaussian initial conditions}
\label{section:initial_Gauss}
%
The situation described in the previous Section corresponds to the typical case where the hydrodynamic evolution is initiated from a source profile generated by the Glauber model, with the initial central temperature or energy serving as a free parameter. The initial density and possibly the initial flow profiles may be also provided by the early partonic dynamics, for instance by the Color Glass Condensate (CGC) \cite{McLerran:1993ka,McLerran:1993ni}. In practice, however, the theory of the partonic stage carries some uncertainty in its parameters, which influences our knowledge of the details of early dynamics. Having in mind such uncertainties we decided to try another class of the initial conditions, i.e., in our calculations we also  include  the case where the initial energy profiles in the transverse plane have the Gaussian shape.  
\par The modified Gaussian parameterization of the initial energy density profile at the initial proper time $\tau_{\rm i}$ has the following form
\begin{eqnarray}
  \varepsilon({\bf x}_\bot) = \varepsilon_{\rm i} \exp \left( -\frac{x^2}{2a^2} -\frac{y^2}{2 b^2} \right).
  \label{eqn:initial_profile}
\end{eqnarray} 
The energy density profile (\ref{eqn:initial_profile}) determines the initial temperature profile that is used in the hydrodynamic code, 
\begin{equation}
  T (\tau_{\rm i},{\bf x}_\bot) = T_{\rm \varepsilon} \left[ \varepsilon_{\rm i} \exp \left( -\frac{x^2}{2a^2} -\frac{y^2}{2 b^2} \right)\right],
  \label{eqn:initial_Tt02}
\end{equation}
The initial central temperature $T (\tau_{\rm i},0)$, which may depend on the centrality, is denoted by $T_{\rm i}$ and is a free parameter of our approach. \\
\medskip\\
{\bf The results obtained with such initial conditions are presented and discussed in Sect. \ref{section:resrhic2-early-start}.}
%
\section{Free streaming}
\label{section:initial_free_stream}
%
The very early start of hydrodynamics means that the matter equilibrates very fast. Such a short thermalization time is difficult to explain on the microscopical grounds and inspires hot discussion about the nature of matter produced at RHIC. To avoid the problem with the early thermalization we consider also the scenario where the hydrodynamic evolution is preceded by the free-streaming stage. In this version of our calculations we assume that partons behave as free particles in the proper time interval $0.25 \leq \tau \leq 1$~fm. At $\tau_{\rm i} = 1$ fm the sudden equilibration takes place which is described with the help of the Landau matching conditions. The global picture is as follows: early phase (CGC) generating partons at time $\tau_0 = 0.25$ fm, partonic free streaming until $\tau_{\rm i} =1$ fm, hydrodynamic evolution until the freeze-out at temperature $T_{\rm f}$, free streaming of hadrons and decay of resonances. We note that similar ideas have been described by Sinyukov et al. in Refs.~\cite{Sinyukov:2006dw,Gyulassy:2007zz}.
\par In the remaining part of this Section we describe in more detail the free-streaming stage. We assume that massless partons are formed at the initial proper time  \mbox{$\tau_0=\sqrt{t_0^2-z_0^2}$} and move along straight lines at the speed of light until the proper time when free streaming ends, \mbox{$\tau=\sqrt{t^2-z^2}$}. We introduce the space-time rapidities 
\begin{equation}
  \eta_0=\frac{1}{2}\log\frac{t_0-z_0}{t_0+z_0}
  \label{eqn:initial_eta0}
\end{equation}
and
\begin{equation}
  \eta_\parallel=\frac{1}{2}\log{\frac{t-z}{t+z}}.
  \label{eqn:initial_etai}
\end{equation}
Elementary kinematics \cite{Sinyukov:2006dw} links the positions of a parton on the initial and final hypersurfaces and its four-momentum 
\begin{equation}
  p^\mu=(p_T \cosh y, p_T \cos \phi_p, p_T \sin \phi_p, p_T \sinh y),
  \label{eqn:initial_pmu_fs}
\end{equation}
where $y$ and $p_T$ are the parton's rapidity and transverse momentum:
\begin{equation}
  \begin{array}{l}
    \tau \sinh(\eta_\parallel - y)=\tau_0 \sinh(\eta_0 - y), \\
    x = x_0 + \Delta \cos \phi_p,	\;\;	y = y_0 + \Delta \sin \phi_p, \\
    \Delta = \tau\, \cosh(y-\eta_\parallel) - \sqrt{\tau_0^2 + \tau^2 \sinh^2(y-\eta_\parallel)}.
  \end{array}
  \label{eqn:initial_kinem}
\end{equation}
Thus the phase-space density of partons at the proper times $\tau_0$ and $\tau$ are related 
\begin{eqnarray}
  \frac{d^6N}{dy d^2p_T d\eta_\parallel dx dy} &=& \int d\eta_0 dx_0 dy_0 \frac{d^6N}{dy d^2p_T d\eta_0 dx_0 dy_0} \times	\nonumber \\ 
  & & \delta \left(\eta_0 - y - {\rm arcsinh} \left[ \frac{\tau}{\tau_0} \sinh(\eta_\parallel - y)\right] \right) \times 
  \label{eqn:initial_fs} \\
  & & \delta(x - x_0 - \Delta \cos \phi_p)\, \delta( y - y_0 - \Delta \sin \phi_p).
  \nonumber
\end{eqnarray}  
In our approach we restrict ourselves to the boost-invariant systems, hence it is reasonable to assume the following factorized form of the initial distribution of partons, 
\begin{eqnarray}
 \frac{d^6N}{dy d^2p_T d\eta_0 dx_0 dy_0} = n(x_0,y_0) F(y-\eta_0,p_T),
\end{eqnarray}
where $n$ is the transverse density of partons obtained again from the GLISSANDO model, namely
\begin{eqnarray}
  n(x_0,y_0)= n_0 \exp \left( -\frac{x_0^2}{2a^2} -\frac{y_0^2}{2 b^2} \right).
  \label{eqn:initial_profile_n}
\end{eqnarray} 
When the rapidity emission profile $F$ is focused near $y=\eta_0$, for instance if we have  \mbox{$F \sim \exp[(y-\eta_0)^2/(2\, \delta y^2)]$}, with $\delta y \sim 1$, and if $\tau \gg \tau_0$, then  the kinematic condition (\ref{eqn:initial_kinem}) effectively transforms it into
\begin{eqnarray}
  F \sim \exp \left( \frac{{\rm arcsinh}^2 \left[ \frac{\tau}{\tau_0} \sin(y-\eta_\parallel) \right]}{2\, \delta y^2} \right) \sim \delta(y-\eta_\parallel).
\end{eqnarray}
Then Eq.~(\ref{eqn:initial_fs}) yields
\begin{eqnarray}
\frac{d^6N}{dy d^2p_T d\eta_\parallel dx dy} = 
n (x - \Delta \tau \cos \phi_p, y - \Delta \tau \sin \phi_p) \delta(y-\eta_\parallel) f(p_T), 
\end{eqnarray}
where $\Delta \tau = \tau -\tau_0$ and $f(p_T)$ is the transverse momentum distribution. 
\begin{figure}[t]
  \begin{center}
    \includegraphics[width=1.0 \textwidth]{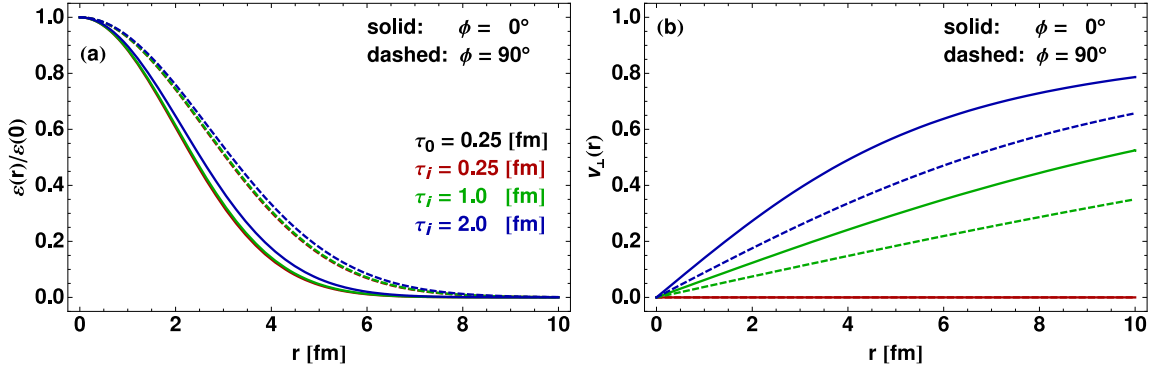}
  \end{center}
  \vspace{-4mm}
  \caption{\small Sections of the energy-density profile $\varepsilon$ (Gaussian-like curves) normalized to unity at the origin, and of the velocity profile $v_\perp=\sqrt{v^2_x+v^2_y}$ (curves starting at the origin), cut along the $x$ axis (solid lines) and $y$-axis (dashed lines). The initial profile is from Eq.~(\ref{eqn:initial_profile_n}) for centrality 20-40\% at $\tau=\tau_0$. The $\varepsilon$ profiles are for $\tau=\tau_0=0.25$, $1$, and $2$~fm, and the velocity profiles are for $\tau = 0.25, 1$ and $2$~fm, all from bottom to top. We note that the flow is azimuthally asymmetric and stronger along the $x$ axis.}
  \label{fig:initial_fs}
\end{figure}
The energy-momentum tensor at the proper time $\tau$, rapidity $\eta_\parallel$, and transverse position $(x,y)$ is given by the formula
\begin{eqnarray}
  & & T^{\mu \nu} = \int dy d^2p_T  \frac{d^6N}{dy d^2p_T d\eta_\parallel dx dy} p^\mu p^\nu 				\label{eqn:initial_tmunu} \\
  & & = \frac{\varepsilon_0}{2 \pi n_0} \int_0^{2 \pi} d\phi_p\, n \left(x - \Delta \tau \cos \phi_p,y - \Delta \tau \sin \phi_p\right) \times	\nonumber \\
  & & \left( \begin{array}{cccc} 
    \cosh^2 \eta_\parallel            & \cosh  \eta_\parallel \cos  \phi_p & \cosh  \eta_\parallel \sin  \phi_p & \cosh   \eta_\parallel \sinh \eta_\parallel \\ 
    \cosh   \eta_\parallel \cos  \phi_p & \cos^2 \phi_p            & \cos   \phi_p \sin  \phi_p & \cos    \phi_p \sinh \eta_\parallel \\ 
    \cosh   \eta_\parallel \sin  \phi_p & \cos   \phi_p \sin  \phi_p & \sin^2 \phi_p            & \sin    \phi_p \sinh \eta_\parallel \\
    \cosh   \eta_\parallel \sinh \eta_\parallel & \cos   \phi_p \sinh \eta_\parallel & \sin   \phi_p \sinh \eta_\parallel & \sinh^2 \eta_\parallel 
  \end{array} \right ), \nonumber 
\end{eqnarray}
where the factor $\varepsilon_0/(2 \pi n_0)$ is a constant from the $p_T$ integration \footnote{Note that the structure of Eq. (\ref{eqn:initial_tmunu}) implies that for $\Delta=0$ the initial energy density given by $T^{00}$ and the initial parton density $n$ have the same profiles in the transverse plane.}  Due to boost invariance the further calculation may be carried for simplicity of notation at $\eta_\parallel=0$. Next, we assume that at the proper time $\tau$ the system equilibrates rapidly. We thus use the Landau matching condition,  
\begin{eqnarray}
  T^{\mu \nu}(x,y) u_\nu(x,y) =\varepsilon(x,y) g^{\mu \nu} u_\nu(x,y), 
  \label{eqn:initial_landau}
\end{eqnarray}
to determine the position dependent four-velocity of the fluid,
\begin{equation}
  u^\mu(\tau,x,y) = \frac{1}{\sqrt{1-v_\bot^2}} (1,v_x,v_y,0)
  \label{eqn:initial_vel}
\end{equation}
and its position dependent energy density $\varepsilon(\tau,x,y)$, which is then identified with the energy-density profile. The obtained in this way the flow and energy profiles are used as an input for the hydrodynamic calculations. The time $\tau$ coincides with the starting time for hydrodynamics, $\tau = \tau_{\rm i}$, and we again use the formula
\begin{equation}
  T (\tau_{\rm i},{\bf x}_\bot) = T_{\rm \varepsilon} \left[ \varepsilon(\tau_{\rm i},0,0) \right]
  \label{eqn:initial_Tt03}
\end{equation}
to define the initial temperature profile in the hydrodynamic code. The initial central temperature $T_{\rm i} = T_{\rm \varepsilon}(\varepsilon_{\rm i})$ is again used as a free parameter (connected with the freedom of choosing $\varepsilon_0$). 
\par The results of solving Eq.~(\ref{eqn:initial_landau}) with $T^{\mu \nu}$ from (\ref{eqn:initial_tmunu}) for $\Delta \tau =0.75$~fm and the initial profile (\ref{eqn:initial_profile_n}) are shown in Fig.~\ref{fig:initial_fs}. The presented results  correspond to the centrality class 20-40\%. The curves normalized to unity at the origin, $r=0$, show the sections along the $x$ and $y$ axes of the energy-density profile at the proper times $\tau_0$ (no free streaming) and $\tau$. Obviously, at $\tau_0$ we find $\varepsilon(x,y) = \varepsilon_0 n(x,y)/n_0$. We note that the profile spreads out as the time progresses. Importantly, this effect is faster along the shorter axes, $x$. This is clearly indicated by the velocity profiles along the $x$ and $y$ axes (the curves starting from 0 at $r=0$). Thus the flow generated by free streaming and sudden equilibration is azimuthally asymmetric. This asymmetry is also seen from the expansion in the parameters $\Delta \tau, x,y$, where straightforward algebra gives (for $x \Delta \tau << a^2$ and $y\Delta \tau << b^2$)
\begin{equation}
  {\bf v}(x,y) = -\frac{\Delta \tau}{3} \frac{\nabla n(x,y)}{n(x,y)} = \frac{\Delta \tau}{3} \left( \frac{x}{a^2}, \frac{y}{b^2} \right).
\end{equation}
\medskip\\
{\bf The results of the hydrodynamic calculations preceded by the free-streaming stage are presented and discussed in Section \ref{section:resrhic2-late-start}.}

\chapter{Freeze-out prescription}
\label{chapter:freeze}
%
As the system expands, its density decreases and the mean free path of particles becomes larger and larger. When the average time between the collisions becomes compatible with the characteristic time of the system expansion, that is given approximately by the inverse of the divergence of four-velocity, $\tau_{\rm exp} \sim (\partial_\mu u^\mu)^{-1}$, the hadrons decouple and may be regarded as free particles. The hydrodynamic stage of the evolution of matter is then changed into the free streaming of hadrons which are later registered by various detectors.  The transition from strongly interacting hydrodynamic system into the non-interacting system of hadrons is called the freeze-out process. In fact one can distinguish two types of freeze-outs: the chemical freeze-out (when the inelastic  processes stop and the hadronic abundances are frozen) and the kinetic freeze-out (when all interactions stop and the momentum spectra are frozen). Of course, the kinetic freeze-out may be regarded as the true freeze-out, after which all interactions cease. The concept of the chemical freeze-out is however very useful since this is the stage when the hadronic abundances are fixed (by definition, the subsequent evolution is dominated by elastic collisions which do not change the hadronic yields). 
\par In our approach we use the simplification that the two freeze-outs discussed above happen at the same time. This approach is known as the single-freeze-out assumption and several earlier studies showed that it leads to the correct description of the data \cite{Baran:2003nm,Broniowski:2001we,Broniowski:2001uk,Broniowski:2002nf,Prorok:2007xp,Prorok:2006ve}. Of course in the hydrodynamic approach the freeze-out happens locally, i.e., at different times for different space positions. In our studies we assume that freeze-out takes place at a given final temperature $T_{\rm f}$. At the beginning of the evolution the hadrons are emitted usually from the edge (surface) of the system where the temperature is always low. At the end of the evolution, the hadrons are emitted mainly from the whole volume, since the temperature eventually drops down to $T_{\rm f}$ everywhere due to the hydrodynamic expansion.  
%
\section{Cooper-Frye formula}
\label{section:freeze_Cooper}
%
The collection of the spacetime points where the freeze-out process takes place is called the freeze-out hypersurface. This hypersurface is a three dimensional manifold in the four-dimensional Minkowski space. The approach which is used to calculate the spectra of the emitted hadrons is based on the Cooper-Frye formula \cite{Cooper:1974mv,Cooper:1974ak}
\begin{equation}
\frac{dN}{dy d^2p_\perp} = \int d\Sigma_\mu p^\mu f(p \cdot u).
\label{eqn:freeze_cf-for}
\end{equation}
In Eq. (\ref{eqn:freeze_cf-for}) $d\Sigma_\mu$ is the element of the freeze-out hypersurface and $f$ is the equilibrium distribution function depending on the scalar product of the particle four-momentum $p^\mu$ and the fluid four-velocity $u^\mu$.
\begin{figure}[!t]
  \begin{center}
    \includegraphics[width=1.0 \textwidth]{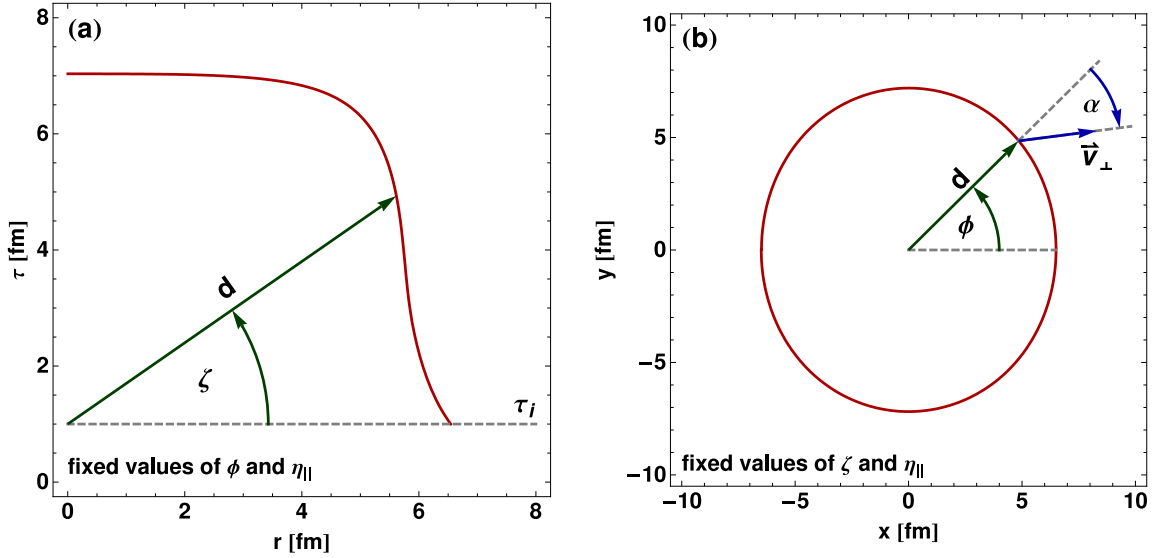}
  \end{center}
  \caption{\small Parameterization of the freeze-out hypersurface. The part ({\bf a}) represents the view in the $\tau - r$ plane with the fixed values of $\phi$ and $\eta_\parallel$. The part ({\bf b}) shows the view in the $x -y$ plane with the fixed values of $\zeta$ and $\eta_\parallel$. }
  \label{fig:freeze_hsdef}
\end{figure}
In our approach the form of the freeze-out hypersurface is delivered by the hydrodynamic code. The criterion $T = T_{\rm f}$ is used to find the appropriate space-time points, whose positions are parameterized in the following general way
\begin{eqnarray}
     t &=& \left[ \tau_{\rm i} + d \left(\phi, \zeta, \eta_\parallel \right) \sin \zeta \right]\, \cosh \eta_\parallel, 
 \quad z = \left[ \tau_{\rm i} + d \left(\phi, \zeta, \eta_\parallel \right) \sin \zeta \right]\, \sinh \eta_\parallel, \nonumber \\
    x &=& d \left(\phi, \zeta, {\eta_\parallel} \right) \cos \zeta\, \cos  \phi, 
\quad y = d \left(\phi, \zeta, {\eta_\parallel} \right) \cos \zeta\, \sin  \phi. 
  \label{eqn:freeze_cooperfryeparam}
\end{eqnarray}
Here $\phi$ is the standard azimuthal angle in the $y-x$ plane, while the variable $\eta_\parallel$ is the space-time rapidity defined as 
\begin{equation}
  \eta_\parallel = \frac{1}{2} \ln \frac{t+z}{t-z} = \tanh^{-1} \left(\frac{z}{t}\right).
\end{equation}
The parameterization (\ref{eqn:freeze_cooperfryeparam}) yields the compact expressions for the proper time $\tau$ and the transverse distance $r$, 
\begin{eqnarray}
   \tau-\tau_{\rm i} &=& \sqrt{t^2 - z^2} -\tau_{\rm i} =  d\left( \phi ,\zeta, {\eta_\parallel} \right) \sin\zeta, \nonumber \\
     r &=& \sqrt{x^2 + y^2} =\, d\left( \phi ,\zeta, {\eta_\parallel} \right) \cos \zeta.
  \label{eqn:freeze_tauandrho}
\end{eqnarray}
For fixed $\phi$ and $\eta_\parallel$ the quantity $d\left(\phi, \zeta, {\eta_\parallel} \right)$ is the distance between the hypersurface point with  coordinates  $(\phi, \zeta, {\eta_\parallel})$ and the spacetime point $(\tau=\tau_{\rm i},x=0,y=0)$, see Fig. \ref{fig:freeze_hsdef}. The variable $\zeta$, restricted to the range $0 \leq \zeta \leq \pi/2$,  is an angle in the $\tau - r$ space.  We have introduced the angle $\zeta$ because in most cases the freeze-out curves in the $\tau-r$ plane  may be treated as functions of this parameter. The use of the transverse distance $r$ is inconvenient, since very often two freeze-out points correspond to one value of $r$.
\par If the parameterization of the freeze-out hypersurface in terms of the variables $\phi$, $\zeta$, and $\eta_\parallel$ is given, then the element of the freeze-out hypersurface may be obtained from the formula known from the differential geometry
\begin{equation}
  d\Sigma_\alpha = \epsilon_{\alpha\beta\gamma\delta}\,\, \begin{array}{|ccc|}
    \frac{\partial x^\beta}{\partial \phi} &  \frac{\partial x^\beta}{\partial \zeta}   
                             &  \frac{\partial x^\beta}{\partial \eta_\parallel} \\ 
    \frac{\partial x^\gamma}{\partial \phi} & \frac{\partial x^\gamma}{\partial \zeta}    
                             & \frac{\partial x^\gamma}{\partial \eta_\parallel} \\ 
    \frac{\partial x^\delta}{\partial \phi} & \frac{\partial x^\delta}{\partial \zeta} 
                             & \frac{\partial x^\delta}{\partial \eta_\parallel}
  \end{array} \,d\phi \, d\zeta \, d\eta_\parallel,
  \label{eqn:freeze_Jacobian}
\end{equation}
where $\epsilon_{\alpha\beta\gamma\delta}$ is the totally antisymmetric tensor with $\epsilon_{0123} = +1$. The direct differentiation gives:
\begin{equation}
  \begin{array}{l}
    d\Sigma^0 = d \, \cos\zeta \left[ \left( \tau_{\rm i} + d \sin\zeta \right) \left( d \, \sin\zeta - \frac{\partial d}{\partial \zeta} \, \cos\zeta \right) \cosh\eta_\parallel + \frac{\partial d}{\partial_{\eta_\parallel}}  \, \sinh\eta_\parallel \right], \\ \\
    d\Sigma^1 = - d \, \left( \tau_{\rm i} + d \, \sin\zeta \right) 
      \left[ \left( d \, \cos\zeta + \frac{\partial d}{\partial \zeta} \, \sin\zeta \right) \cos\zeta \cos\phi + \frac{\partial d}{\partial \phi} \, \sin\phi \right], \\ \\
    d\Sigma^2 = - d \, \left( \tau_{\rm i} + d \, \sin\zeta \right) 
    \left[ \left( d \, \cos\zeta + \frac{\partial d}{\partial \zeta} \, \sin\zeta \right) \cos\zeta \sin\phi + \frac{\partial d}{\partial \phi} \, \cos\phi \right], \\ \\
    d\Sigma^3 = d \, \cos\zeta \left[ \left( \tau_{\rm i} + d \, \sin\zeta \right) \left( d \, \sin\zeta - \frac{\partial d}{\partial \zeta} \, \cos\zeta \right) \sinh{\eta_\parallel} + \frac{\partial d}{\partial_{\eta_\parallel}}  \, \cosh\eta_\parallel \right].
  \end{array}
  \label{eqn:freeze_dSigmaMu}
\end{equation}
With Eqs. (\ref{eqn:freeze_dSigmaMu}) and the standard definition of the four-momentum expressed in terms of the rapidity and transverse momentum, 
\begin{equation}
  p^\mu = \left( m_T \cosh y, p_T \cos\phi_p, p_T \sin\phi_p, m_T \sinh y \right),
\end{equation}
where $m_T = \sqrt{m^2+p_T^2}$ is the transverse mass, we find the explicit form of the Cooper-Frye integration measure \cite{Cooper:1974mv}
\begin{eqnarray}
  d\Sigma_\mu\, p^\mu & = & d \, \left( \tau_{\rm i} + d \, \sin\zeta \right) \left[ 	\vphantom{\frac{\partial}{\partial}}
    d \cos\zeta \left( m_T \sin\zeta \cosh\left({\eta_\parallel}-y\right) + p_T \cos\zeta \cos\left(\phi-\phi_p\right)\right) \right.  \nonumber \\
    & + & \left. \frac{\partial d}{\partial \zeta} \cos\zeta \left(-m_T \cos\zeta \cosh\left({\eta_\parallel}-y\right) + p_T \sin\zeta \cos\left(\phi-\phi_p\right) \right) \right. \nonumber \\ 
    & + & \left. \frac{\partial d}{\partial \phi } p_T \sin\left(\phi-\phi_p\right) + \frac{\partial d}{\partial {\eta_\parallel}} m_T \cot\zeta \sinh\left({\eta_\parallel}-y\right) 
  \right] d{\eta_\parallel} d\phi d\zeta.
  \label{eqn:freeze_dSigma1}
\end{eqnarray}
This equation, when used in the Cooper-Frye formula \cite{Cooper:1974mv}, leads to the six- dimensional particle density at freeze-out
\begin{eqnarray}
& &  \frac{dN}{dy d\phi_p p_T dp_T d{\eta_\parallel} d\phi d\zeta}  \nonumber \\
& = &  \frac{g}{(2\pi)^3}\, d \, \left( \tau_{\rm i} + d \, \sin\zeta \right) \left[ \vphantom{\frac{\partial}{\partial}}
    d \cos\zeta \left( m_T \sin\zeta \cosh\left({\eta_\parallel}-y\right) + p_T \cos\zeta \cos\left(\phi-\phi_p\right) \right) \right. \nonumber \\
    & + & \left. \frac{\partial d}{\partial\zeta} \cos\zeta \left(-m_T \cos\zeta \cosh\left({\eta_\parallel}-y\right) + p_T \sin\zeta \cos\left(\phi-\phi_p\right) \right) \right. \nonumber \\
    & + & \left. \frac{\partial d}{\partial\phi } p_T \sin\left(\phi-\phi_p\right) 
  \right] \nonumber \\
    & \times & \left\{ \exp\left[ 
    \frac{\beta}{\sqrt{1-v_\bot^2}} \left( m_T \cosh(y-{\eta_\parallel}) - p_T v_\bot \cos(\phi+\alpha-\phi_p) \right) - \beta\mu 
  \right] \pm 1 \right\}^{-1}. \nonumber \\
  \label{eqn:freeze_6cf1}
\end{eqnarray}
In the transition from (\ref{eqn:freeze_dSigma1}) to (\ref{eqn:freeze_6cf1}) we used the assumption of boost-invariance and removed the term $\partial d/ \partial {\eta_\parallel}$ -- for boost-invariant systems the function $d$ depends only on $\phi$ and $\zeta$. The last line in (\ref{eqn:freeze_6cf1}) contains the equilibrium statistical distribution function, with  $-1$ for bosons and $+1$ for fermions. The argument is the Lorentz-invariant product $p^\mu u_\mu$. The parameter \mbox{$\beta = 1/T$} is the inverse temperature and $g$ denotes the spin degeneracy factor. The four-velocity field has been expressed in terms of the transverse flow $v_\perp(\phi,\zeta)$ and the dynamical angle $\alpha(\phi,\zeta)$ which depend on the space-time positions on the freeze-out hypersurface.
\par Although in the hydrodynamic evolution we neglected the effects of the chemical potential $\mu$, it is introduced in the generation of the particles at freeze-out. Thus, the equilibrium distributions include the term
\begin{eqnarray}
  \mu =\mu_{\rm B} B + \mu_{\rm S} S+ \mu_{\rm I_3} I_3,
  \label{eqn:freeze_mus}
\end{eqnarray}
with $B$, $S$, and $I_3$ denoting the baryon number, strangeness, and isospin of the particle. The baryonic, strange, and isospin chemical potentials assume the values: $\mu_{\rm B}=28.5~{\rm MeV}$, $\mu_{\rm S}=9~{\rm MeV}$, $\mu_{\rm I_3}=-0.9~{\rm MeV}$ at the highest RHIC energies \cite{ Baran:2003nm}, and $\mu_{\rm B}= 0.8~{\rm MeV}$ and $\mu_{\rm S}=\mu_{\rm I_3}=0~{\rm MeV}$ at the LHC \cite{Andronic:2005yp}.\\
%
\section{Calculation of observables}
\label{section:freeze_observables}
%
In order to calculate the physical observables we first run our hydrodynamic code and extract the functions describing the freeze-out hypersurface and flow: $d=d(\phi,\zeta)$, $v_\perp=v_\perp(\phi,\zeta)$, and $\alpha=\alpha(\phi,\zeta)$. In the next step these functions are used as input for the {\tt THERMINATOR}\cite{Kisiel:2005hn} code, which generates, according to the formula (\ref{eqn:freeze_6cf1}), the distributions of the primordial particles. The primordial particles include the stable hadrons as well as all hadronic resonances. The resonances decay through strong, electromagnetic, or weak interactions at the random proper time controlled by the particle life-time, and at the location following from the kinematics. As the final result we obtain the stable particle distributions including the feeding from the resonance decays. Since the memory on all decays is kept, one may apply the experimental cuts or the weak-decay feeding policy, which facilitates the more accurate comparison of the model to the data. For example, one may extract easily the model proton spectrum which does not include the feeding from the $\Lambda$ decays.\\
The momentum distribution functions are obtained by integrating over $\phi$, ${\eta_\parallel}$,  and $\zeta$,
\begin{equation}
  \frac{dN}{dy d^2 p_T} = \int \limits_{-\infty}^{\infty} d{\eta_\parallel} \int \limits_{0}^{2 \pi} d\phi \int \limits_{0}^{\pi/2} d\zeta 
  \frac{dN}{dy d\phi_p p_T dp_T d{\eta_\parallel} d\phi d\zeta}.
  \label{eqn:freeze_cf1}
\end{equation}
For cylindrically asymmetric collisions and midrapidity, $y=0$, the transverse-momentum spectrum has the following expansion in the azimuthal angle of the emitted particles
\begin{equation}
  \frac{dN}{dy d^2 p_T} = \frac{dN}{dy\, 2 \pi p_T \, dp_T} \left( 1 + 2 v_2(p_T) \cos(2 \phi_p) + ...\right).
  \label{eqn:freeze_v2def}
\end{equation}
Equation (\ref{eqn:freeze_v2def}) defines the elliptic flow coefficient $v_2$. 
\par The well-known problem of the freeze-out description in the Cooper-Frye formulation is that the hypersurface from hydrodynamics  typically contains a non-causal piece, where particles are emitted back to the hydrodynamic region \cite{Bugaev:2004kq}. In our approach we follow the usual strategy of including only that part of the hypersurface where $d\Sigma_\mu p^\mu \ge 0$. To estimate the effect from the non-causal part we compute the ratio of particles flowing backwards ({\em i.e.} where $d\Sigma_\mu p^\mu < 0$) to all particles. For the cases studied in the following sections we find this ratio to be a fraction of a percent. Such very small values show that the known conceptual problem is not of practical importance for our study.
\par The  correlation function for identical pions is obtained with the two-particle Monte-Carlo method discussed in 
detail in Ref.~\cite{Kisiel:2006is,Kisiel:2006yv}. In this approach the evaluation of the correlation function is reduced to the 
calculation of the following expression
\begin{eqnarray}
  C({\bf q}, {\bf k}) = \frac{\sum\limits_{i} \sum\limits_{j \neq i}
    \delta_\Delta \left( {\bf q} -                  {\bf p}_i + {\bf p}_j        \right) \,\,
    \delta_\Delta \left( {\bf k} - \frac{1}{2}\left[{\bf p}_i + {\bf p}_j \right]\right) \,
    \left|\Psi({\bf k}^{*}, {\bf r}^{*}) \right|^2
  }{\sum\limits_i \sum\limits_{j \neq i}
    \delta_\Delta \left( {\bf q} -                  {\bf p}_i + {\bf p}_j        \right) \,\,
    \delta_\Delta \left( {\bf k} - \frac{1}{2}\left[{\bf p}_i + {\bf p}_j \right]\right)
  },
  \label{eqn:freeze_cfbysum}
\end{eqnarray}
where $\delta_{\Delta}$ denotes the box function
\begin{equation}
  \delta_{\Delta}({\bf p}) =
  \begin{cases}
    1, & \mbox{if}\,\, |p_x| \leq \frac{\Delta}{2}, |p_y| \leq \frac{\Delta}{2}, |p_z| \leq \frac{\Delta}{2} \\
    0, & \mbox{otherwise}.
  \end{cases}
  \label{eqn:freeze_deltadelta}
\end{equation}
In the numerator of Eq. (\ref{eqn:freeze_cfbysum}) we include the sum of the squares of modules of the wave function calculated for all pion pairs with the relative momentum $\bf q$ (we use the bin resolution $\Delta = 5~{\rm MeV})$ and the pair average momentum $\bf k$. For non-central collisions we only provide azimuthally integrated HBT radii. For each pair the wave function $\Psi({\bf k}^{*}, {\bf r}^{*})$, including the Coulomb interaction, is calculated in the rest frame of the pair; ${\bf k}^{*}$ and ${\bf r}^{*}$ denote the relative momentum and the relative distance in the pair rest frame, respectively. In the denominator of Eq. (\ref{eqn:freeze_cfbysum}) we put the number of pairs with the relative momentum $\bf q$ and the average momentum $\bf k$. The correlation function (\ref{eqn:freeze_cfbysum}) is then expressed with the help of the Bertsch-Pratt coordinates $k_T, q_{\rm out}, q_{\rm side}, q_{\rm long}$ and approximated by the Bowler-Sinyukov formula \cite{Bowler:1991vx,Sinyukov:1998fc}
\begin{eqnarray}
  C({\bf q}, {\bf k}) &=& (1 - \lambda) + \lambda\, K_{\rm coul} (q_{\rm inv}) \times \nonumber \\
  &\times& \left[ 1 + \exp \left(
    -R_{\rm out }^2\, q_{\rm out }^2 
    -R_{\rm side}^2\, q_{\rm side}^2
   - R_{\rm long}^2\, q_{\rm long}^2
  \right) \right],
  \label{eqn:freeze_cffitbs}
\end{eqnarray}
where $K_{\rm coul}(q_{\rm inv})$ with $q_{\rm inv} = 2 k^*$ is the squared Coulomb wave function integrated over a static Gaussian source. We use, following the STAR procedure \cite{Adams:2004yc}, the static Gaussian source characterized by the widths of 5 fm in all three directions. Four $k_{T}$ bins, $(0.15-0.25)$, $(0.25-0.35)$, $(0.35-0.45)$, and $(0.45-0.60)$~GeV, are considered. The 3-dimensional correlation function with the exact treatment of the Coulomb interaction is then fitted with this approximate formula and the HBT radii $R_{\rm out}$, $R_{\rm side}$, and $R_{\rm long}$ are obtained. They can be compared directly to the experimental radii.

\chapter{Soft-hadronic observables at RHIC}
\label{chaper:resrhic}
%
In this Section we apply our model to describe soft hadron production in the relativistic heavy-ion collisions at the highest RHIC energy, \mbox{$\sqrt{s_{NN}}$ = 200 GeV}. We analyze the data delivered by the PHENIX \cite{Adler:2003cb,Adler:2003kt} and STAR Collaborations \cite{Adams:2004yc}. We consider the most central events, defined by the centrality class \mbox{$c$ = 0 - 5\%}, where we use the impact parameter $b=2.26$~fm. We also use the data from the centrality classes \mbox{$c$ = 20 - 30\%} (transverse-momentum spectra and the HBT radii) and \mbox{$c$ = 20 - 40\%} (elliptic flow), for which we take (for simplicity) the same value of the impact parameter, $b=7.16$~fm.
\par The link between the centrality class and the impact parameter involves the total Au+Au inelastic cross section. To a very good accuracy for not-too-peripheral collisions we may use the formula \cite{Broniowski:2001ei}
\begin{equation}
  \langle c\rangle\, \simeq \frac{\pi b^2}{\sigma_{\rm in}^{\rm Au Au}}.
  \label{eqn:resrhic_centdef}
\end{equation}
The total inelastic Au+Au cross section $\sigma_{\rm in}^{\rm Au Au}$ is not directly measured. Typically, it is obtained from Glauber simulations and carries model uncertainties \cite{Miller:2007ri}. With total inelastic $pp$ cross section $\sigma_{\rm in}$ = 42 mb and for gold nuclei used at RHIC, the {\tt GLISSANDO} Monte-Carlo Glauber simulation \cite{Broniowski:2007nz} gives $\sigma_{\rm in}^{\rm Au Au} \simeq$ 6.4 b, so $b$ = 2.26 fm for $\langle c\rangle$ = 2.5\% and \mbox{$b$ = 7.16 fm} for $\langle c\rangle$ = 25\%.
%
\section{Central collisions}
\label{section:resrhic-central}
%
\begin{figure}[!t]
  \begin{center}
    \includegraphics[width=0.4 \textwidth]{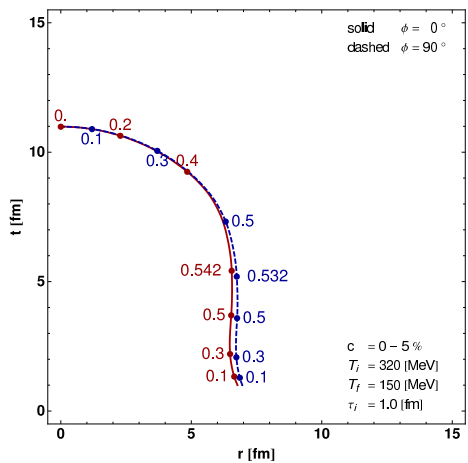}
  \end{center}
  \caption{\small In-plane and out-of-plane freeze-out curves, i.e., the intersections of the freeze-out hypersurface with the planes \mbox{$y=0 \, (\phi = 0^o)$} and $x=0 \, (\phi = 90^o)$, obtained for central RHIC collisions; $c = 0 - 5 \%$, $b = 2.26~{\rm fm}$, $\tau_{\rm i} = 1.0~{\rm fm}$, $T_{\rm i} = 320~{\rm MeV}$, and $T_{\rm f} = 150~{\rm MeV}$. The two curves overlap, indicating that the system at freeze-out is almost azimuthally symmetric in the transverse plane. }
  \label{fig:resrhicTi320c0005-hs}
\end{figure}
\begin{figure}[!t]
  \begin{center}
    \includegraphics[width=0.5 \textwidth]{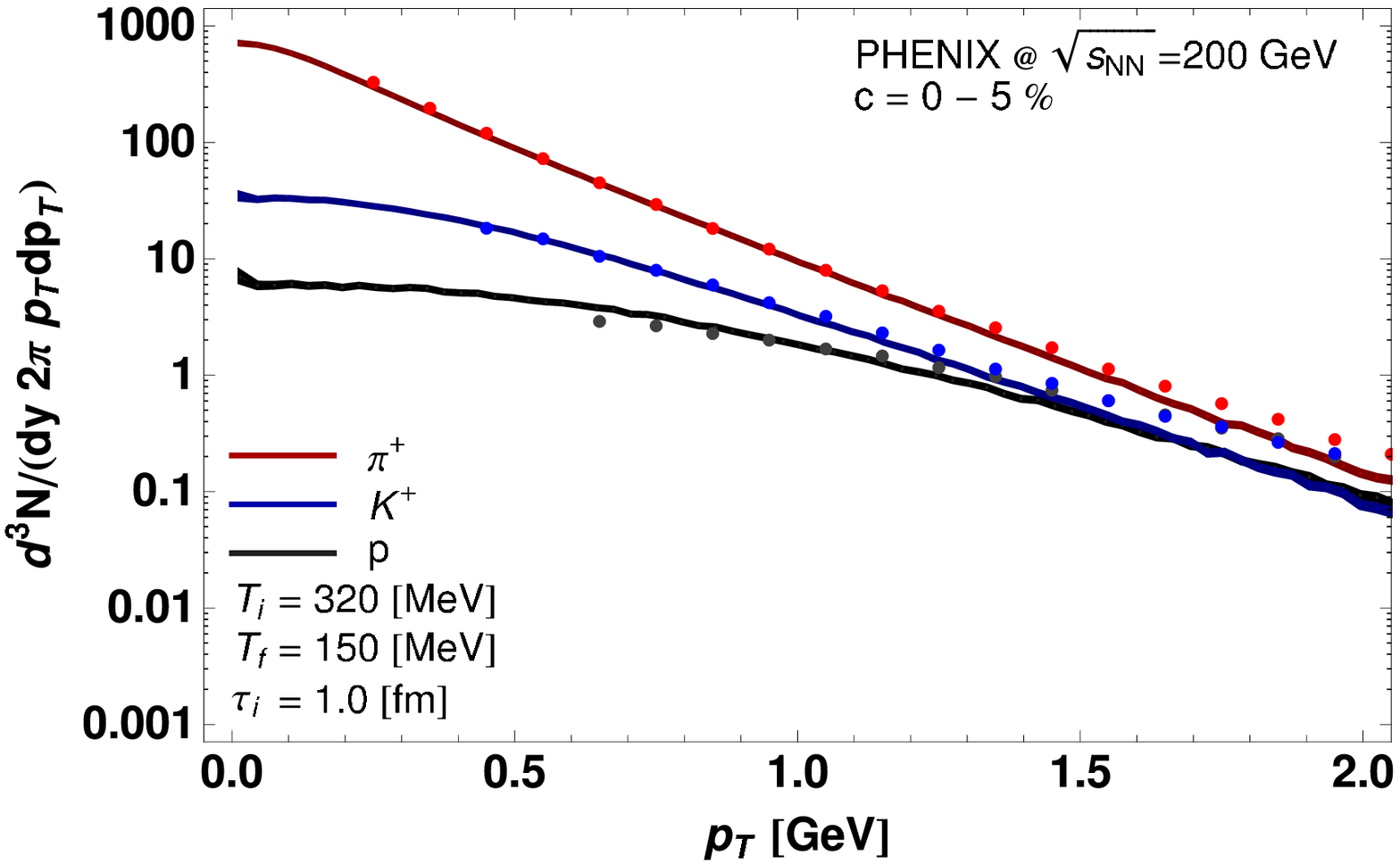}
  \end{center}
  \caption{\small Transverse-momentum spectra of $\pi^+$, $K^+$, and protons. The PHENIX experimental results \cite{Adler:2003cb} for Au+Au collisions at $\sqrt{s_{NN}}= 200~{\rm GeV}$ and the centrality class 0-5\% (points) are compared to the model calculations (solid lines) with the same parameters as in Fig.~\ref{fig:resrhicTi320c0005-hs}. }
  \label{fig:resrhicTi320c0005-sppt}
\end{figure}
\begin{figure}[tb]
  \begin{center}
    \includegraphics[width=0.5\textwidth]{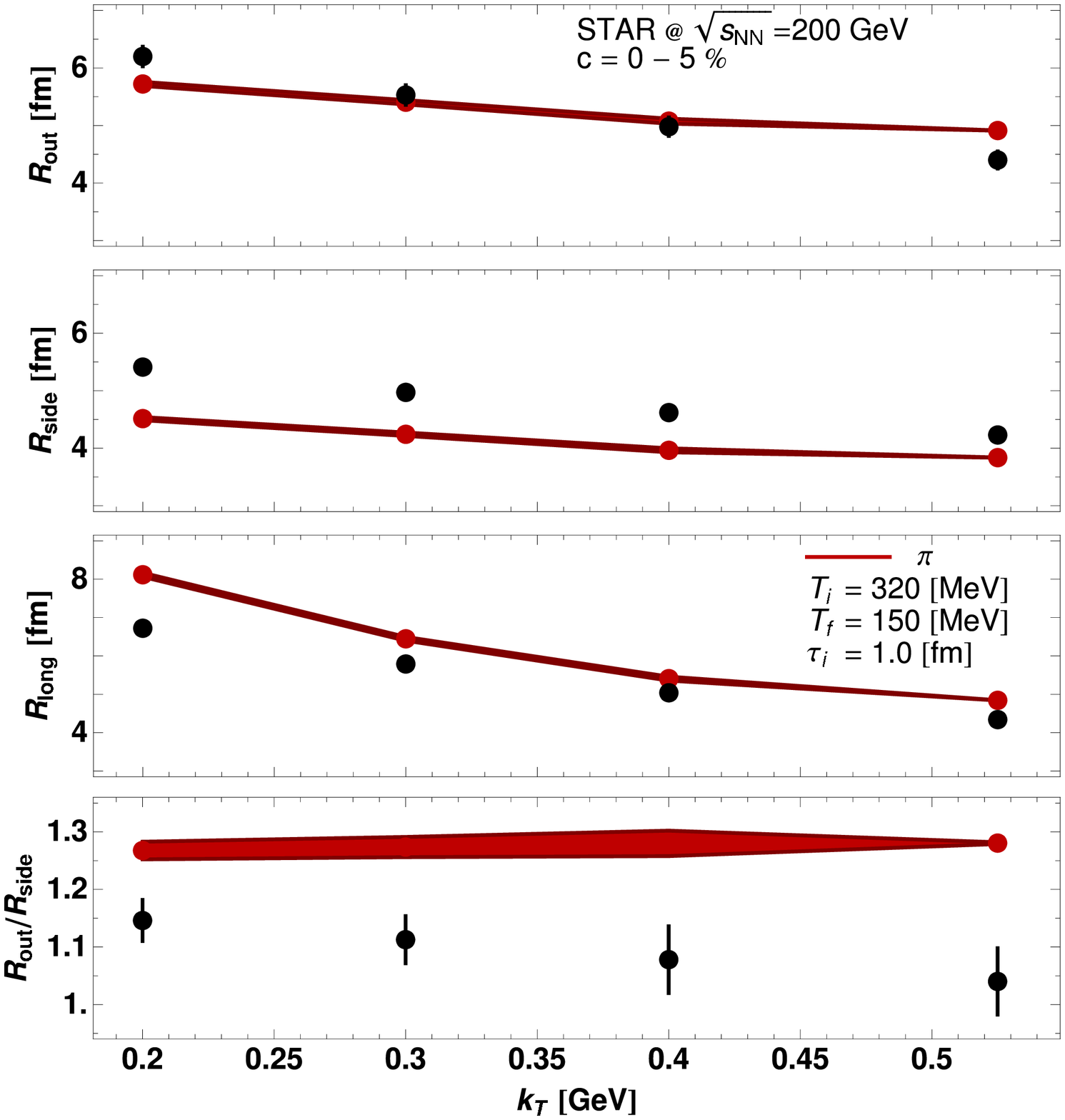}
  \end{center}
  \caption{\small The pionic HBT radii plotted as functions of the average transverse momentum of the pair compared to the STAR data \cite{Adams:2004yc} at the centrality 0-5\%. The calculation uses the two-particle method and includes the Coulomb effects.  The values of the model parameters are the same as in Fig.~\ref{fig:resrhicTi320c0005-hs}.}
  \label{fig:resrhicTi320c0005-hbt}
\end{figure}
\par First, we consider the centrality class 0 - 5\% with the corresponding value of the impact parameter \mbox{$b = 2.26$~fm}. The in-plane and out-of-plane freeze-out curves are defined as the intersections of the freeze-out hypersurface with the planes \mbox{$y=0 \,(\phi = 0^o)$} and $x=0 \, (\phi = 90^o)$. They are obtained with the initial starting time for hydrodynamics $\tau_{\rm i} = 1$ fm, the initial central temperature $T_{\rm i} = 320~{\rm MeV}$, the final (freeze-out) temperature $T_{\rm f} = 150~{\rm MeV}$, and are shown in Fig.~\ref{fig:resrhicTi320c0005-hs}. The two freeze-out curves practically overlap, indicating that the expansion of the system is almost azimuthally symmetric in the transverse plane.  This feature is certainly expected for the almost central collisions, where the impact parameter is very small. We note that the shape of the isotherms is consistent with the result presented in Fig.~2 of Ref.~\cite{Eskola:2005ue}, where the same decoupling temperature of 150~MeV was considered.
\par In Fig.~\ref{fig:resrhicTi320c0005-sppt} we present our results for the hadron transverse-momentum spectra with the same values of the parameters. The dots show the PHENIX data \cite{Adler:2003cb} for positive pions, positive kaons, and protons, while the solid lines show the results obtained from our hydrodynamic code linked to {\tt THERMINATOR}\cite{Kisiel:2005hn}. Our model describes the data well up to the transverse-momentum values of about 1.5~GeV. For larger values of $p_T$ the model underpredicts the data. This effect may be explained by the presence of the semi-hard processes, not included in our approach.
\par In our approach, the values of $T_{\rm i}$ and $b$ control the overall normalization. On the other hand, the value of the freeze-out temperature $T_{\rm f}$ determines mainly the relative normalization and the slopes of the spectra. We note that the correct slope for the pions and kaons is recovered at a relatively high value of $T_{\rm f}$. This is possible, since the spectra of the observed hadrons contain the contributions from all well established hadronic resonances. This single-freeze-out picture \cite{Florkowski:2001fp,Broniowski:2001we,Broniowski:2001uk}, where the same temperature is used to describe the ratios of hadronic abundances and the spectra, was first tested for hadronic spectra in Ref.~\cite{Broniowski:2001we}, see also \cite{Rafelski:2000by,Broniowski:2002nf}. We note that the value \mbox{$T_{\rm f}$ = 150 MeV} agrees with the recent results obtained in the framework of the single-freeze-out model in Ref.~\cite{Prorok:2007xp}. In the calculation of the proton spectra, in order to be consistent with the PHENIX experimental procedure, we removed the protons coming from the $\Lambda$ decays. 
\par In Fig. \ref{fig:resrhicTi320c0005-hbt} the model results and the STAR data \cite{Adams:2004yc} for the HBT radii are presented. Again, for the same values of the parameters, a quite reasonable agreement is found. Discrepancies at the level of 10-15\% are observed in the behavior of the $R_{\rm side}$, which is too small, and $R_{\rm long}$, which is too large, probably due to the assumption of strict boost invariance. The ratio $R_{\rm out}/R_{\rm side}\simeq 1.25$ is larger than one, which is a typical discrepancy of hydrodynamic studies. Nevertheless, when compared to other hydrodynamic calculations, our ratio $R_{\rm out}/R_{\rm side}$ is significantly closer to the experimental value. The ratio is almost constant as a function 
of $k_T$, contrary to the decrease observed in the data. 
%
\section{Non-central collisions}
\label{section:resrhic-noncentral}
%
\begin{figure}[!t]
  \begin{center}
    \includegraphics[width=0.4 \textwidth]{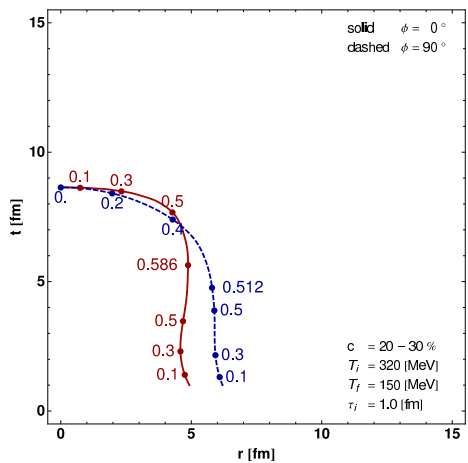}
  \end{center}
  \caption{\small The freeze-out curves for $c = 20 - 30 \%$, $\tau_{\rm i} = 1.0~{\rm fm}$, \mbox{$T_{\rm i} = 320~{\rm MeV}$}, \mbox{$T_{\rm f} = 150~{\rm MeV}$}, and \mbox{$b = 7.16~{\rm fm}$}. The solid line describes the in-plane profile, while the dashed line describes the out-of-plane profile.}
  \label{fig:resrhicTi320c2030-hs}
\end{figure}
\begin{figure}[!t]
  \begin{center}
    \includegraphics[width=0.5 \textwidth]{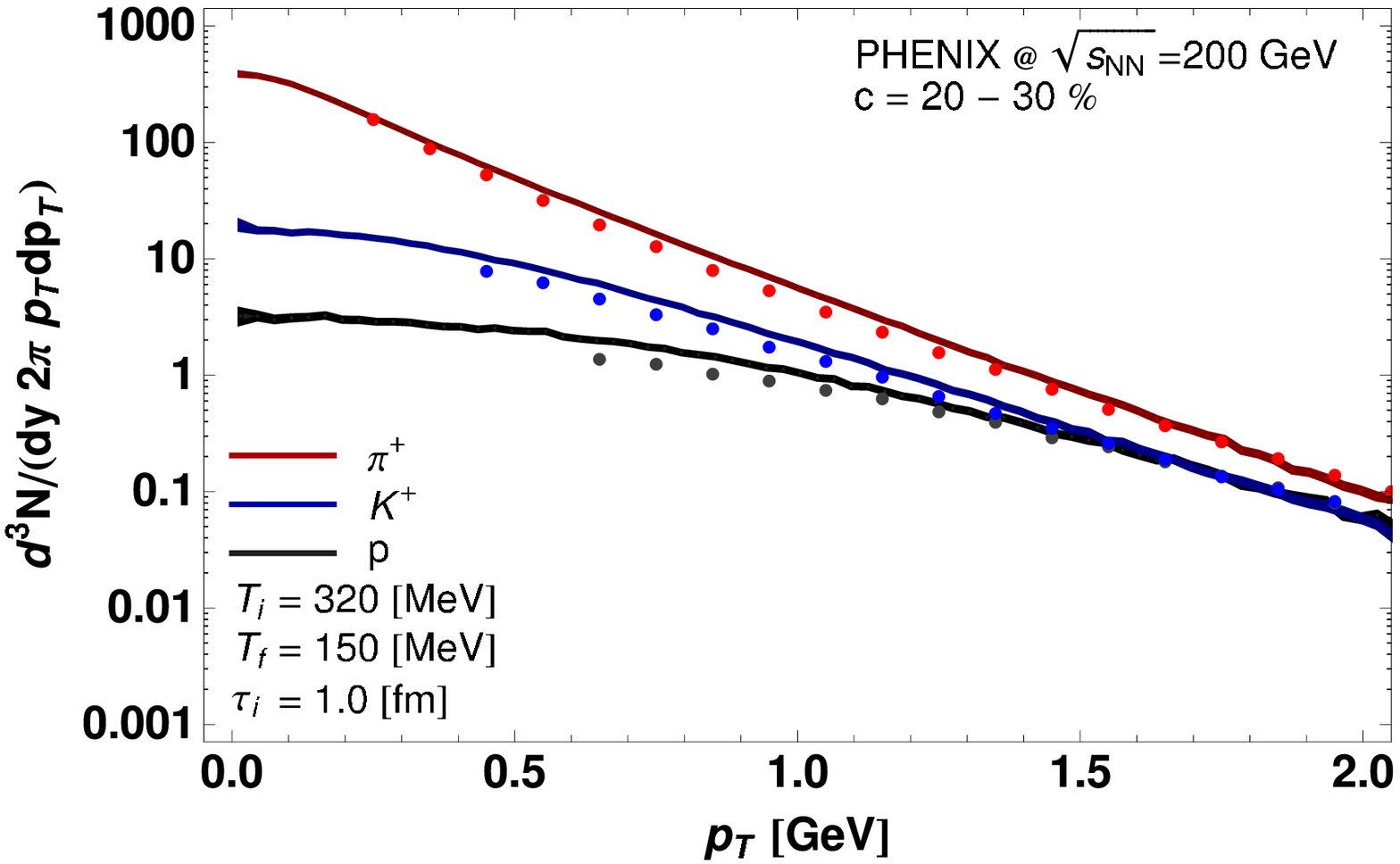}
  \end{center}
  \caption{\small Transverse-momentum spectra of $\pi^+$, $K^+$, and protons. The PHENIX experimental results \cite{Adler:2003cb} for Au+Au collisions at $\sqrt{s_{NN}}= 200~{\rm GeV}$ and the centrality class 20-30\% (points) are compared to the model calculations (solid lines). The values of the model parameters are the same as in Fig. \ref{fig:resrhicTi320c2030-hs}. }
  \label{fig:resrhicTi320c2030-sppt}
\end{figure}
\begin{figure}[tb]
  \begin{center}
    \includegraphics[width=0.5 \textwidth]{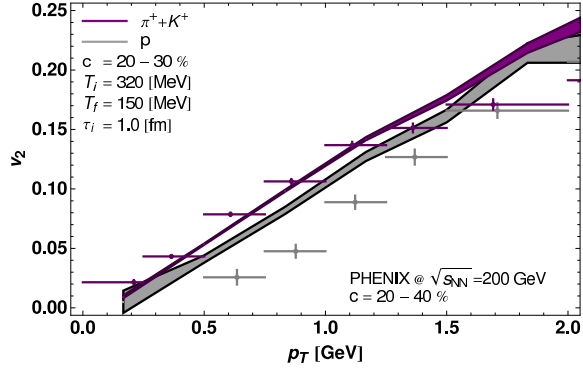}
  \end{center}
  \caption{\small The elliptic flow coefficient $v_2$. The values measured by PHENIX \cite{Adler:2003kt} at $\sqrt{s_{NN}}= 200~{\rm GeV}$ and the centrality class 20-40\% are indicated by the upper (pions + kaons) and lower (protons) points, with the horizontal bars indicating the $p_T$ bin. The corresponding model calculations are indicated by the solid lines, with the bands displaying the statistical error of the Monte-Carlo method. The parameters are the same as in Fig.~\ref{fig:resrhicTi320c2030-hs}.}
  \label{fig:resrhicTi320c2030-v2pt}
\end{figure}
\begin{figure}[!tb]
  \begin{center}
    \includegraphics[width=0.5 \textwidth]{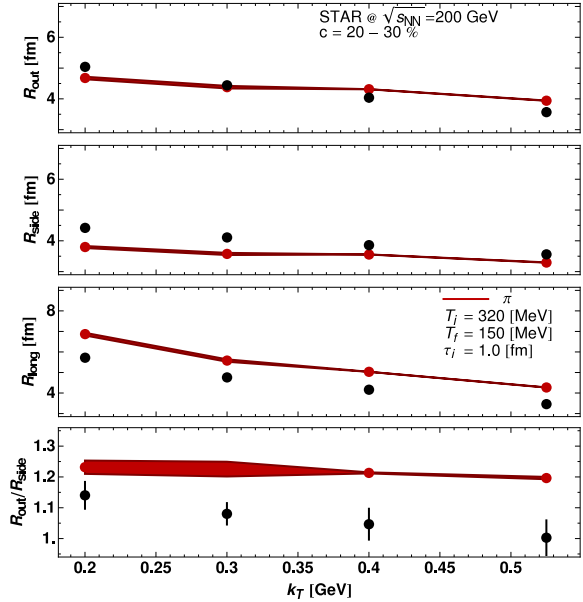}
  \end{center}
  \caption{\small The pionic HBT radii plotted as functions of the average transverse momentum of the pair compared to the STAR data \cite{Adams:2004yc} at centrality 20-30\%. As in the case of the central collisions, the calculation uses the two-particle method and includes the Coulomb effects. The values of the model parameters are the same as in Fig. \ref{fig:resrhicTi320c2030-hs}. }
  \label{fig:resrhicTi320c2030-hbt}
\end{figure}
Next, we consider the centrality classes \mbox{20 - 30\%} and \mbox{20 - 40\%}. The data are compared with the model results obtained with the impact parameter \mbox{$b$ = 7.16 fm}. The values of the initial central temperature and the final temperature are the same as in the case of the central collisions, i.e., $T_{\rm i} = 320~{\rm MeV}$ and $T_{\rm f} = 150~{\rm MeV}$. The two freeze-out curves are shown in Fig.~\ref{fig:resrhicTi320c2030-hs}.
We observe that the out-of-plane profile is wider than the in-plane profile. This difference indicates that the system is elongated along the $y$ axis at the moment of freeze-out. This feature is in qualitative agreement with the HBT measurements of the azimuthal dependence of  $R_{\rm side}$. The comparison of the experimental and model transverse-momentum spectra is presented in Fig.~\ref{fig:resrhicTi320c2030-sppt}. In this case we also find a reasonable agreement between the data and the model results.  
\par In Fig. \ref{fig:resrhicTi320c2030-v2pt} the model results and the PHENIX data \cite{Adler:2003kt} on the elliptic flow coefficient $v_2(p_T)$ are compared. We observe that the $v_2$ of pions+kaons agrees with the data. Taking into account the uncertainly in the initial eccentricity, discussed in Sect.~\ref{section:initial_Glauber}, which is at the level of 10-20\%, the obtained agreement is reasonable. On the other hand, the model prediction for proton $v_2$ is too large. The discrepancy is probably caused by the final-state elastic interactions, not included in our approach \cite{Eskola:2007zc}. 
\par The HBT radii for non-central RHIC collisions are shown in Fig. \ref{fig:resrhicTi320c2030-hbt}. Similarly to the central collisions, we observe that $R_{\rm side}$ is slightly too small and $R_{\rm long}$ is slightly too large. Still, the ratio $R_{\rm out}/R_{\rm side}$ is very close to the data. Comparing our values of $R_{\rm side}$ with other hydrodynamic calculations we find that our values are larger. This effect is caused by the halo of decaying resonances which increases the system size by about 1 fm, see Ref.~\cite{Kisiel:2006is}.

\chapter{Predictions for LHC}
\label{chaper:reslhc}
%
\par We expect that the increase of the initial beam energy from $\sqrt{s_{NN}}=$ 200 GeV at RHIC to $\sqrt{s_{NN}}=$ 5.5 TeV at LHC results essentially in a higher initial temperature $T_{\rm i}$ used as the input for the hydrodynamic calculation (in the midrapidity region studied here). Therefore, to make predictions for the collisions at the LHC energies we use a set of values for $T_{\rm i}$ which are higher than those used at RHIC, namely, \mbox{$T_{\rm i}$ = 400, 450, and 500 MeV}. Of course, it is not obvious which value of $T_{\rm i}$ will be realized at LHC. Estimates based on extrapolations \cite{Abreu:2007kv} suggest an increase of the multiplicity by a factor of 2 compared to the highest RHIC energies, which would favor $T_{\rm i}$ around 400~MeV. However, to investigate a broader range of possibilities, we take into account much higher temperatures as well. In the case of LHC we also use a larger value of the nucleon-nucleon cross section in the definition of the initial conditions ($\sigma_{\rm in} =63~{\rm mb}$, $\kappa=0.2$, $A=208$) and different values of the chemical potentials in the hadronization made by {\tt THERMINATOR} ($\mu_{\rm B} = 0.8$ MeV and $\mu_{\rm s} = \mu_{\rm I_3} = 0$). The changes of $\sigma_{\rm in}$ and the nuclear profile imply  that for LHC we use \mbox{$b$ = 2.4 fm} for $c$ = 2.5\% and $b$ = 7.6 fm for $c$ = 25\%.
\par In Tables~\ref{tab:reslhc_cent} and \ref{tab:reslhc_pery} we list our results for the following quantities: the total $\pi^+$ multiplicity $dN/dy$, the inverse-slope parameter for pions $\lambda$, the pion+kaon elliptic flow $v_2$ at \mbox{$p_T= 1$~GeV}, and the three HBT radii calculated at the pion pair momentum $k_T= 300$~MeV. The inverse-slope parameter given in Tables~\ref{tab:reslhc_cent} and \ref{tab:reslhc_pery} is obtained from the formula
\begin{equation}
  \lambda = - \left[ \frac{d}{dp_T} \ln 
    \left( 
      \frac{dN_\pi}{2\pi p_T dp_T dy}
    \right)
  \right]^{-1}.
  \label{eqn:reslhc_lambda}
\end{equation}
The tables show several expected qualitative features. Obviously, as the initial temperature increases, the multiplicity grows. This is due a larger initial entropy, which causes a larger size of the freeze-out hypersurface. We find that the following approximate parameterizations work very well for the multiplicity of $\pi^+$ at LHC for the two considered centrality cases:
\begin{equation}
  \begin{array}{ll}
    \frac{dN}{dy}=12000\, (T_{\rm i}/{\rm GeV})^{3.4}, & (b=2.4~{\rm fm})\\
    \\
    \frac{dN}{dy}= 6000\, (T_{\rm i}/{\rm GeV})^{3.4}, & (b=7.6~{\rm fm}).
  \end{array}
  \label{eqn:reslhc_dNform}
\end{equation} 
The power 3.4 works remarkably well. This behavior reflects the dependence of the initial entropy on $T$ as shown in top left panel of Fig.~\ref{fig:thermo_thermo} of Chapter \ref{chapter:thermo}, where for the relevant temperature range of $300-500$~MeV we have the approximate scaling $s/T^3 \sim T^{0.4}$.
\par Similarly, for the slopes in the studied domain we have 
\begin{equation}
  \begin{array}{ll}
    \lambda(1~{\rm GeV}) = 0.639\, T_{\rm i} + 0.033~{\rm GeV}, & (b=2.4~{\rm fm})\\
    \\
    \lambda(1~{\rm GeV}) = 0.661\, T_{\rm i} + 0.033~{\rm GeV}, & (b=7.6~{\rm fm}).
  \end{array}
  \label{eqn:reslhc_lamform}
\end{equation} 
The HBT radii increase rather moderately with $T_{\rm i}$, as can be seen from the tables.
\begin{table}[!ht]
  \begin{center}
    \begin{small}
      \begin{tabular}{c|r|c|c|c|c}
        $\displaystyle \vphantom{\frac{1}{2}}$
        $T_{\rm i}$ [MeV] &  $\frac{dN}{dy}$ & $\lambda$ [MeV] &  $R_{\rm side}$ [fm] & 
        $R_{\rm out}$ [fm] & $R_{\rm long}$ [fm]  \\ \hline 
        $\displaystyle \vphantom{\frac{1}{2}}$ 320   &   274  &  237  &  4.2 & 5.4 & 6.5 \\ \hline 
        $\displaystyle \vphantom{\frac{1}{2}}$ 400   &   543  &  288  &  5.2 & 5.9 & 7.7 \\ 
        $\displaystyle \vphantom{\frac{1}{2}}$ 450   &   802  &  320  &  5.6 & 6.1 & 8.2 \\
        $\displaystyle \vphantom{\frac{1}{2}}$ 500   &  1136  &  352  &  6.0 & 6.2 & 8.8
      \end{tabular}
    \end{small}
  \end{center}
  \caption{\small Central collisions at RHIC corresponding to the results from Chapter \ref{section:resrhic-central} (the second row) and LHC (the three lower rows): A set of our results obtained for four different values of the initial temperature: $T_{\rm i}$ = 320, 400, 450, and 500 MeV. The columns contain the following information: $dN/dy$ -- the total pion multiplicity  (positive pions only), $\lambda$ -- the inverse-slope parameter for positive pions at \mbox{$p_T$ = 1 GeV},  \mbox{$R_{\rm side}, R_{\rm out}, R_{\rm long}$} -- the three HBT radii calculated at the average momentum \mbox{$k_T$ = 300 MeV.}}
  \label{tab:reslhc_cent}
\end{table}
\begin{table}[!ht]
  \begin{center}
    \begin{small}
      \begin{tabular}{c|c|c|c|c|c|c}
      $\displaystyle \vphantom{\frac{1}{2}}$ 
      $T_{\rm i}$ [MeV] &  $\frac{dN}{dy}$ & $\lambda$ [MeV] & $v^{\pi+K}_2$ & $R_{\rm side}$ [fm] & 
      $R_{\rm out}$ [fm] & $R_{\rm long}$ [fm]  \\ \hline 
      $\displaystyle \vphantom{\frac{1}{2}}$ 320   & 152   &  244  & 0.14  & 3.6  & 4.5  & 5.6 \\ \hline
      $\displaystyle \vphantom{\frac{1}{2}}$ 400   & 272   &  299  & 0.16  & 4.2  & 4.5  & 6.2 \\ 
      $\displaystyle \vphantom{\frac{1}{2}}$ 450   & 401   &  331  & 0.16  & 4.6  & 4.6  & 6.7 \\ 
      $\displaystyle \vphantom{\frac{1}{2}}$ 500   & 569   &  363  & 0.15  & 5.2  & 4.9  & 7.4
    \end{tabular}
    \end{small}
  \end{center}
  \caption{\small Non-central collisions at RHIC corresponding to the results from Chapter \ref{section:resrhic-noncentral} (the second row) and LHC (the three lower rows): The same quantities shown as in Table~\ref{tab:reslhc_cent} with the additional information on the pion elliptic flow $v_2$ at \mbox{$p_T$ = 1 GeV}.}
  \label{tab:reslhc_pery}
\end{table}
%
\section{Central collisions}
\label{section:reslhc-central}
%
\begin{figure}[!t]
  \begin{center}
    \subfigure{\includegraphics[width=0.32 \textwidth]{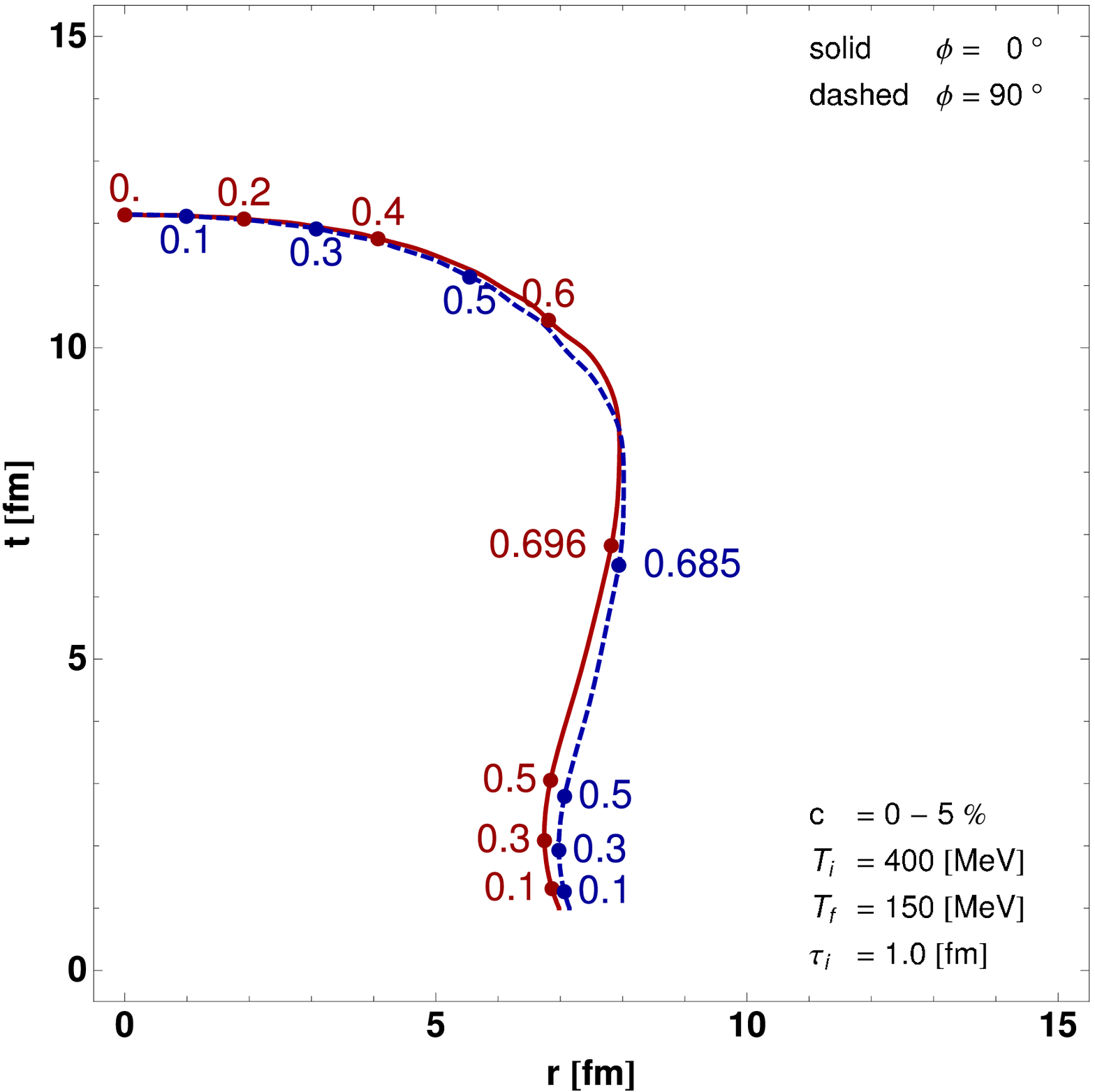}}
    \subfigure{\includegraphics[width=0.32 \textwidth]{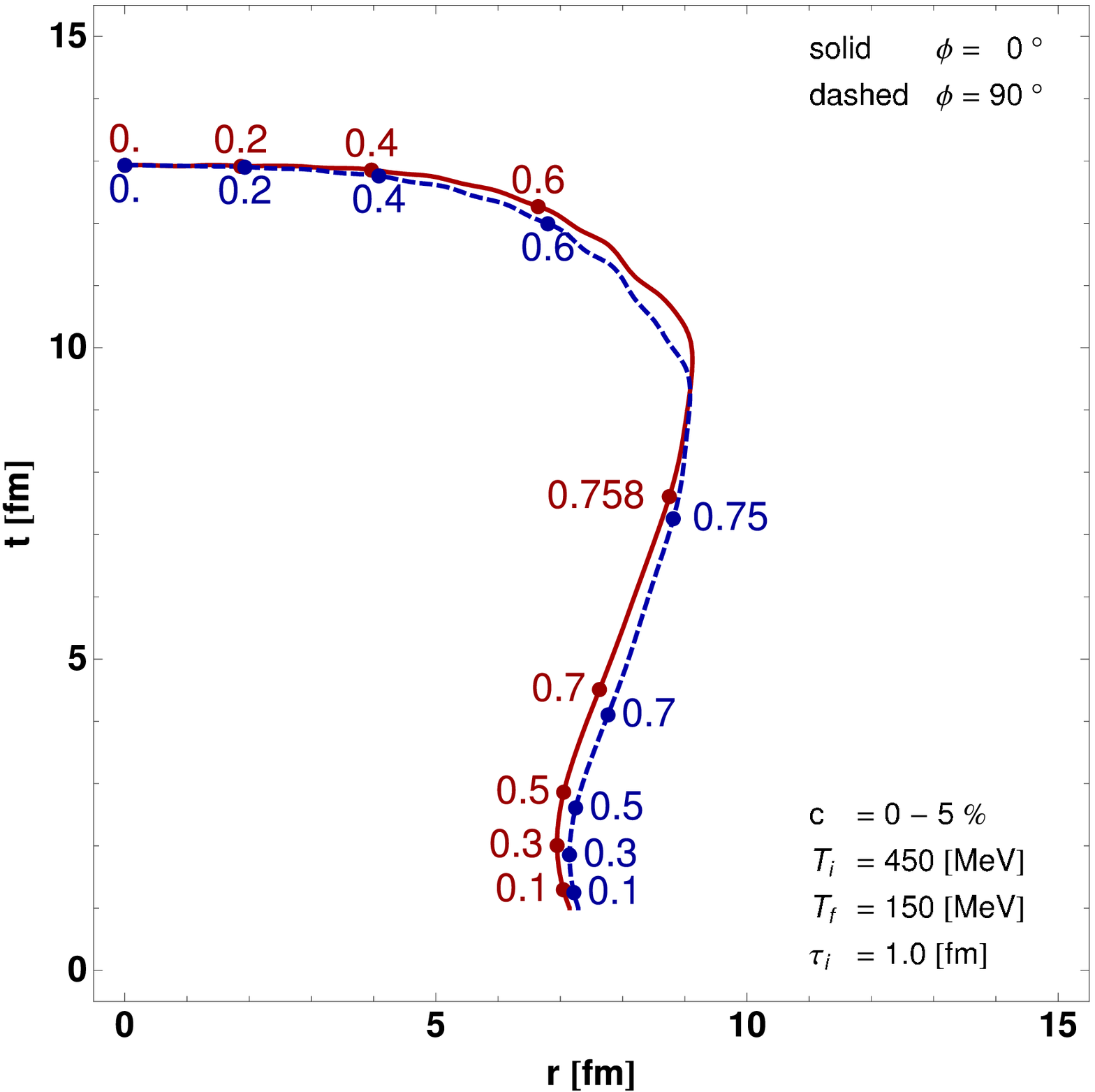}}
    \subfigure{\includegraphics[width=0.32 \textwidth]{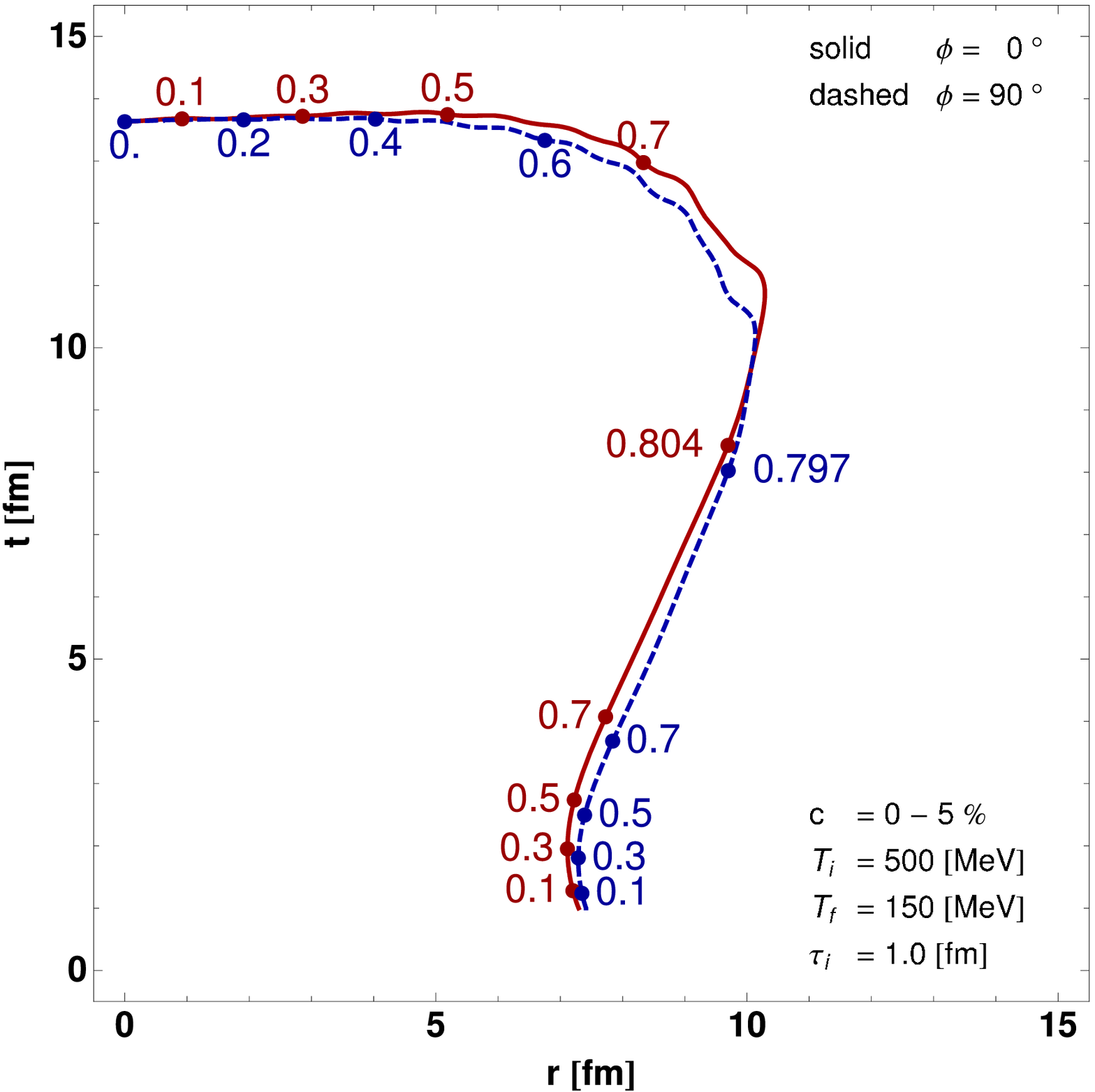}}
  \end{center}
  \caption{\small The freeze-out curves for central LHC collisions with $c = 0 - 5 \%$, $b = 2.4~{\rm fm}$, $\tau_{\rm i} = 1.0~{\rm fm}$ for three initial temperatures $T_{\rm i} = 400, 450 \hbox{ and } 500~{\rm MeV}$, and $T_{\rm f} = 150~{\rm MeV}$. Similarly to the central RHIC collisions, the two curves overlap indicating that the system at freeze-out is symmetric in the transverse plane. }
  \label{fig:reslhcc0005-hs}
\end{figure}
\begin{figure}[!t]
  \begin{center}
    \subfigure{\includegraphics[width=0.45 \textwidth]{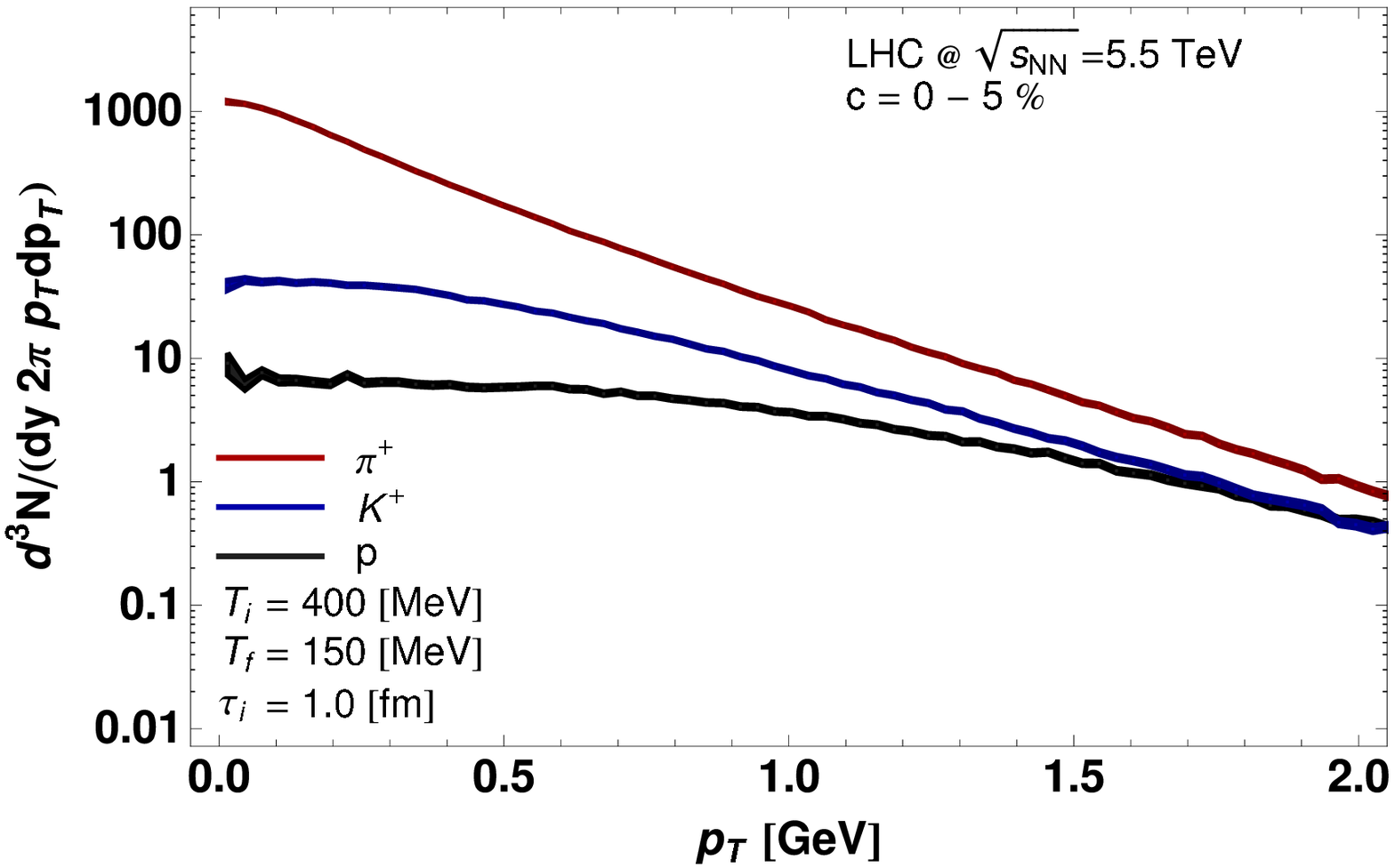}}
    \subfigure{\includegraphics[width=0.45 \textwidth]{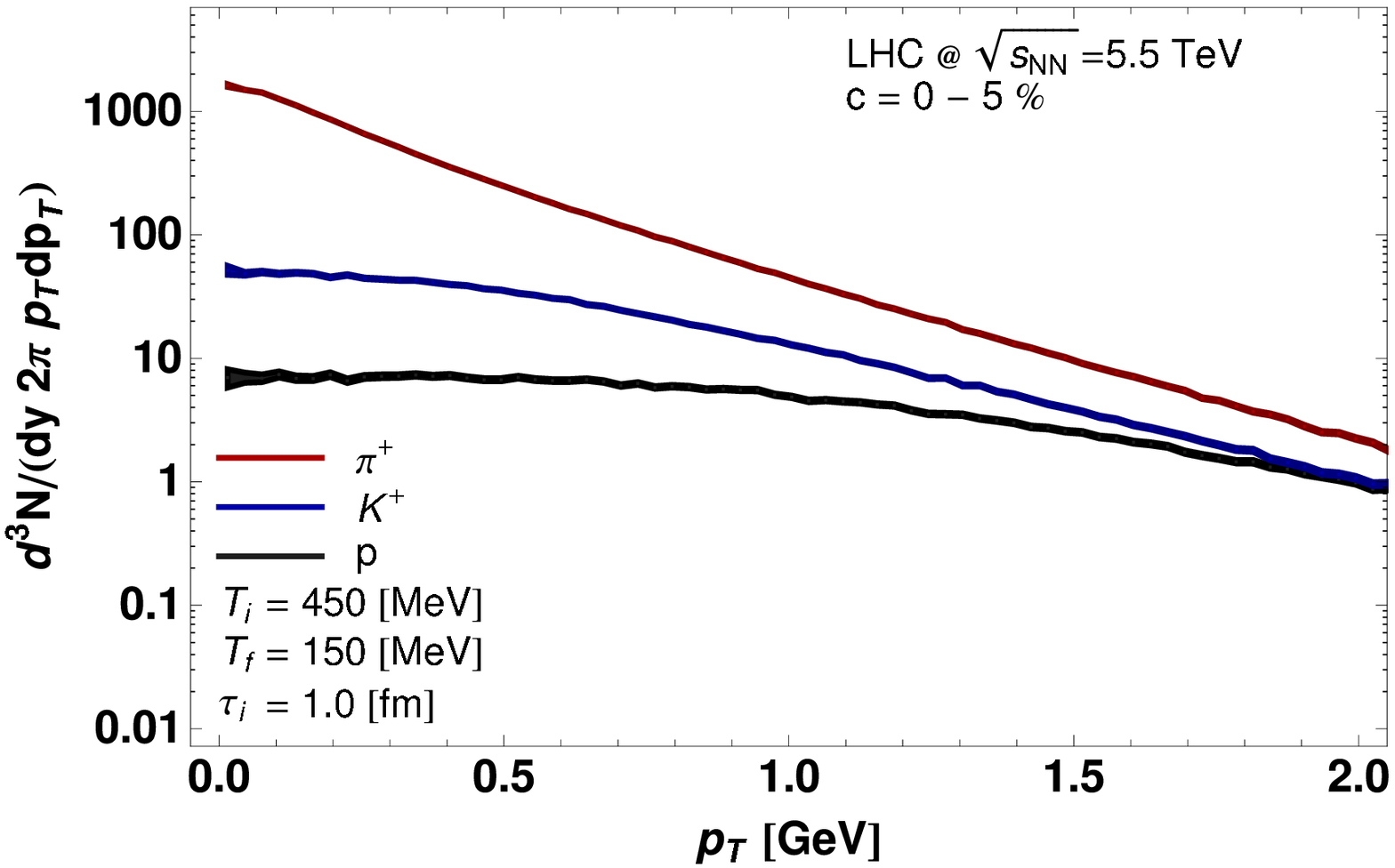}}\\
    \subfigure{\includegraphics[width=0.45 \textwidth]{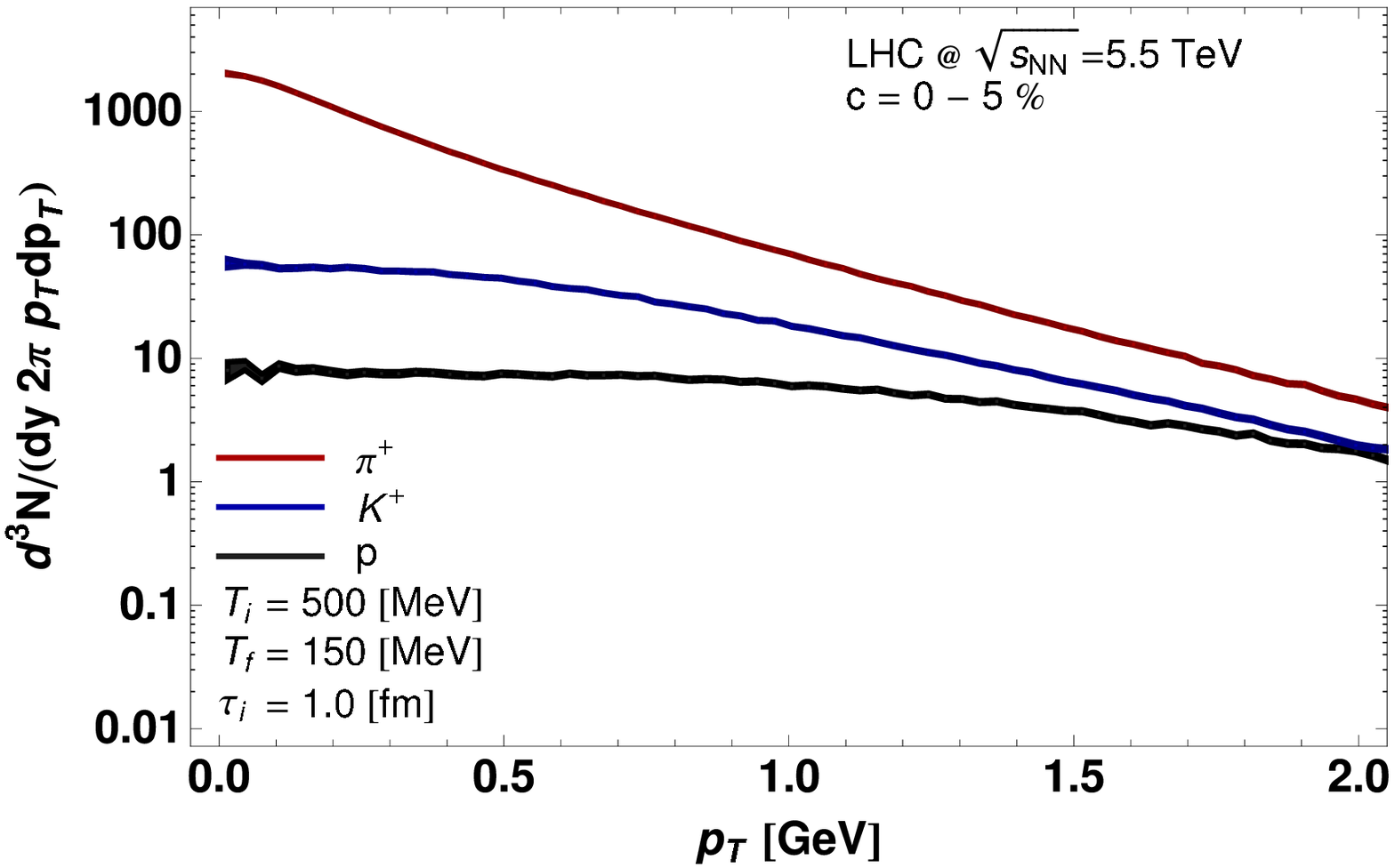}}
  \end{center}
  \caption{\small The model results for the transverse-momentum spectra of $\pi^+$, $K^+$, and protons. The values of the model parameters are the same as in Fig. \ref{fig:reslhcc0005-hs}. }
  \label{fig:reslhcc0005-sppt}
\end{figure}
\begin{figure}[!ht]
  \begin{center}
    \subfigure{\includegraphics[width=0.32\textwidth]{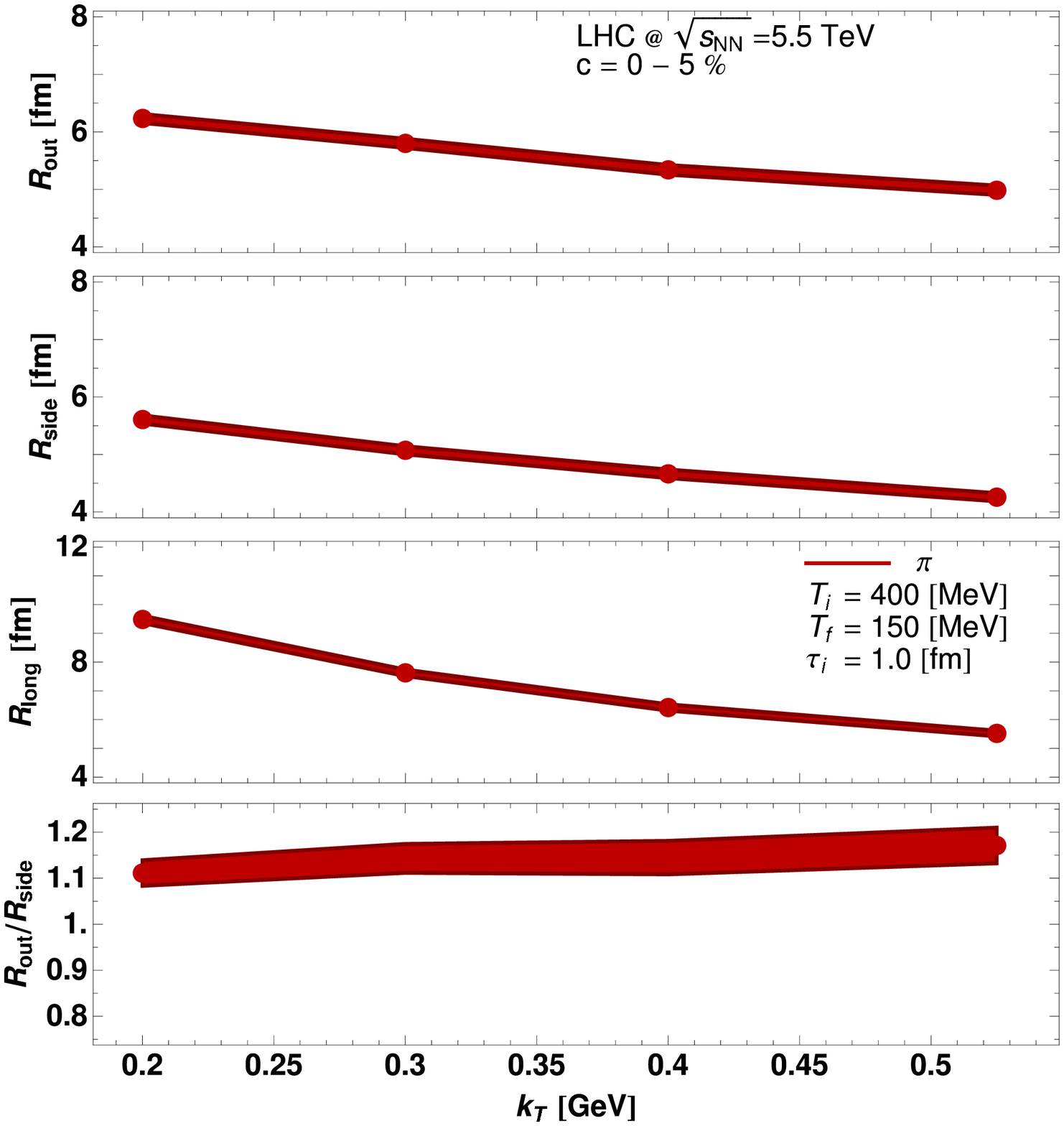}}
    \subfigure{\includegraphics[width=0.32\textwidth]{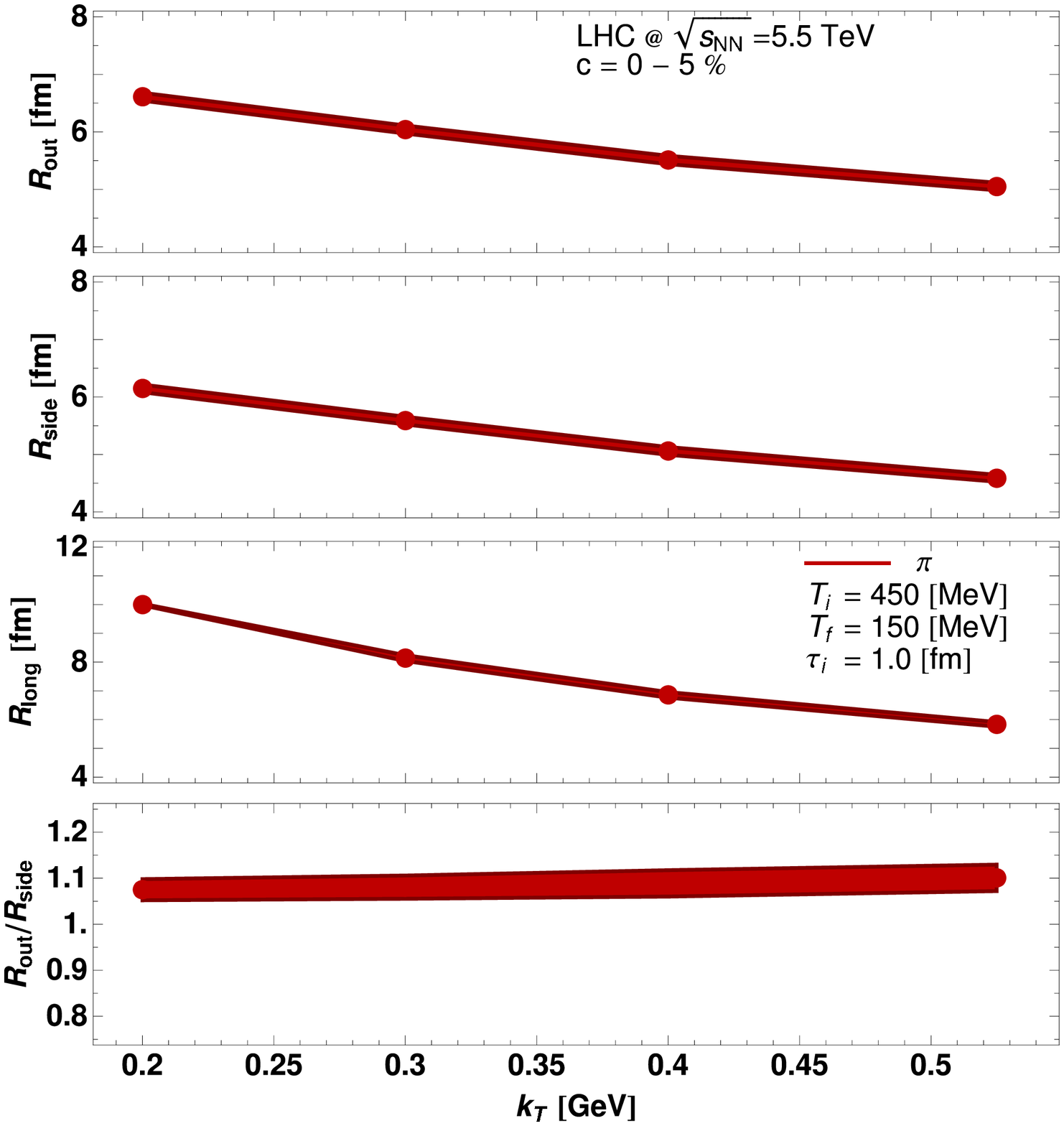}}
    \subfigure{\includegraphics[width=0.32\textwidth]{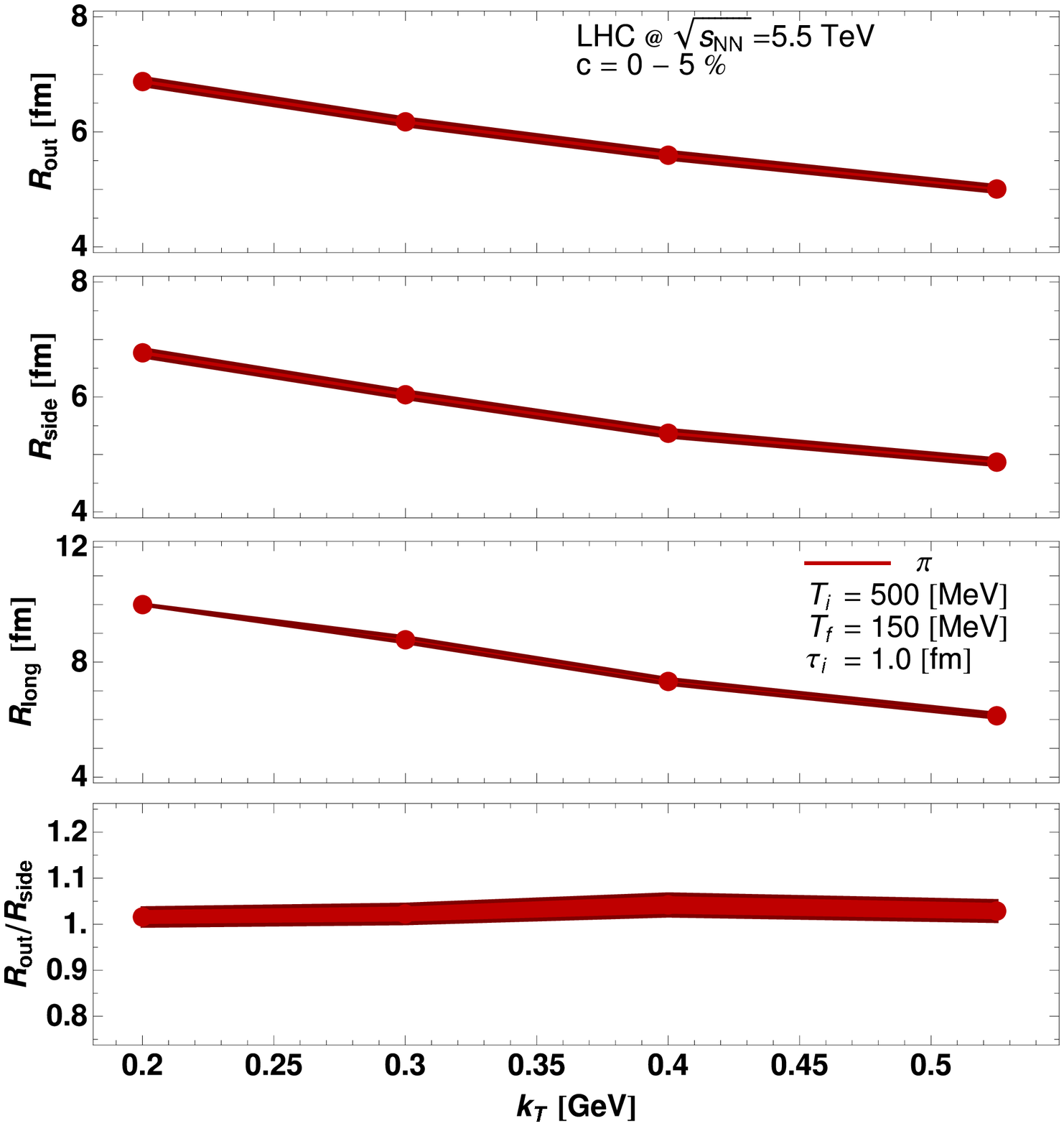}}
  \end{center}
  \caption{\small The model results for the pionic HBT radii. The calculation uses the two-particle method and includes the Coulomb effects.  The values of the model parameters are the same as in Fig. \ref{fig:reslhcc0005-hs}.}
  \label{fig:reslhcc0005-hbt}
\end{figure}
In the following Sections we show more details, discussing the spectra, $v_2$, and the HBT radii obtained for the initial temperature $T_{\rm i}$ = 400, 450 and 500 MeV.\\
In Fig. \ref{fig:reslhcc0005-hs} we show the freeze-out curves obtained from our hydrodynamic code with $T_{\rm i}$ = 400, 450 and 500 MeV. Comparing to the corresponding central RHIC collisions with \mbox{$T_{\rm i}$ = 320 MeV} from Fig. \ref{fig:resrhicTi320c0005-hs}, we observe that the difference in the initial temperature results  in the longer time of the hydrodynamic expansion and a larger transverse size (both increase by about 3 fm). On the other hand, similarly to the RHIC results, we find that the two freeze-out profiles overlap, hence the system at freeze-out is, as expected, azimuthally symmetric in the transverse plane. We also note that the shape of the isotherms is consistent with the result presented in Fig. 4 of Ref. \cite{Eskola:2005ue}.
\par In Fig.~\ref{fig:reslhcc0005-sppt} we give the model transverse-momentum spectra of hadrons. Compared to the RHIC results from Fig.~\ref{fig:resrhicTi320c0005-sppt}, we find much larger multiplicities of the produced hadrons and smaller slopes of the spectra, indicating the larger transverse flow that is caused by the larger initial temperature.\\
Our model calculations of the HBT radii are shown in Fig.~\ref{fig:reslhcc0005-hbt}. The increase of the central temperature from  \mbox{$T_{\rm i}= 320$~MeV} to \mbox{$T_{\rm i} = 400 - 500$~MeV} makes all the radii moderately larger. The ratio $R_{\rm out}/R_{\rm side}$ decreases by about 10\% which is an effect of the larger transverse flow caused, in turn, by the larger initial temperature.
%
\section{Non-central collisions}
\label{section:reslhc-noncentral}
%
\begin{figure}[!ht]
  \begin{center}
    \subfigure{\includegraphics[width=0.32 \textwidth]{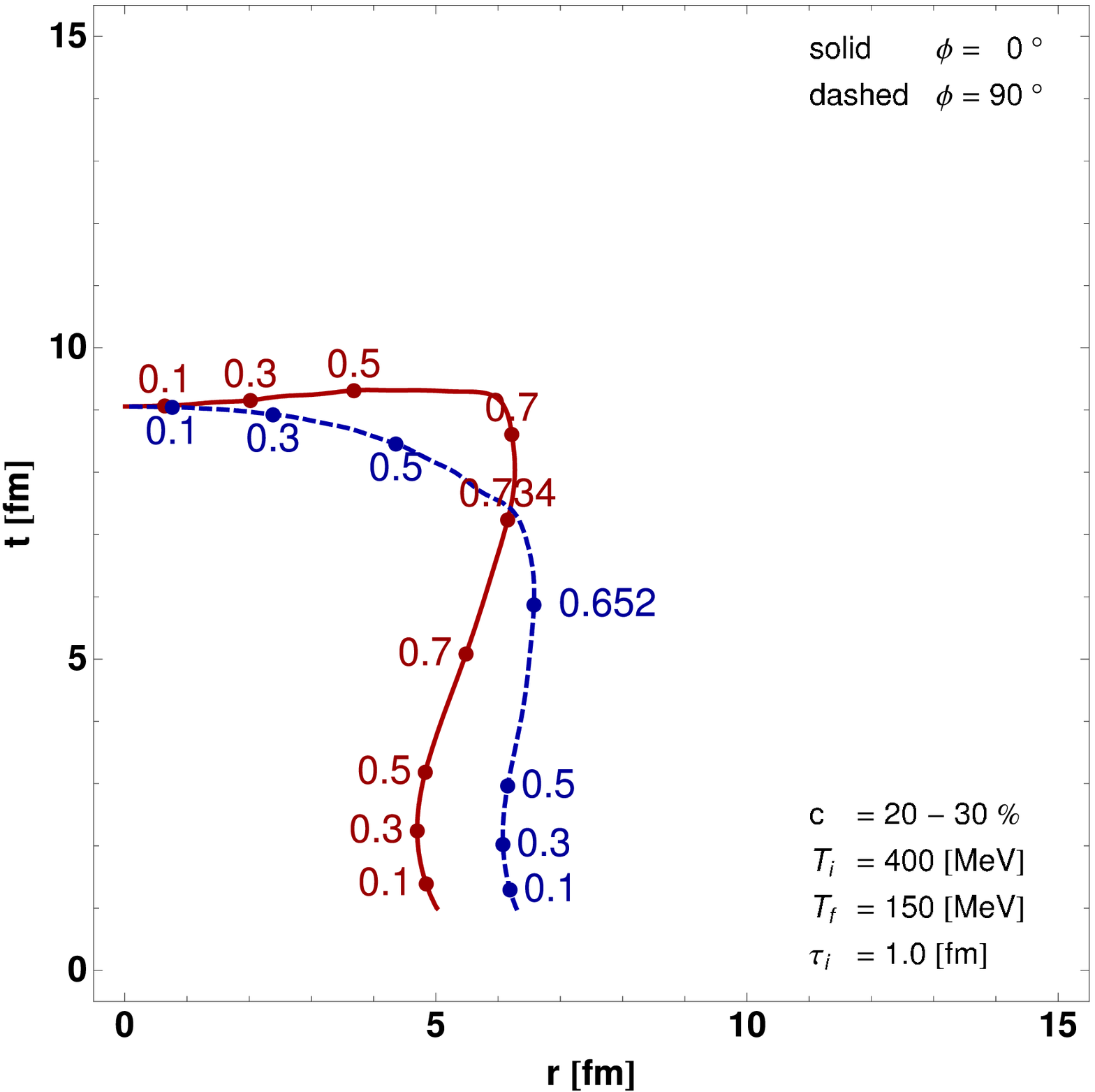}}
    \subfigure{\includegraphics[width=0.32 \textwidth]{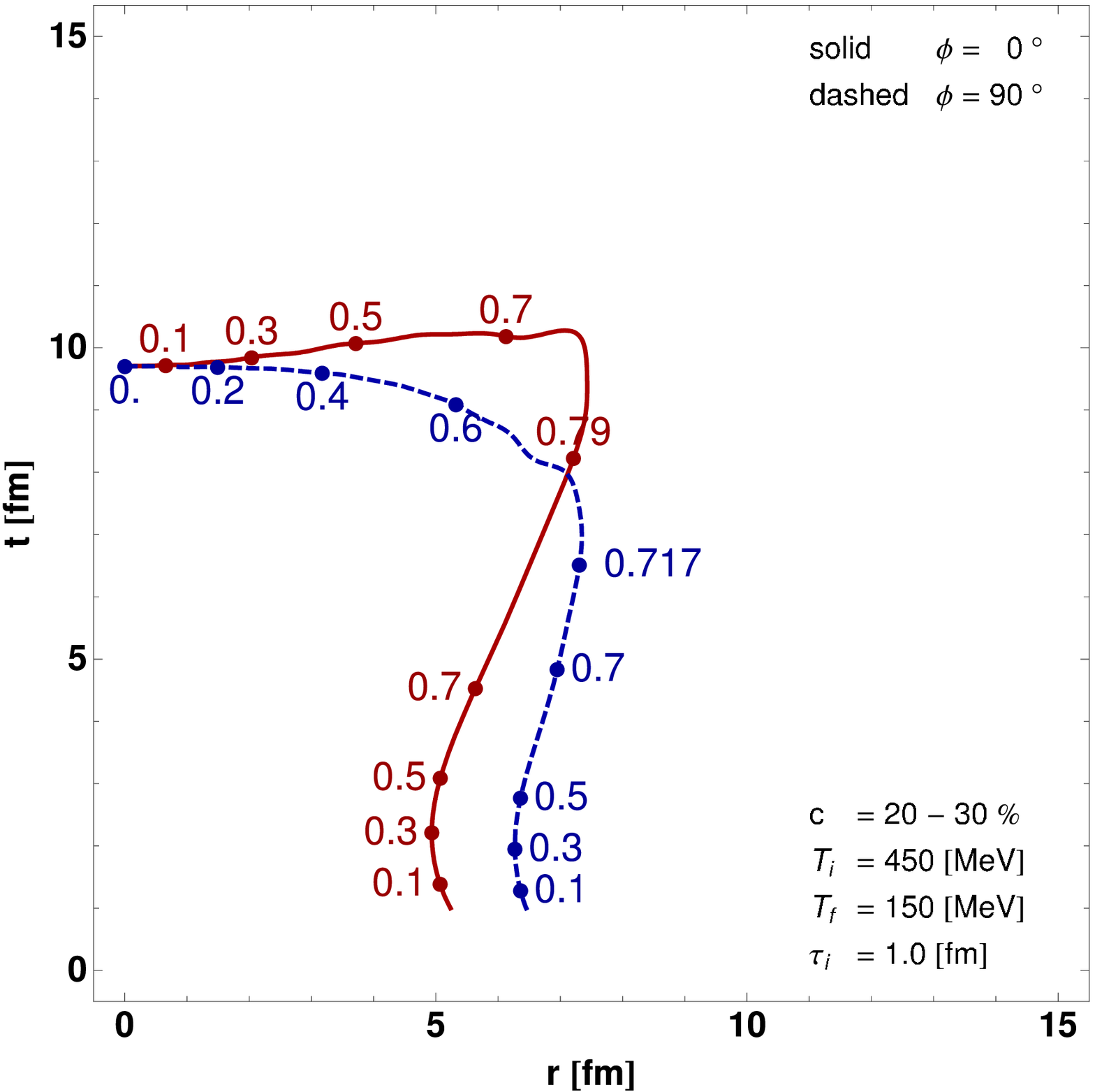}}
    \subfigure{\includegraphics[width=0.32 \textwidth]{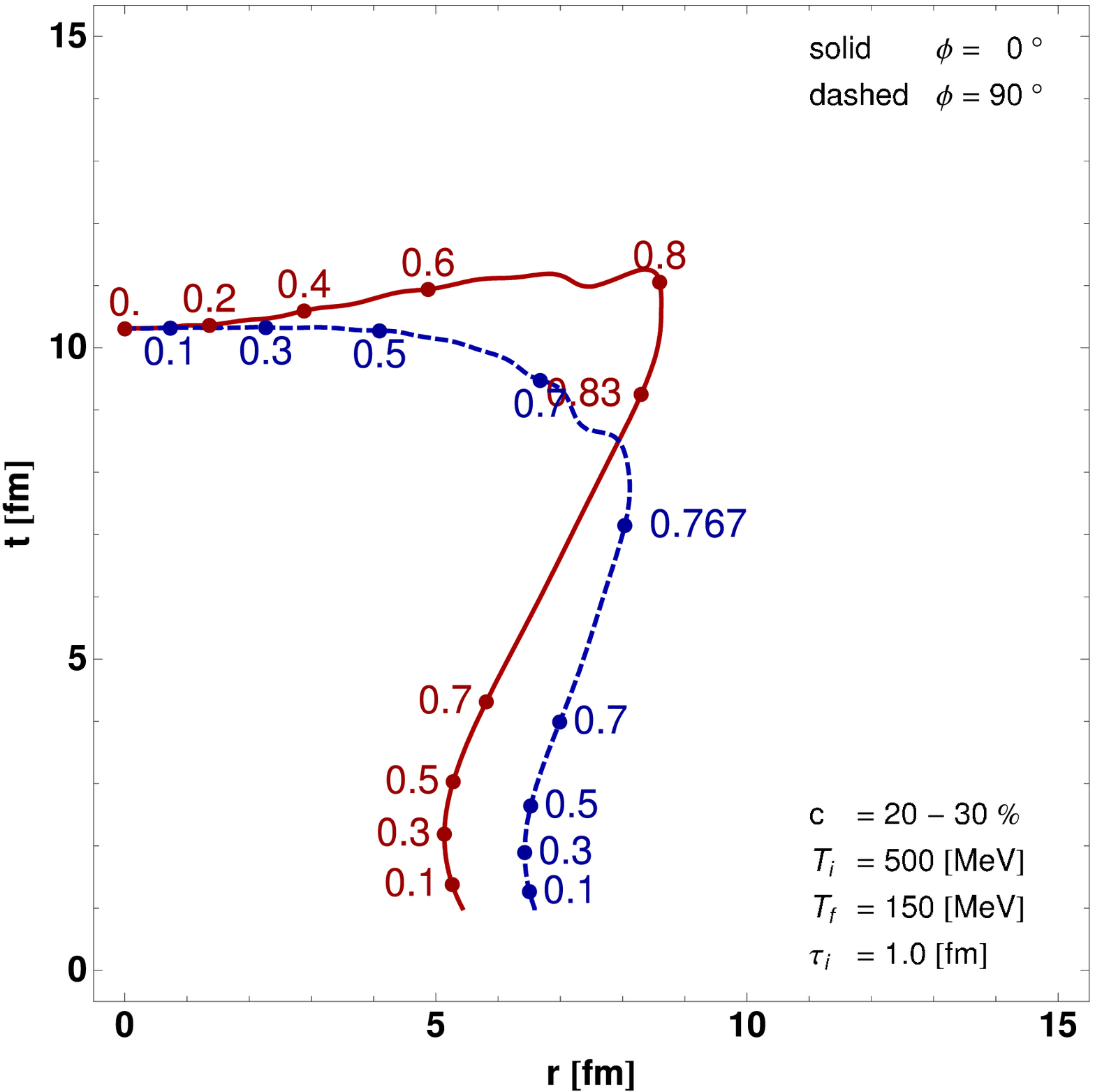}}
  \end{center}
  \caption{\small The freeze-out curves for peripheral collisions at LHC, $c = 20-30 \%$, \mbox{$b = 7.6~{\rm fm}$}, $\tau_{\rm i} = 1.0~{\rm fm}$, \mbox{$T_{\rm i} = 400, 450 \hbox{ and } 500~{\rm MeV}$}, and \mbox{$T_{\rm f} = 150~{\rm MeV}$}. The solid (dashed) line shows the in-plane (out-of-plane) profile.}
  \label{fig:reslhcc2030-hs}
\end{figure}
\begin{figure}[!tb]
  \begin{center}
    \subfigure{\includegraphics[width=0.45 \textwidth]{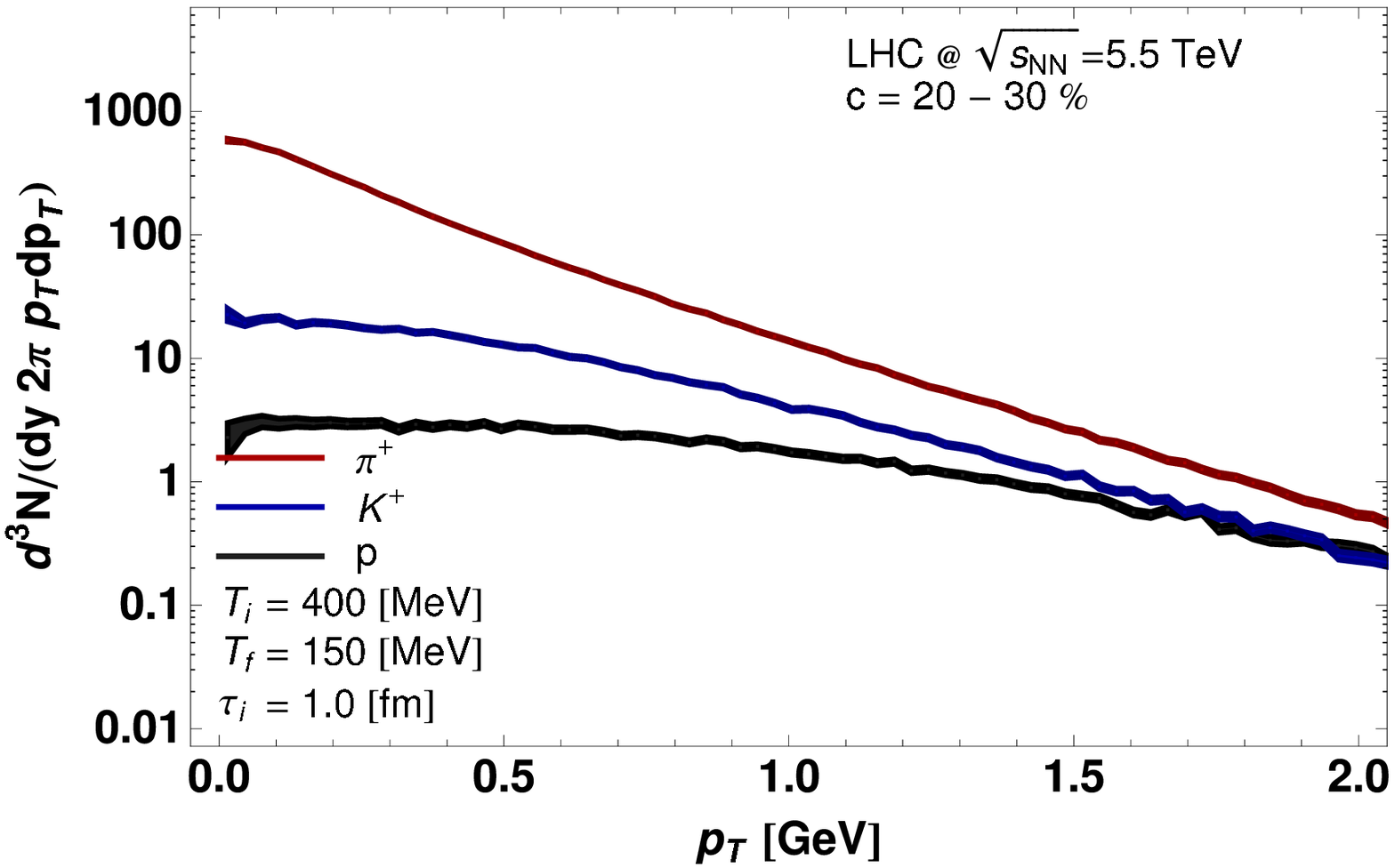}}
    \subfigure{\includegraphics[width=0.45 \textwidth]{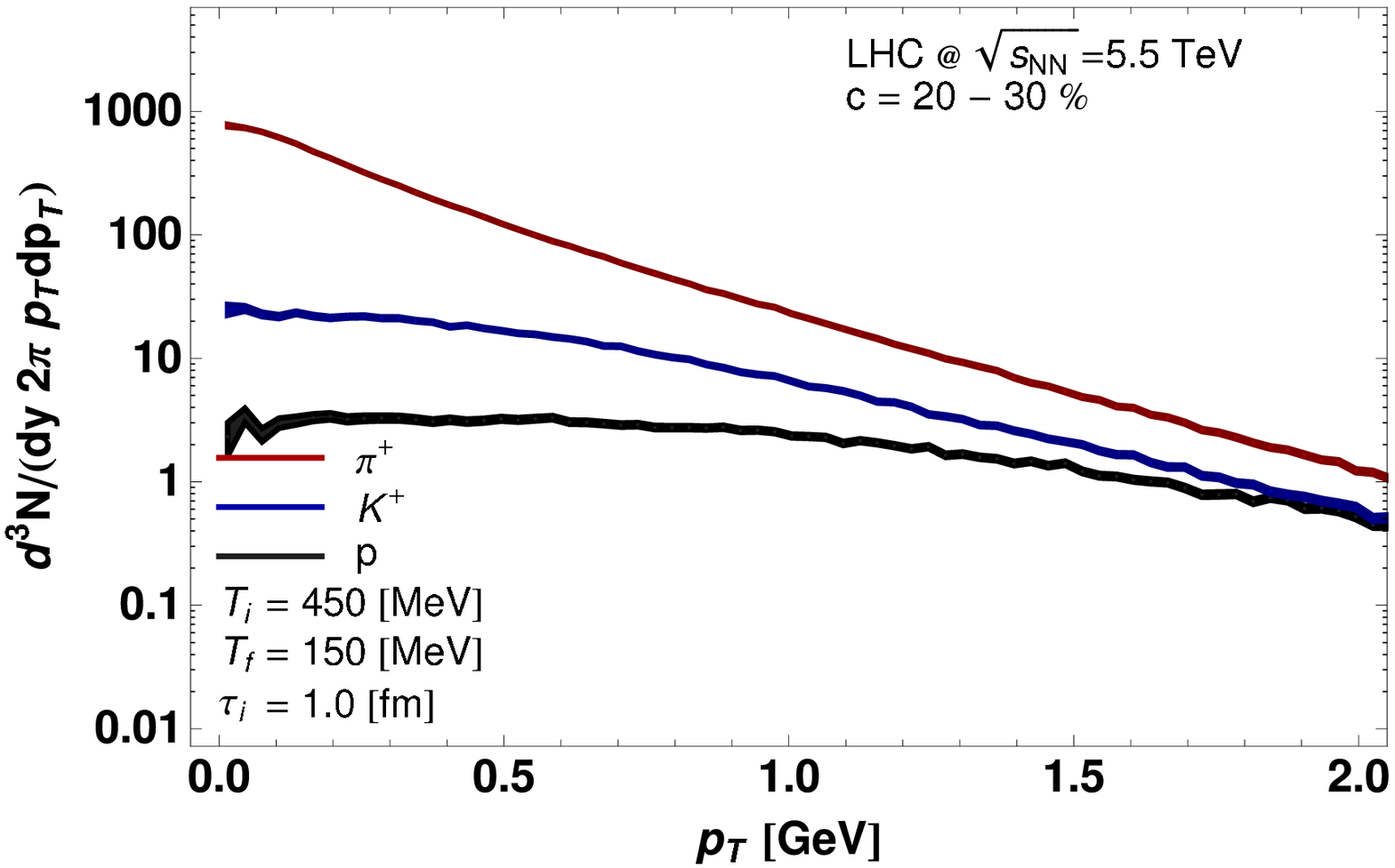}}\\
    \subfigure{\includegraphics[width=0.45 \textwidth]{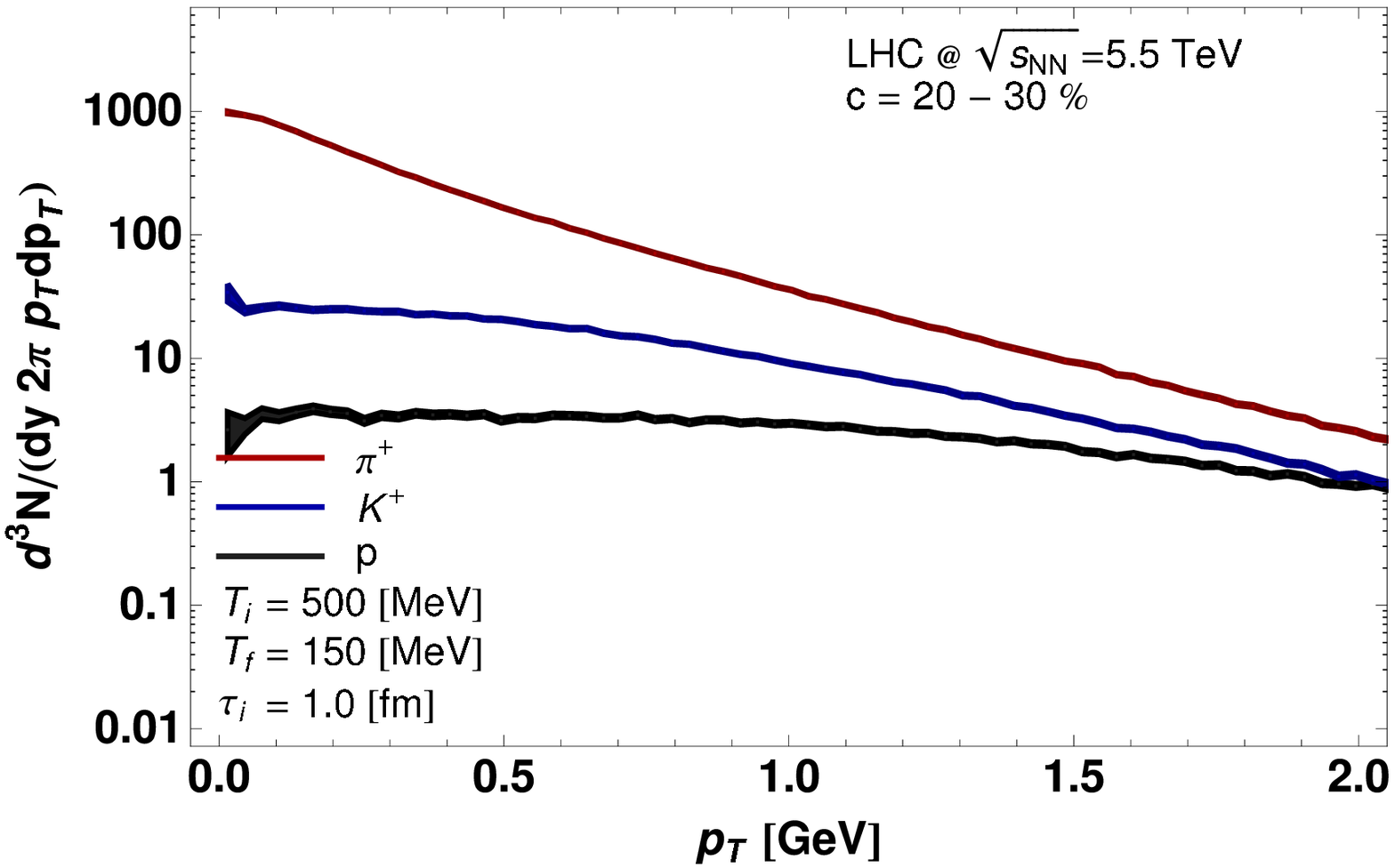}}
  \end{center}
  \caption{\small The model results for the transverse-momentum spectra of $\pi^+$, $K^+$, and protons. The values of the model parameters are the same as in Fig. \ref{fig:reslhcc2030-hs}.}
  \label{fig:reslhcc2030-sppt}
\end{figure}
\begin{figure}[tb]
  \begin{center}
    \subfigure{\includegraphics[width=0.45 \textwidth]{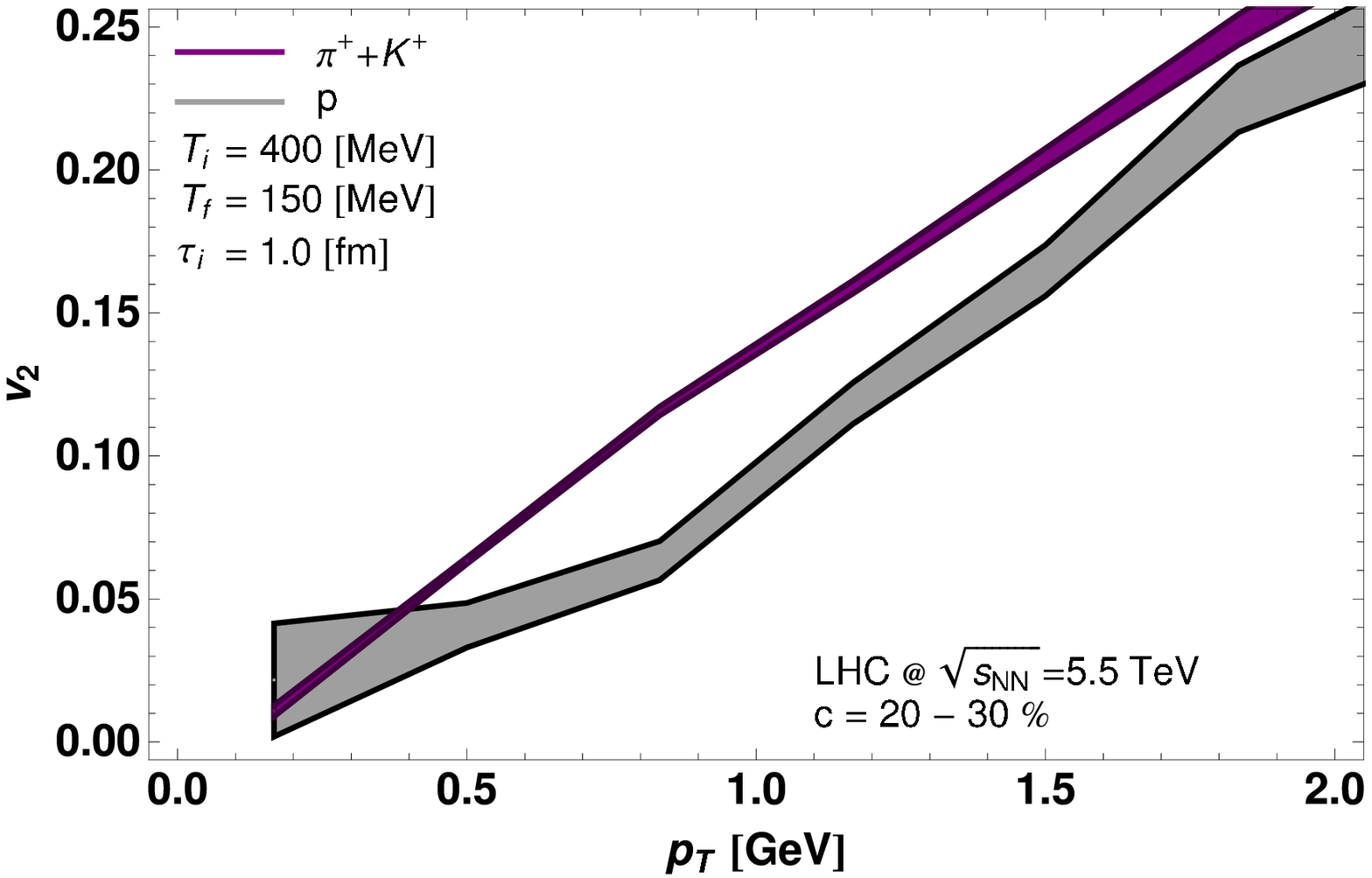}}
    \subfigure{\includegraphics[width=0.45 \textwidth]{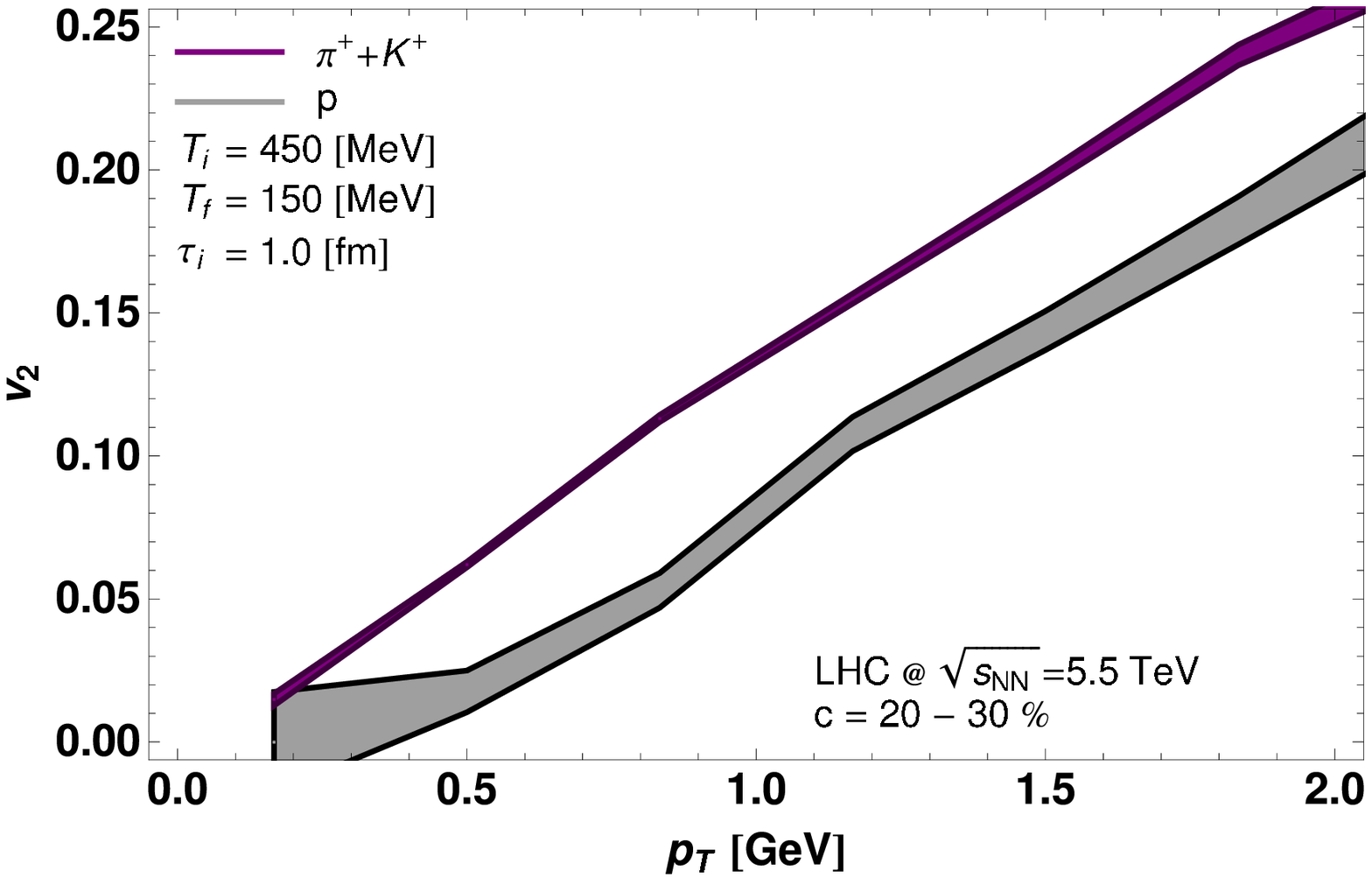}}\\
    \subfigure{\includegraphics[width=0.45 \textwidth]{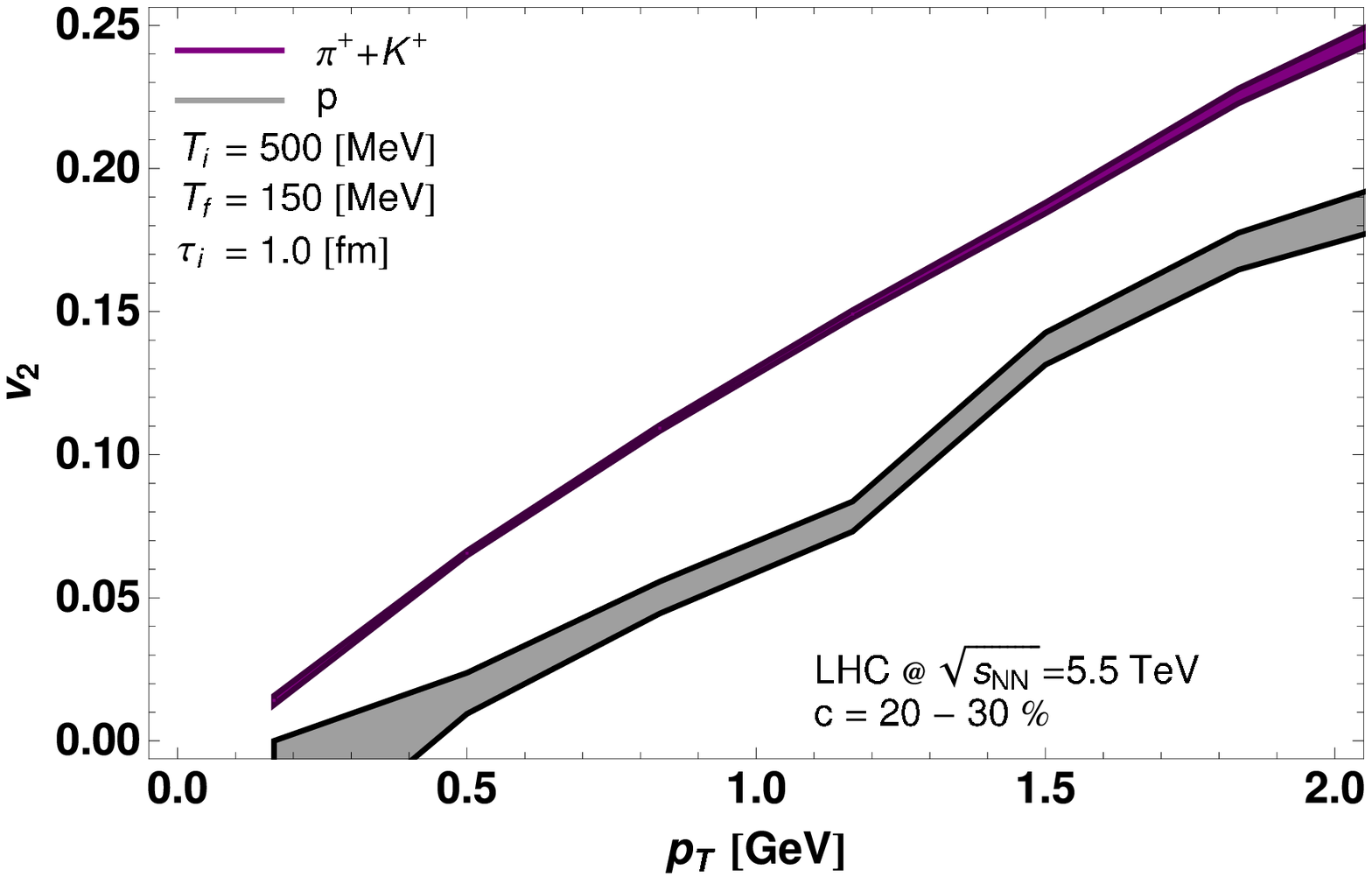}}
  \end{center}
  \caption{\small The elliptic flow coefficient $v_2$. The parameters are the same as in Fig.~\ref{fig:reslhcc2030-hs}.  }
  \label{fig:reslhcc2030-v2pt}
\end{figure}
\begin{figure}[!t]
  \begin{center}
    \subfigure{\includegraphics[width=0.32 \textwidth]{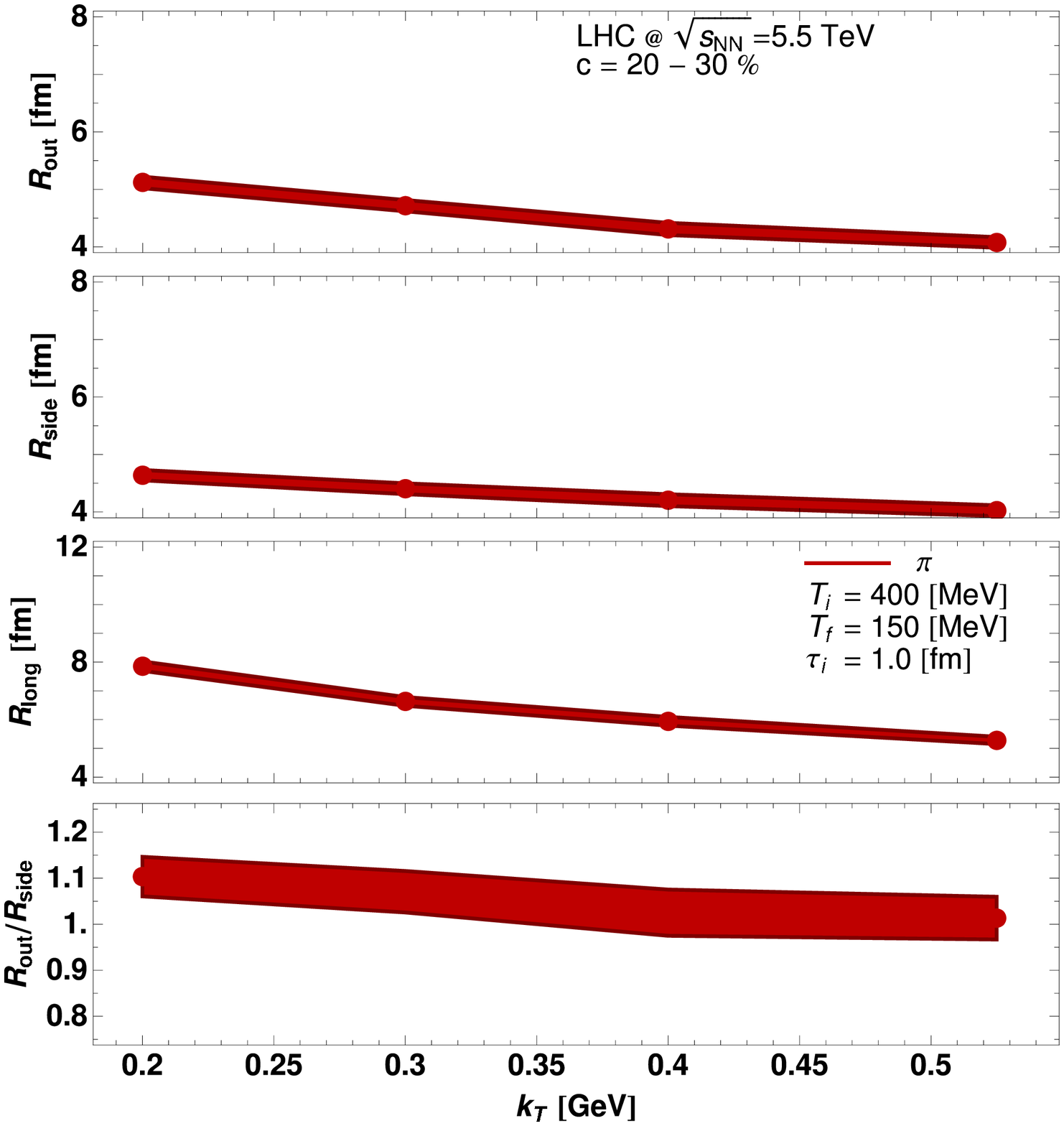}}
    \subfigure{\includegraphics[width=0.32 \textwidth]{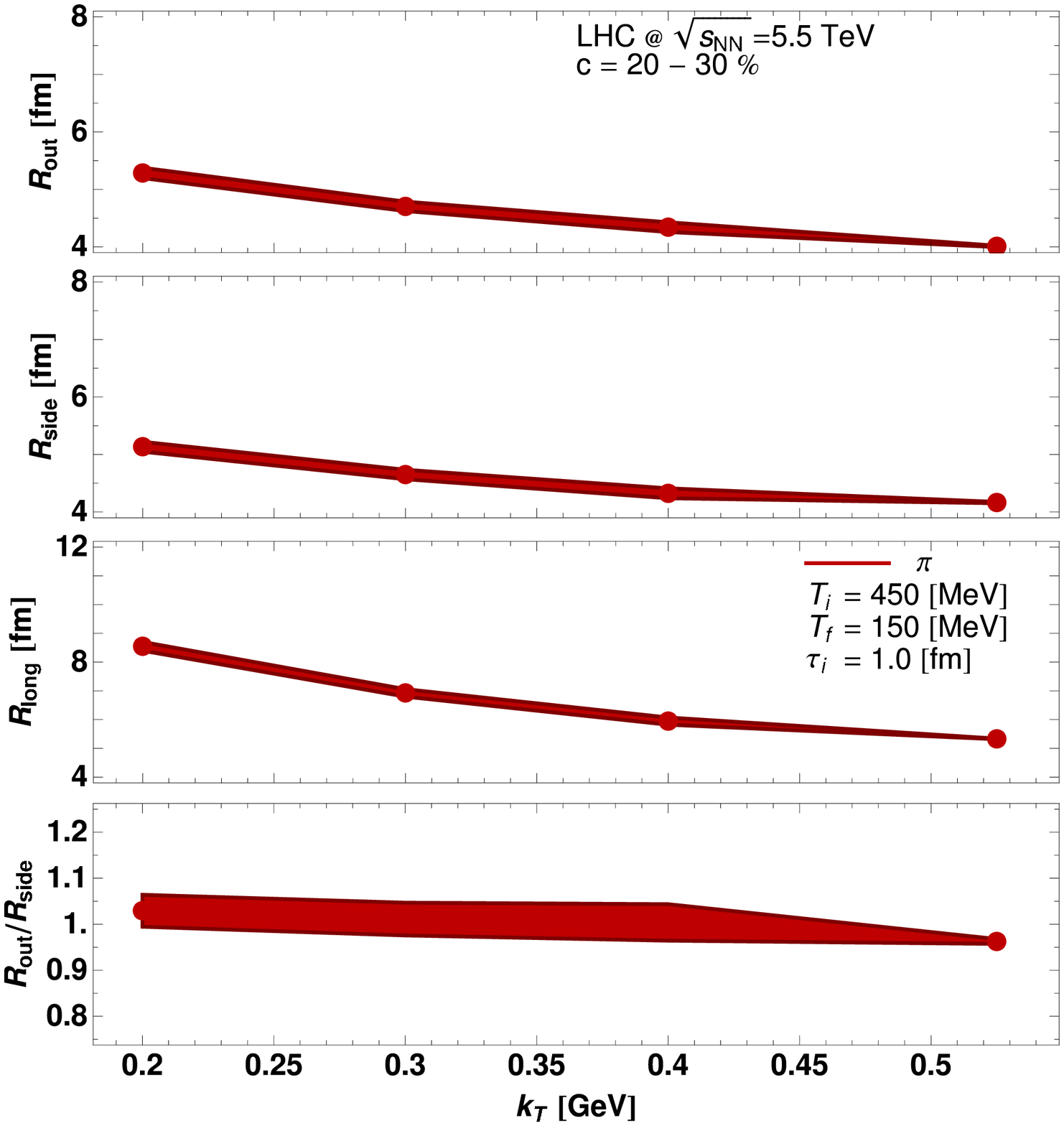}}
    \subfigure{\includegraphics[width=0.32 \textwidth]{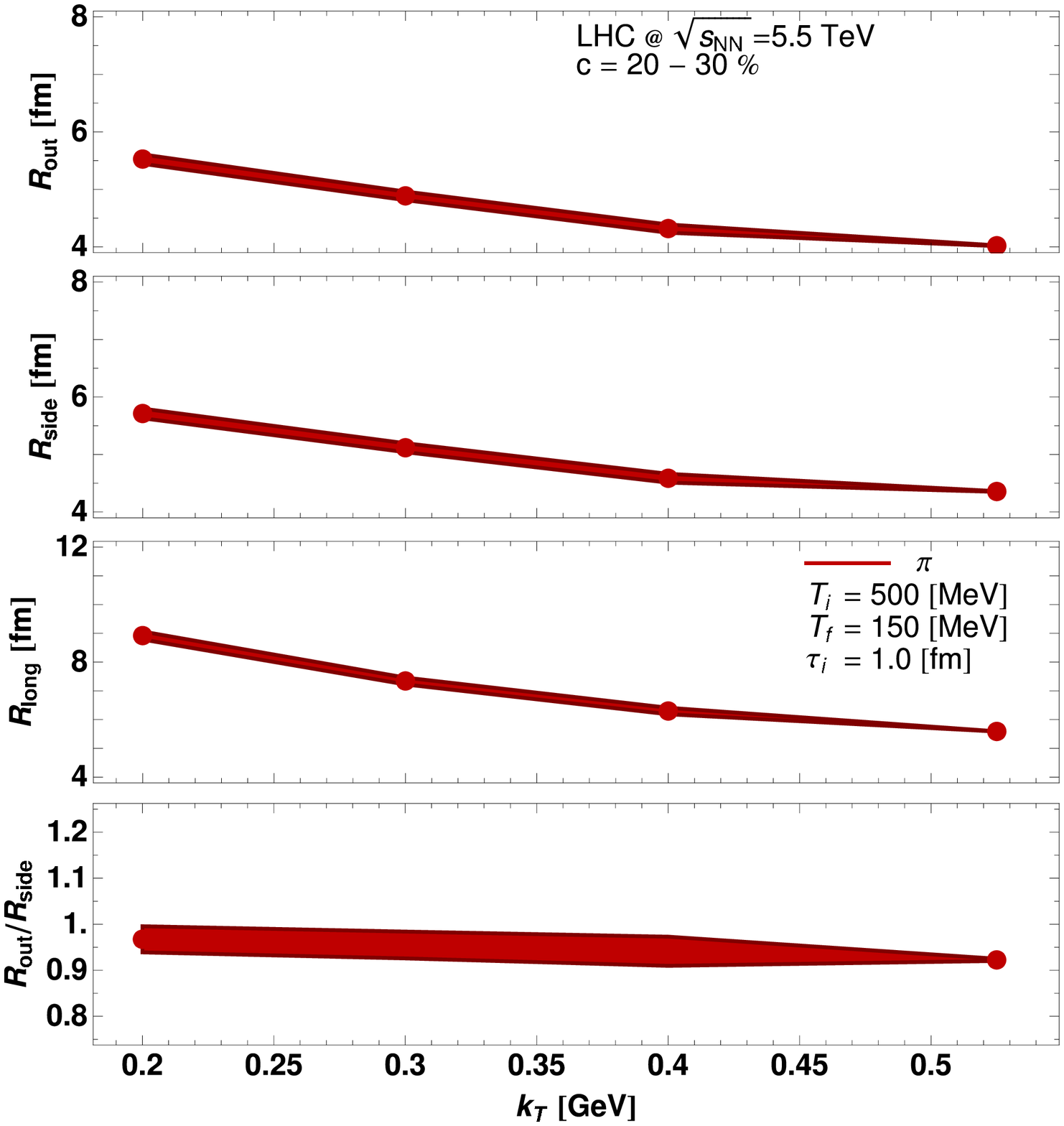}}
  \end{center}
  \caption{The pionic HBT radii. The values of the model parameters are the same as in Fig. \ref{fig:reslhcc2030-hs}. }
  \label{fig:reslhcc2030-hbt}
\end{figure}
In this Section we present our results describing the peripheral collisions with \hbox{$T_{\rm i} = 400$}, 450 and 500  MeV, $T_{\rm f} = 150~{\rm MeV}$, and $b = 7.6~{\rm fm}$. In Fig. \ref{fig:reslhcc2030-hs} we show the freeze-out curves. One can observe that the system is initially elongated along the $y$ axis, but in the end of the evolution it becomes elongated along the $x$ axis. This behavior is indicated by the crossing of the freeze-out curves. The change of shape is caused by the strong flow which transforms the initial ``almond'' into a ``pumpkin'' \cite{Heinz:2002sq}. 
Azimuthally sensitive HBT probes such non-trivial behavior and can be used as a precise confirmation test for the existence of such effects in the data.
\par In Fig.~\ref{fig:reslhcc2030-sppt} we show the model transverse-momentum spectra of hadrons for the same values of the parameters. Compared to the RHIC case with $T_{\rm i} = 320~{\rm MeV}$ and the same values of $T_{\rm f}$, we find flatter spectra with higher multiplicity. In Fig.~\ref{fig:reslhcc2030-v2pt} we show our results for the elliptic-flow coefficient. The stronger transverse flow generated in this case induces larger splitting between the pion+kaon $v_2$ and the proton $v_2$, with the values of the pion+kaon elliptic flow very similar to those found in the case $T_{\rm i} = 320~{\rm MeV}$. This result indicates the saturation of the elliptic flow of light particles for a given initial space asymmetry. On the other hand, the proton elliptic flow is significantly reduced. This observation is consistent with the findings of Kestin and Heinz discussed in Ref.~\cite{Abreu:2007kv}. Finally, in Fig.~\ref{fig:reslhcc2030-hbt} we show our model calculations of the HBT radii. We note that the ratio $R_{\rm out}/R_{\rm side}$ is very close to one.

\chapter[Uniform description of the RHIC data]{Uniform description of the RHIC data with Gaussian initial conditions}
\label{chapter:resrhic2}
%
In this Chapter we return to the discussion of the RHIC results obtained at the top beam energy of $\sqrt{s_{NN}}$ = 200 GeV. In the approach presented in Chapter \ref{chaper:resrhic} we used the standard initial conditions and were able to reproduce the experimental data with the accuracy of about 10-20\%. In the following we are going to show that further improvement in the description of the data may be achieved if one uses modified initial conditions, namely,  we suggest to use a two-dimensional Gaussian as an initial profile for the energy density in the transverse plane.
\par Of course, the elaboration of the precise initial conditions for hydrodynamics remains a challenge. Ideally, such initial conditions should be provided by the early-stage dynamics, for instance by the Color Glass Condensate \cite{McLerran:1993ni,McLerran:1993ka,Kharzeev:2001yq}. In practice, the theory of the early stage carries some uncertainty in its parameters. Moreover, different phenomena may be present in the early stage, see e.g. \cite{Mrowczynski:1993qm,Muller:2007rs}, so its precise modeling is difficult. 
%
\section{Early start of hydrodynamics}
\label{section:resrhic2-early-start}
%
In the approach discussed in this Section we use a simple parameterization of the initial energy density profile in the transverse plane. At the proper time of \mbox{$\tau_i=0.25$~fm} we assume that it may be described by a Gaussian
\begin{eqnarray}
  \varepsilon({\bf x}_\bot) = \varepsilon_{\rm i} \exp \left ( -\frac{x^2}{2a^2} -\frac{y^2}{2 b^2} \right ),
  \label{eqn:resrhic2_e0}
\end{eqnarray} 
where the values of the $a$ and $b$ width parameters depend on the centrality class and are obtained by matching to the results for $\langle x^2 \rangle$ and $\langle y^2 \rangle$ from {\tt GLISSANDO} \cite{Broniowski:2007nz}, which implements the eccentricity fluctuations of the system in the Glauber approach \cite{Andrade:2006yh,Hama:2007dq}. The expression $(a^2+b^2)/2$ describes the overall transverse size of the system, whereas $(b^2-a^2)/(b^2+a^2)$ parameterizes its eccentricity. The use of the Gaussian profile (\ref{eqn:resrhic2_e0}) leads to a faster development of the transverse flow as compared to the Glauber initial conditions. This effect is crucial for the success of reproducing the HBT radii. 
\par The values for centrality classes used in this work are collected in Table~\ref{tab:resrhic2_ab}. Since the equation of state is known, the initial central temperature $T_{\rm i}$ defines the initial central energy density $\varepsilon_{\rm i}$. It depends on centrality and has been adjusted to reproduce the total pion multiplicity.
\begin{table}[!t]
  \caption{Initial central temperatures and shape parameters for various centrality classes. 
  \label{tab:resrhic2_ab}}
  \begin{center}
  \begin{tabular}{r|rrrrrr}
      $\displaystyle \vphantom{\frac{1}{2}}$ $c$ [\%] &  0 -  5 &  5 - 10 & 10 - 20 & 20 - 30 & 30 - 40 & 40 - 50 \\ \hline 
      $\displaystyle \vphantom{\frac{1}{2}}$ $T_{\rm i}$ [MeV] & 500 & 490 &    475 &     460 &     430 & 390 \\
      $\displaystyle \vphantom{\frac{1}{2}}$ $a$ [fm] &   2.696 &   2.536 &   2.284 &   2.000 &   1.771 &   1.577 \\
      $\displaystyle \vphantom{\frac{1}{2}}$ $b$ [fm] &   2.925 &   2.849 &   2.738 &   2.591 &   2.448 &   2.305 \\
      \multicolumn{7}{c}{}\\
      \multicolumn{7}{c}{}\\
      $\displaystyle \vphantom{\frac{1}{2}}$ $c$ [\%] & 50 - 60 & 60 - 70 & 70 - 80 & 80 - 90 & 90 - 100 & 30 - 80 \\ \hline
      $\displaystyle \vphantom{\frac{1}{2}}$ $T_{\rm i}$ [MeV] & 345 & 303 &    260 &     --- &    ---   &  330 \\
      $\displaystyle \vphantom{\frac{1}{2}}$ $a$ [fm] &   1.401 &   1.224 &   1.040 &   0.841 &    0.511 &   1.592 \\
      $\displaystyle \vphantom{\frac{1}{2}}$ $b$ [fm] &   2.164 &   2.019 &   1.854 &   1.649 &    1.228 &   2.307 \\
  \end{tabular}
  \end{center}
  \vspace{-5mm}
\end{table}
With the Gaussian initial conditions we continue the hydrodynamic evolution till the freeze-out temperature $T_{\rm f}=145$~MeV, where the system decouples and hadrons (stable and resonances) are generated according to the standard Cooper-Frye formalism \cite{Cooper:1974mv} -- see Chapters \ref{chaper:resrhic} and \ref{chaper:reslhc},  Eq. {\ref{eqn:freeze_6cf1}}. The lower value of $T_{\rm f}$ leads to larger sizes of the system and larger transverse flow. This in turn results in flatter $p_{\rm T}$ spectra and larger splitting of the $v_2$ of pions/kaons and protons.
\par
\begin{figure}[!ht]
  \begin{center}
    \subfigure{\includegraphics[width=0.4\textwidth]{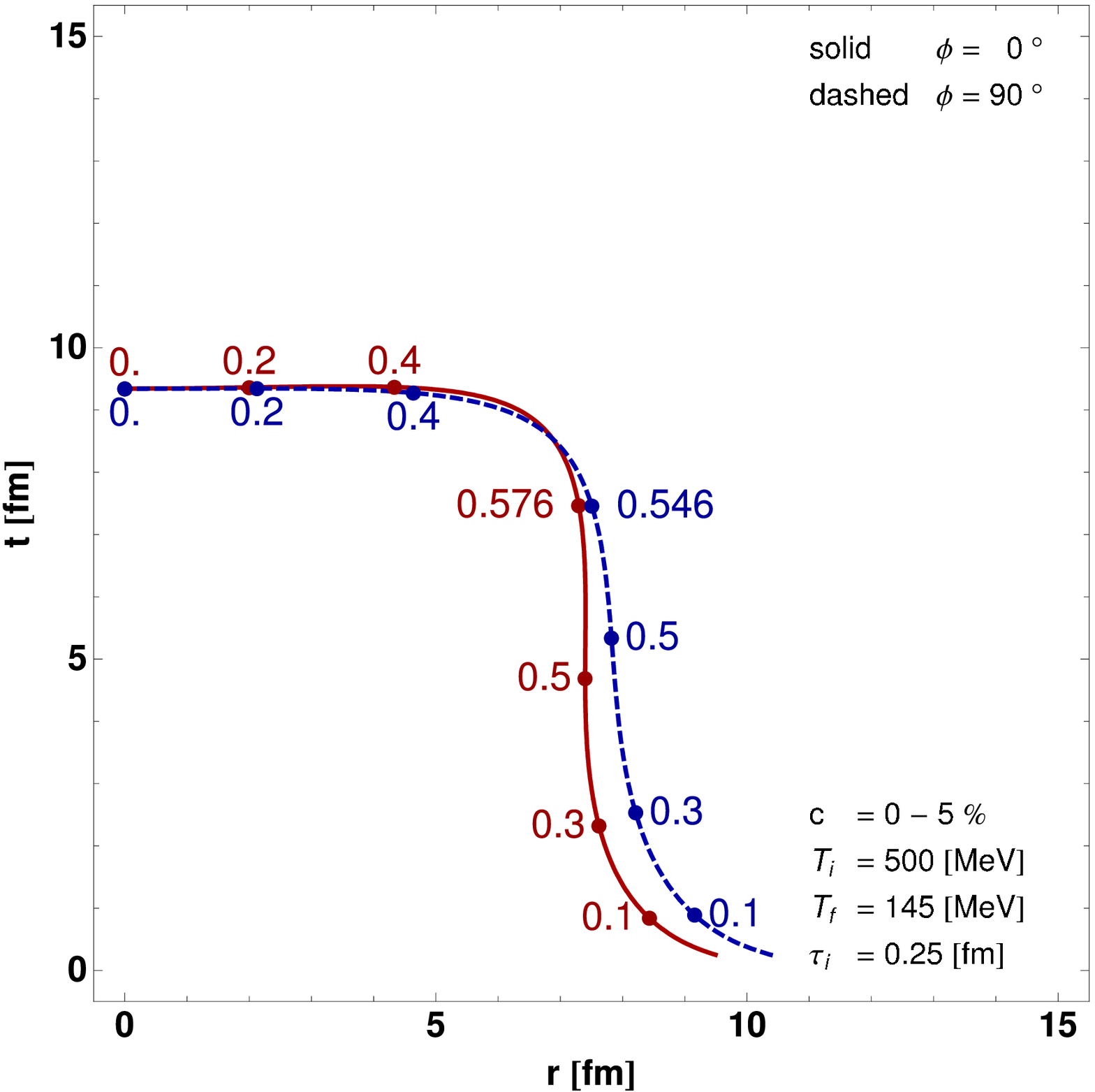}}
    \hspace{1cm}
    \subfigure{\includegraphics[width=0.4\textwidth]{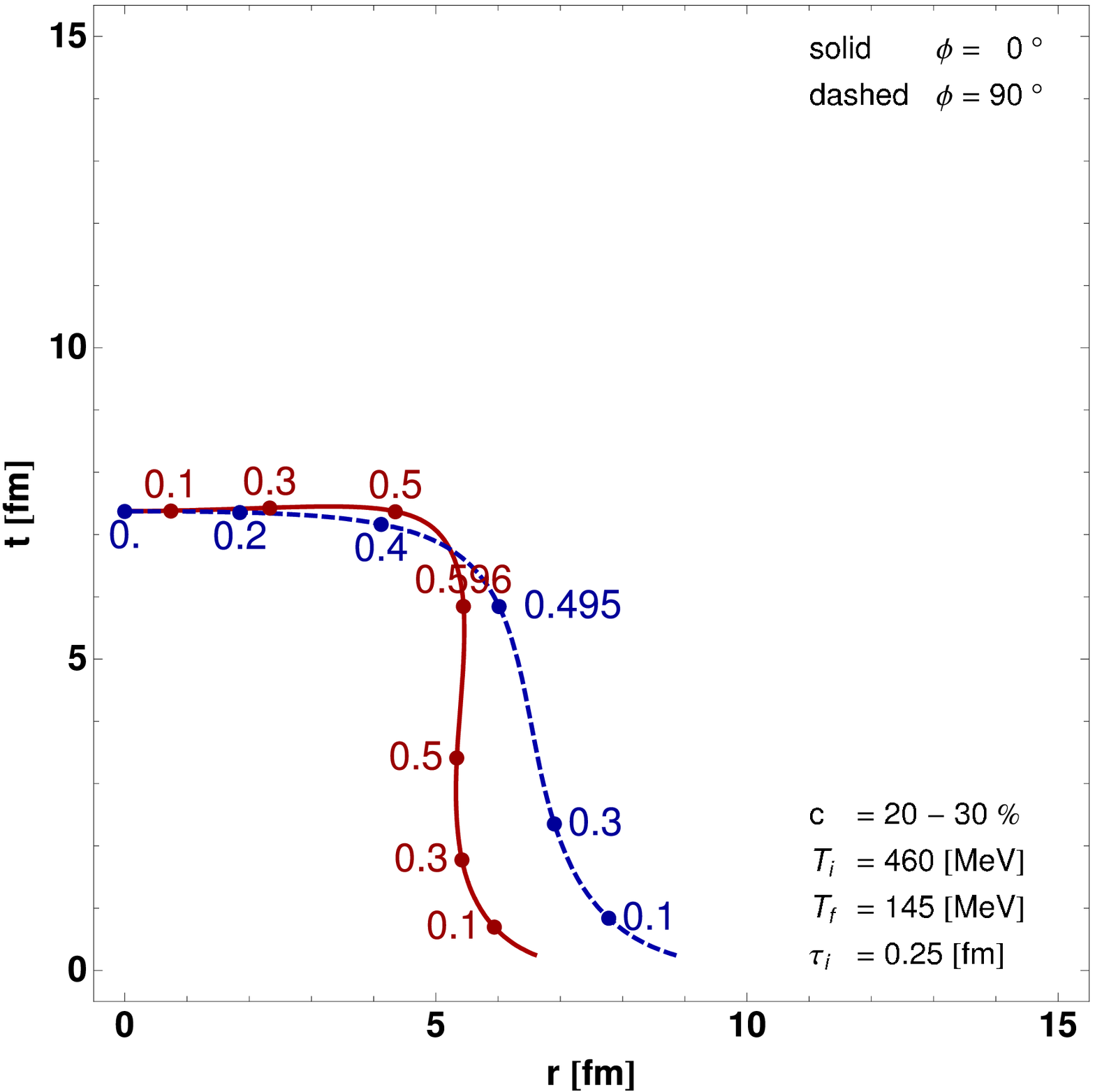}}
  \end{center}
  \caption{\small  Freeze-out hypersurfaces for \mbox{$c=0-5$\%} (left) and \mbox{$c=20-30$\%} (right), obtained with the initial temperature $T_{\rm i}$  = 500 MeV and 460 MeV, for the central and semi-peripheral collisions, respectively, $\tau_{\rm i} = 0.25~{\rm fm}$ and with the final temperature $T_{\rm f}=145$~MeV. The two curves show the in-plane and out-of-plane sections. The labels indicate the transverse flow velocity.}
  \label{fig:resrhic2nofs-hs}
\end{figure}

 The freeze-out hypersurfaces for two centrality classes are shown in Fig.~\ref{fig:resrhic2nofs-hs}. For comparison, in Figs.~\ref{fig:resrhicTi320c0005-hs} and \ref{fig:resrhicTi320c2030-hs} in Chapter \ref{chaper:resrhic} we showed the hypersurfaces obtained in a calculation with the standard Glauber initial condition (and with same equation of state).  We note that the Glauber initial condition results in hypersurfaces of a smaller transverse size and a longer evolution time, which translates into lower $R_{\rm side}$ and larger $R_{\rm out}$.  More generally, we note a quite different shape of the hypersurfaces in Fig.~\ref{fig:resrhic2nofs-hs}, however, the flow values are similar, which results from the fact that the model parameters are adjusted in such a way that the slopes of the $p_T$-spectra and the $v_2$ coefficient are reproduced. This leads to practically the same values of the flow velocity at freeze-out for the two compared calculations.
\par We should also stress out the similarity of the volume-emission parts (with time-like normal vectors) to the blast-wave parameterization \cite{Schnedermann:1993ws}, with a short lifetime: about 9~fm for central \mbox{(c = 0-5\%)} and 7~fm for mid-peripheral (c = 20-30\%) collisions. However, the freeze-out hypersurfaces contain also the surface emission parts (with space-like normal vectors), absent in the traditional blast-wave parameterizations. We have checked that the potential problems from the non-causal surface emission \cite{Bugaev:1996zq,Anderlik:1998et,Borysova:2005ng}, are negligible, as less than 0.5\% of particles are emitted back into the hydrodynamic region. The reason for this very small fraction is the sizable transverse flow velocity at large radii, as indicated by labels in Fig.~\ref{fig:resrhic2nofs-hs}, which pushes the particles outward, as well as the fact that the hypersurfaces are not bent back at low values of $t$. We have determined that about half of the produced particles comes from the volume part and about half from the surface part of the freeze-out hypersurface. The surface emission is crucial for fitting the HBT data from RHIC, as also advocated in \cite{Sinyukov:2006dw,Gyulassy:2007zz}.
\begin{figure}[!ht]
  \begin{center}
    \subfigure{\includegraphics[width=0.49\textwidth]{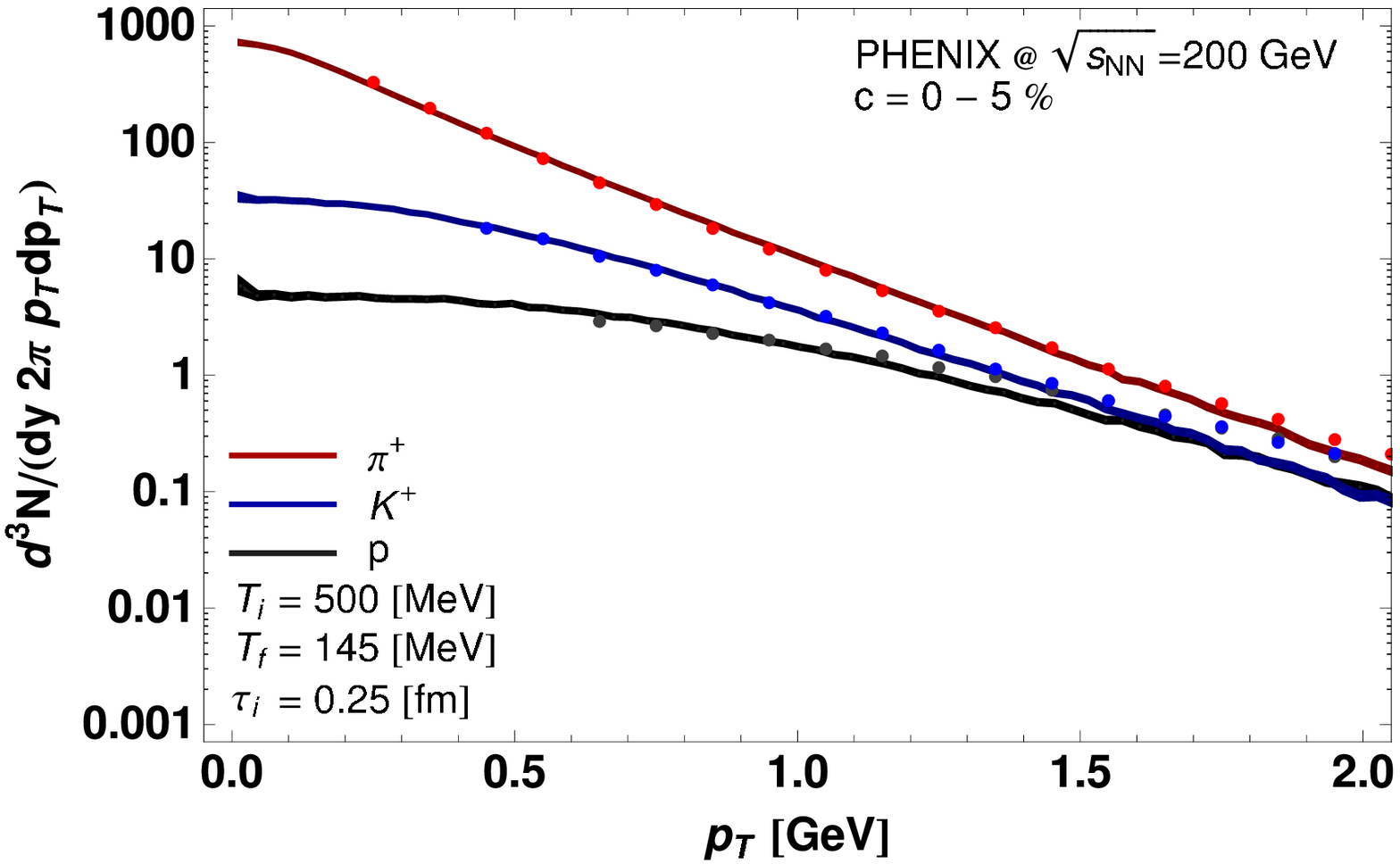}}
    \subfigure{\includegraphics[width=0.49\textwidth]{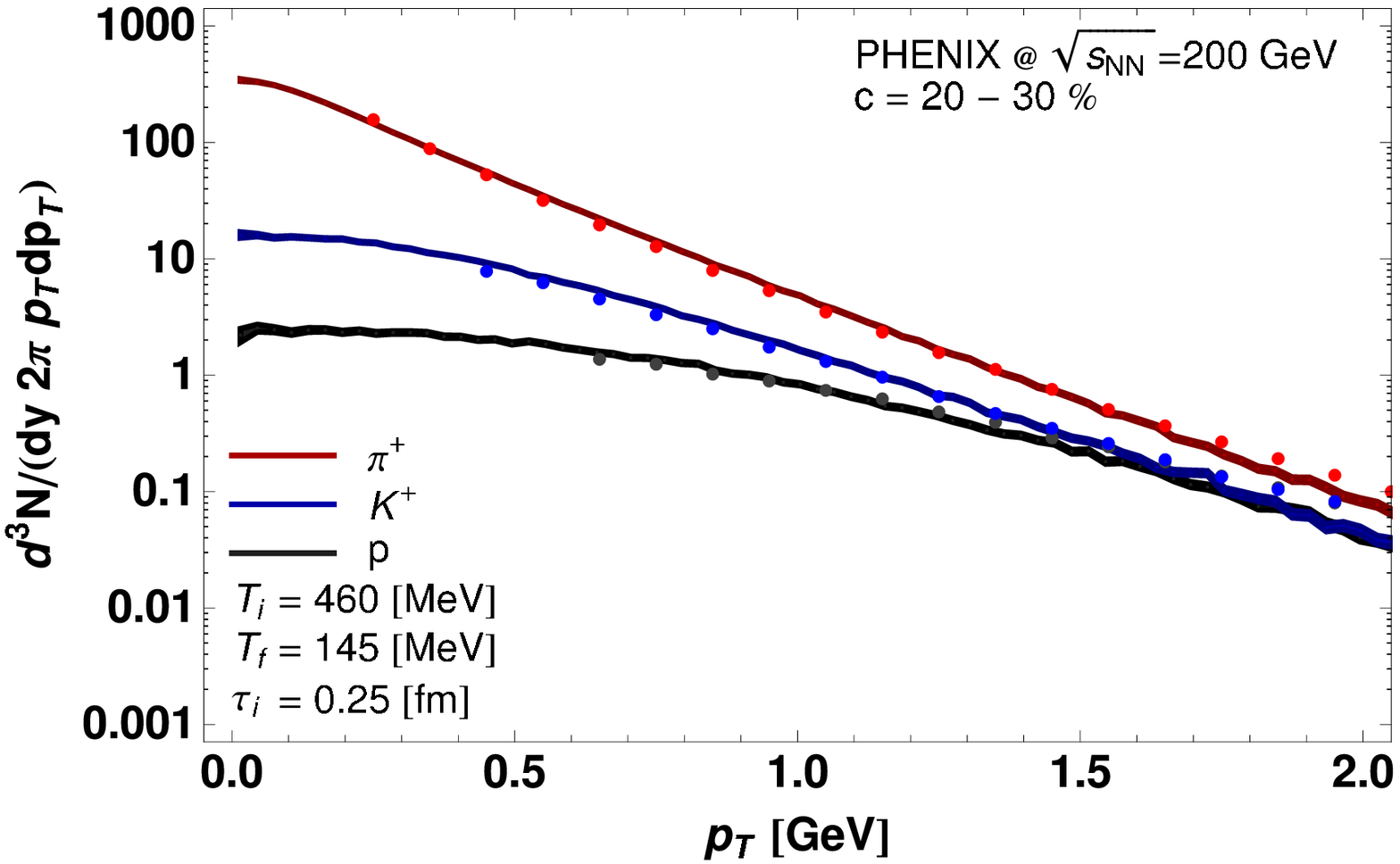}}\\
    \subfigure{\includegraphics[width=0.49\textwidth]{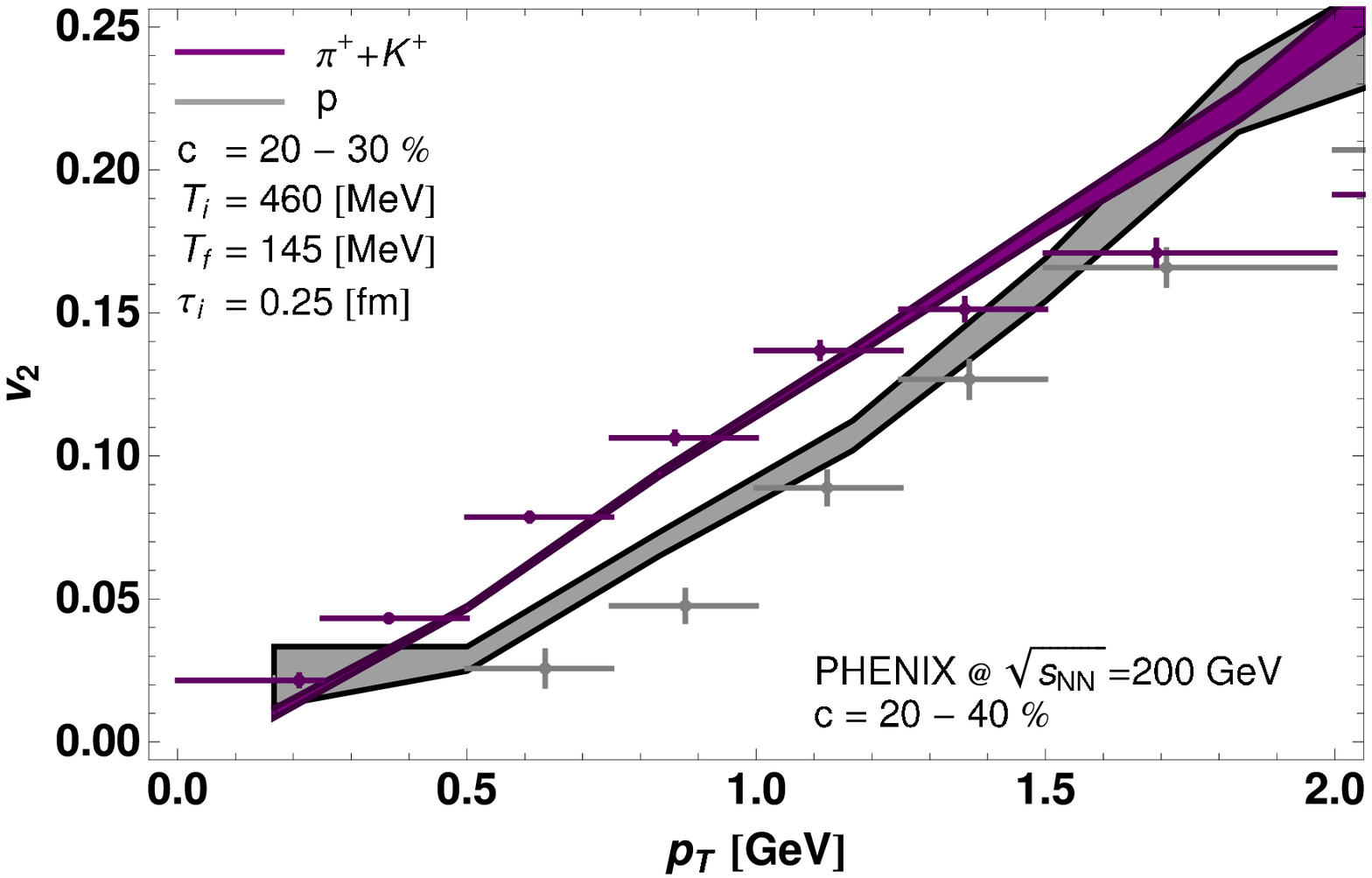}}
  \end{center}
  \caption{\small The transverse-momentum spectra of pions, kaons and protons for $c$=0-5\% (left), $c$=20-30\% (right), and the elliptic flow coefficient $v_2$ for  $c$=20-40\% (bottom), plotted as functions of the transverse momentum and compared to the data from \cite{Adler:2003cb,Adler:2003kt}. Model parameters same as in Fig.~\ref{fig:resrhic2nofs-hs}.}
  \label{fig:resrhic2nofs-spv2pt}
\end{figure}
\par In Fig.~\ref{fig:resrhic2nofs-spv2pt} we present our results for the hadron transverse-momentum spectra for positive pions and  kaons and protons in the central and semi-peripheral collisions with the experimental data from PHENIX \cite{Adler:2003cb}. The bottom plot of Fig.~\ref{fig:resrhic2nofs-spv2pt} represents the elliptic flow coefficient $v_2$ of joined pions + kaons and protons as a function of $p_{\rm T}$  together with the PHENIX \cite{Adler:2003kt} data points. We see a very good agreement of the one-particle observables. The $v_2$ description of protons is better than the one presented in the Chapter \ref{chaper:resrhic}. The splitting between the summed $v_2$ of pions and kaons is larger which is a sign of a larger flow difference that developed on the hypersurface. This description can be further improved by including the elastic rescattering processes among hadrons after freeze-out.
\par
\begin{figure}[!t]
  \begin{center}
    \subfigure{\includegraphics[width=0.49\textwidth]{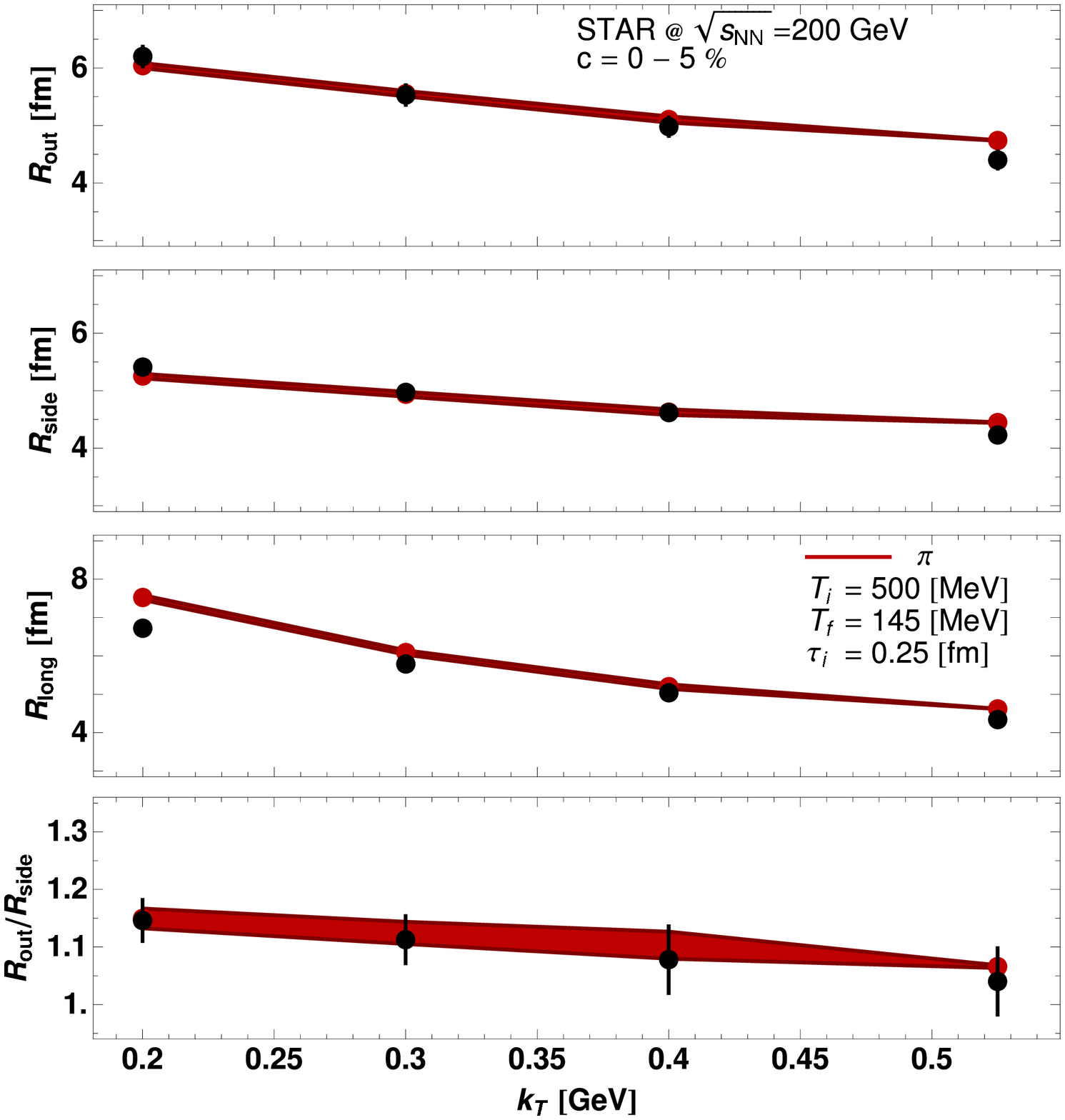}}
    \subfigure{\includegraphics[width=0.49\textwidth]{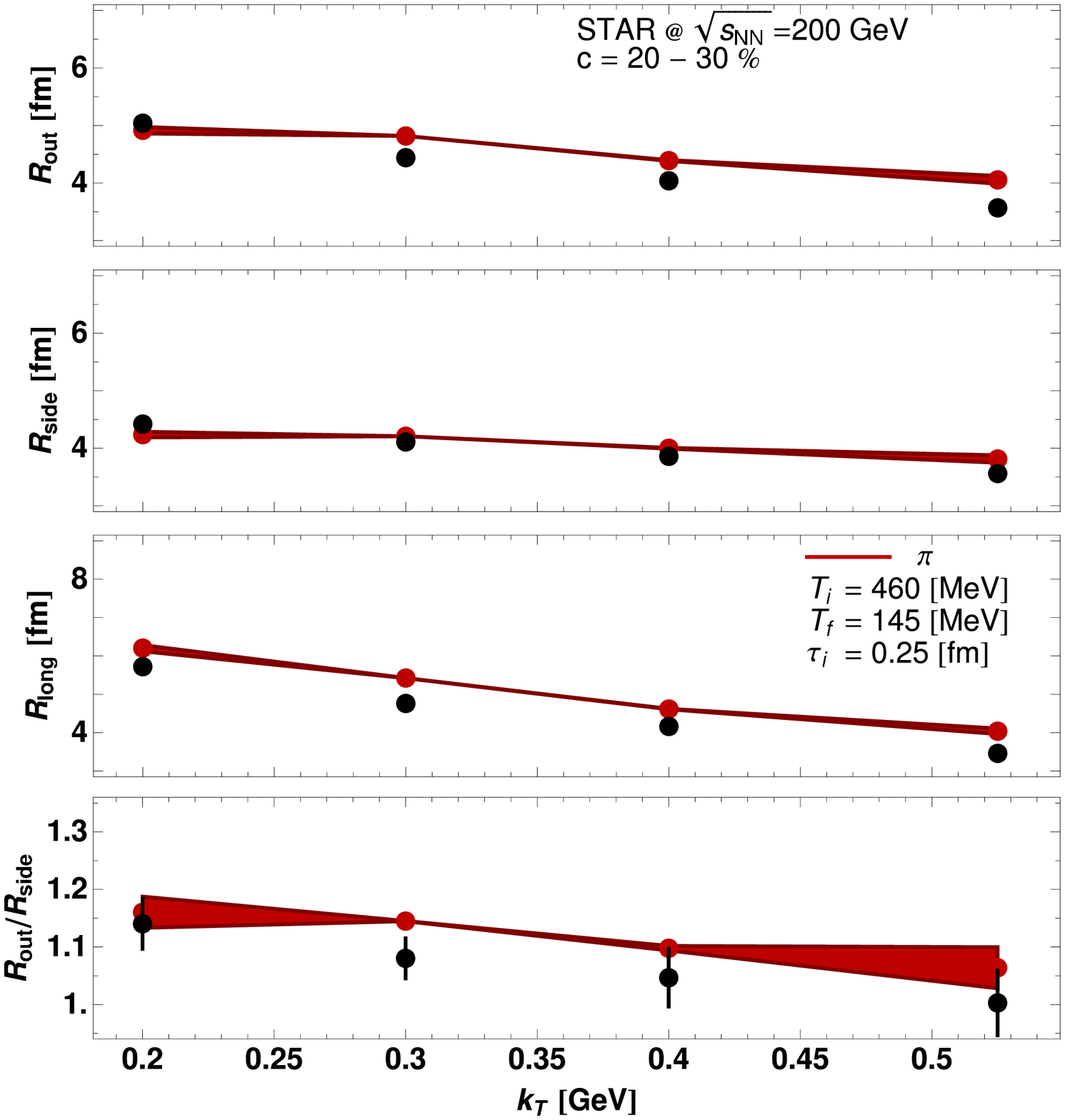}}\\
  \end{center}
  \caption{\small The pion HBT radii $R_{\rm side}$ , $R_{\rm out}$ , $R_{\rm long}$, and the ratio $R_{\rm out}/R_{\rm side}$ for central collisions, compared to the data from \cite{Adams:2004yc}. Model parameters same as in Fig. \ref{fig:resrhic2nofs-hs}.}
  \label{fig:resrhic2nofs-hbt}
\end{figure}

 Figure \ref{fig:resrhic2nofs-hbt} shows the HBT radii $R_{\rm side}$ , $R_{\rm out}$ , $R_{\rm long}$ and the ratio $R_{\rm out}/R_{\rm side}$. For the first time we can present a set of soft observables that are fully consistent with the experimental data. The most significant point is the correct description of the ratio $R_{\rm out}/R_{\rm side}$ in both central and non-central collisions. As in the previous calculation showed in Chapter \ref{chaper:resrhic},  $R_{\rm long}$ exceeds the data which is most likely caused by the boost-invariance of our model and can be fixed by implementing a rapidity dependent initial distribution.
\par Although the success of the use of the Gaussian initial condition is evident, it is clear that a more detailed model describing the initial non-equilibrated stage of the evolution is necessary to address the issue of the microscopic validity of the initial Gaussian conditions. We leave this problem for our future investigations.   
%
\section{Hydrodynamics preceded by free-streaming}
\label{section:resrhic2-late-start}
%
The very successful description of the soft observables that was presented in the previous Section demanded a very early start of the hydrodynamic evolution, \mbox{$\tau_{\rm i} = 0.25$} fm. Such short starting times have become now a common practice \cite{Pratt:2008qv}. Nevertheless, the processes needed to thermalize the system require at least a few collisions to create local equilibrium. In this respect, the starting time of 0.25 fm seems to be very difficult to explain microscopically.
\par In Chapter \ref{section:initial_free_stream} we have introduced the process of parton free-streaming. The basic idea behind it is the approximation of the early, stage evolution of a partonic system formed in heavy-ion collisions.  In our approach the parton free streaming is followed by sudden equilibration (described by the Landau matching conditions), and then by hydrodynamics. This kind of a formalism was proposed by Kolb, Sollfrank, and Heinz \cite{Kolb:2000sd} several years ago in the context of the development of azimuthally asymmetric flow. The approach assumes a sudden but delayed transition from a non-equilibrium initial state to a fully thermalized fluid.
\par The process of free-streaming decreases the spatial asymmetry of the density profile. This effect was the reason for not implementing the free streaming before hydrodynamics. The less asymmetric initial condition for the hydrodynamic stage produces smaller amounts of the elliptic flow. Free streaming itself cannot generate azimuthal asymmetry in the momentum distribution. Interactions among produced particles are needed to achieve this goal. But a sudden equilibration preceded by free-streaming is in fact capable of developing azimuthally asymmetric flow. The energy-momentum tensor of the system changes abruptly into a diagonal form (in the reference frame co-moving with the fluid element). In this way the space-flow correlations are induced, which results in a collective elliptic flow, further enhanced by the subsequent hydrodynamic evolution.
\begin{figure}[!ht]
  \begin{center}
    \subfigure{\includegraphics[width=0.4\textwidth]{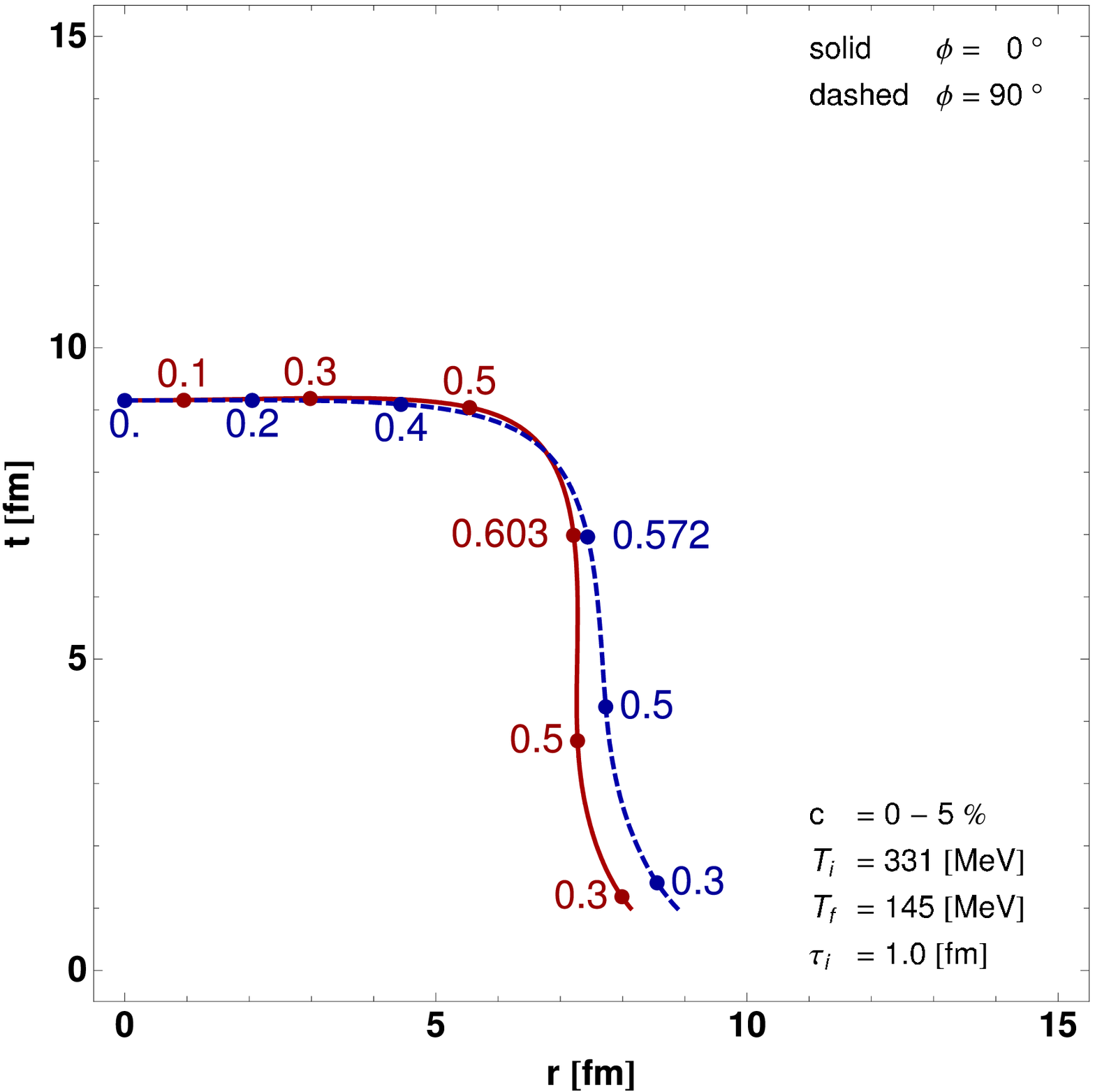}}
    \hspace{1cm}
    \subfigure{\includegraphics[width=0.4\textwidth]{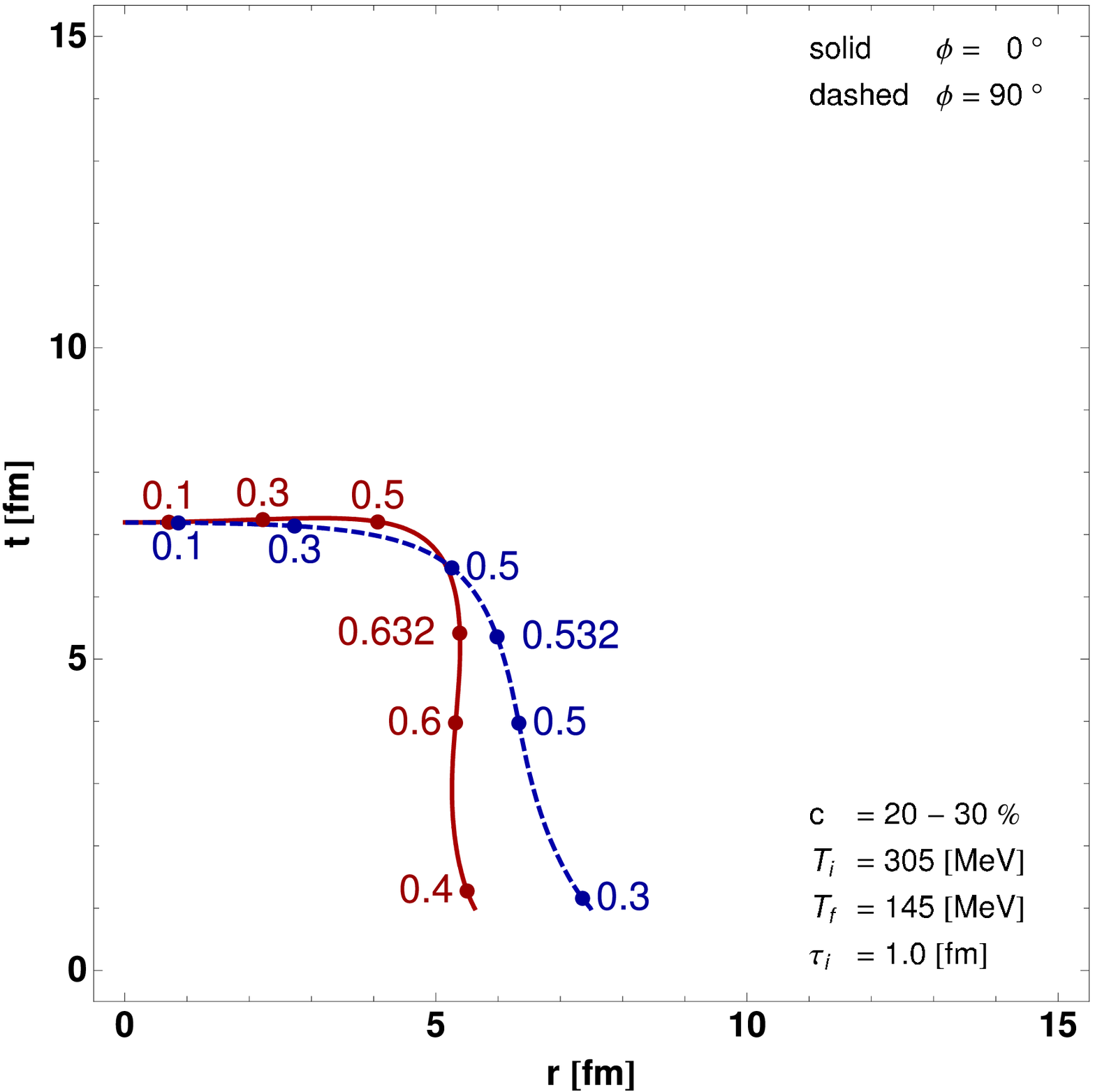}}
  \end{center}
  \caption{\small Freeze-out hypersurfaces for \mbox{$c=0-5$\%} (left) and \mbox{$c=20-30$\%} (right). Calculations include the free-streaming phase from $\tau_0$ = 0.25 fm to $\tau_{\rm i}$ = 1.0 fm. The initial central temperature $T_{\rm i}$ equals 331 and 305 MeV, for central and non-central collisions, respectively. The freeze-out temperature is set to $T_{\rm f}$ = 145 MeV.}
  \label{fig:resrhic2fs-hs}
\end{figure}
\begin{figure}[!hb]
  \begin{center}
    \subfigure{\includegraphics[width=0.49\textwidth]{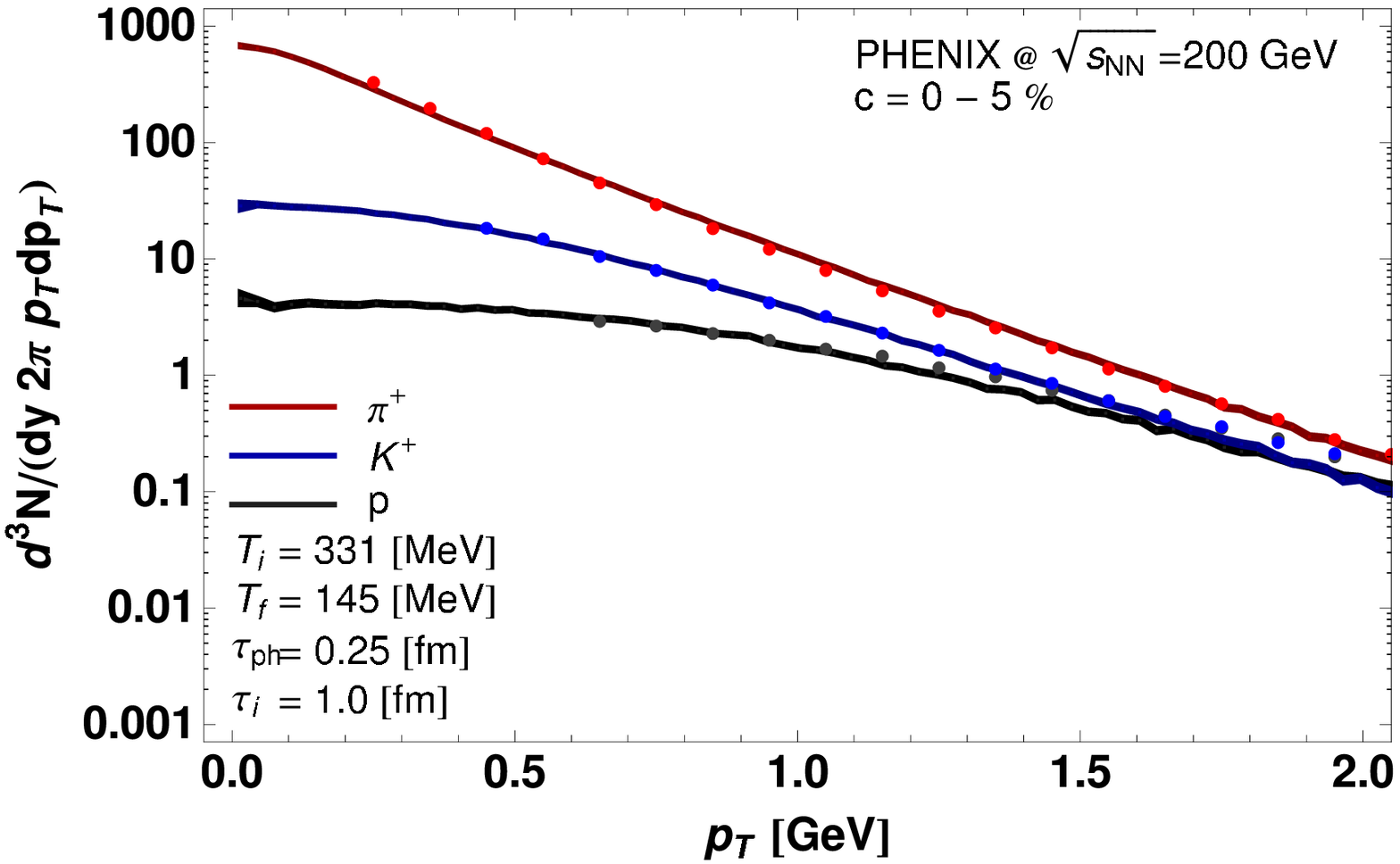}}
    \subfigure{\includegraphics[width=0.49\textwidth]{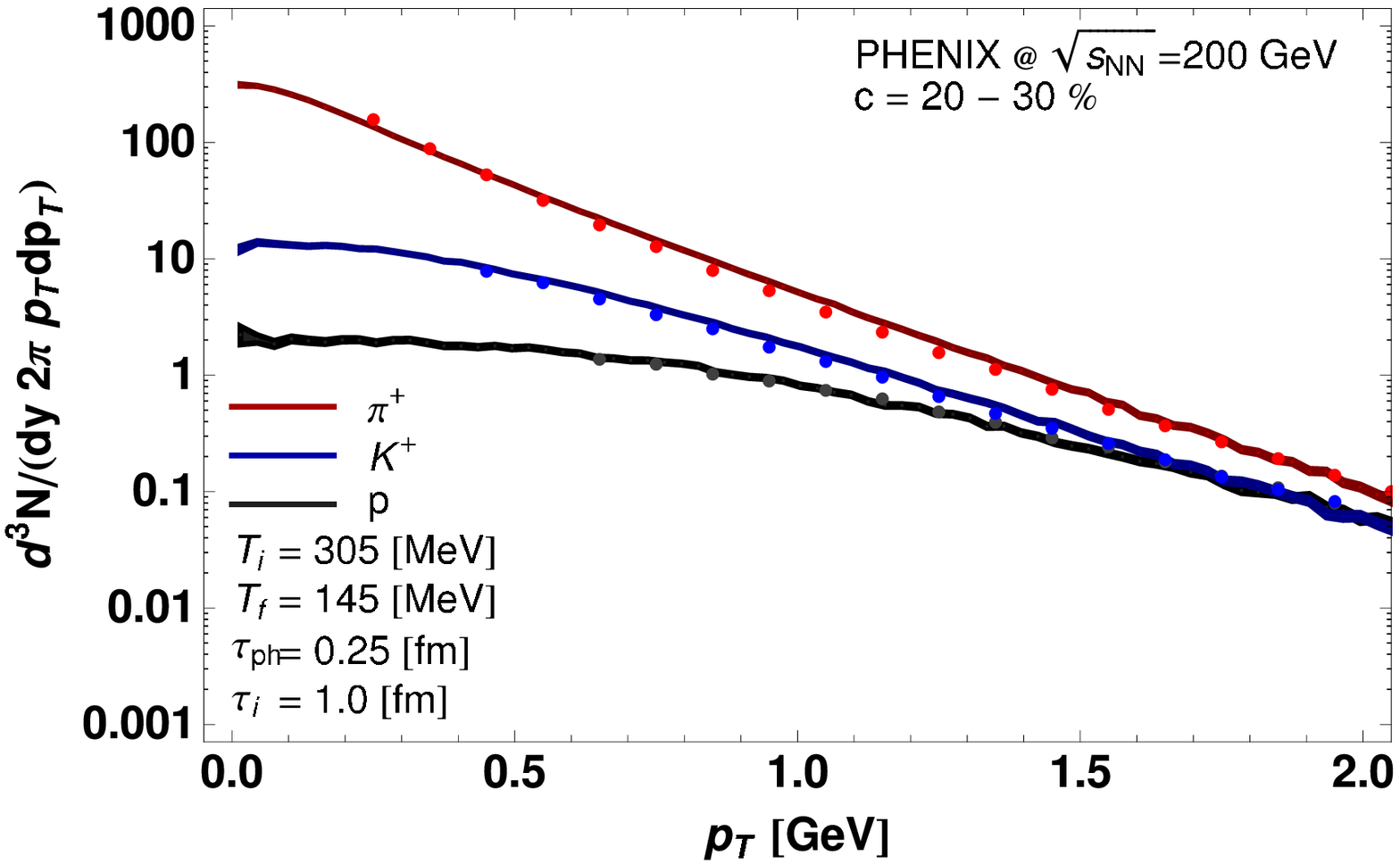}}\\
    \subfigure{\includegraphics[width=0.49\textwidth]{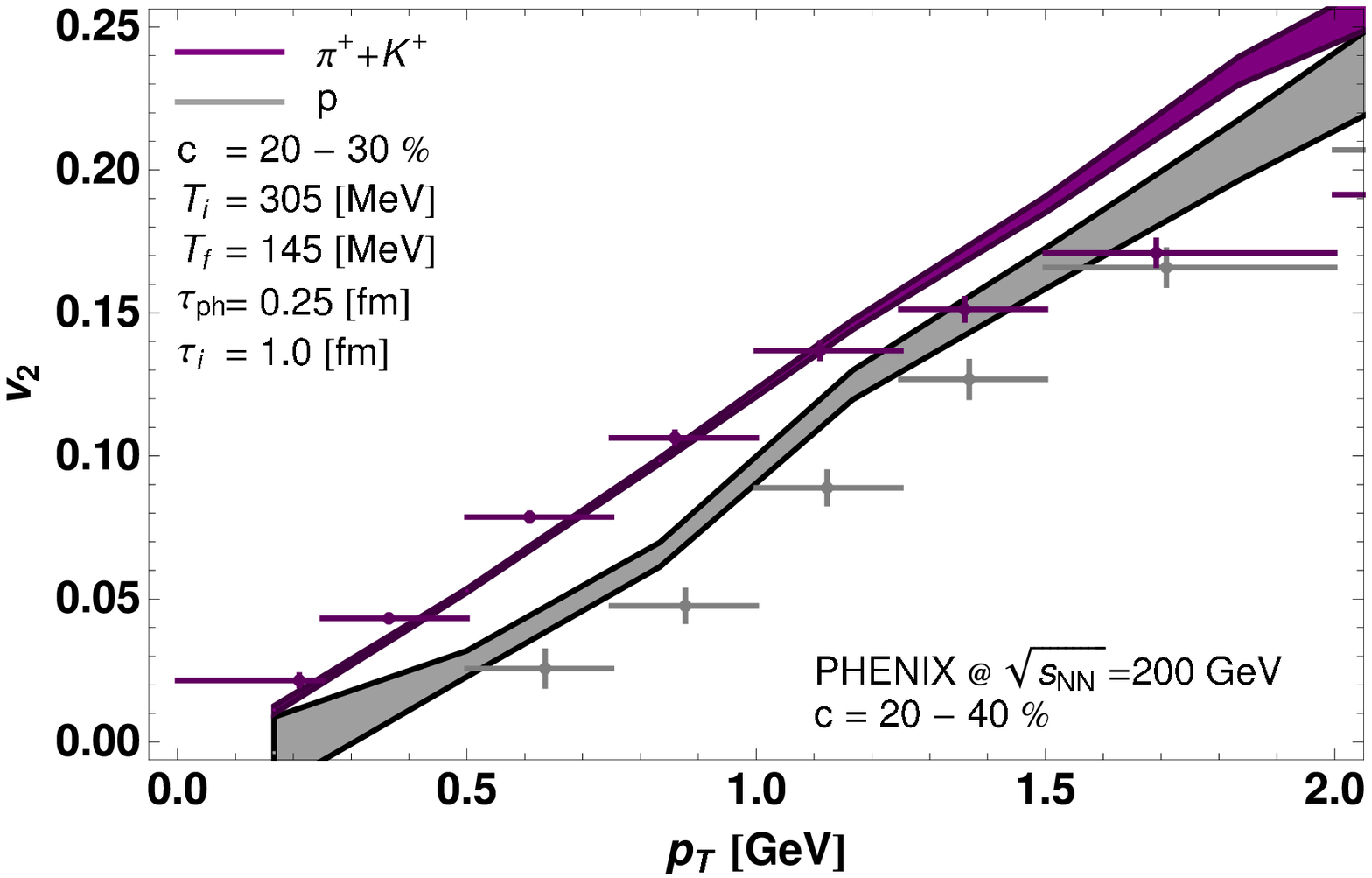}}
  \end{center}
  \caption{\small The transverse-momentum spectra of pions, kaons and protons for $c$=0-5\% (upper left panel), $c$=20-30\% (upper right panel) and the elliptic flow coefficient $v_2$ for  $c$=20-40\% (lower panel). Model parameters as in Fig. \ref{fig:resrhic2fs-hs}. Data from \cite{Adler:2003cb,Adler:2003kt}.}
  \label{fig:resrhic2fs-spv2pt}
\end{figure}
\par Our results presented in this Section were obtained with the free-streaming stage lasting from $\tau_0 = 0.25$ to $\tau_{\rm i}=1.0$ fm, followed by a sudden equilibration and hydrodynamics. The initial hydrodynamic time was shifted to a later time of 1.0 fm. Such a change requires a different normalization of the energy density at the center of the fireball, since during the time of 0.75 fm the system expands and its density decreases. To make an \underline{ estimate} of the initial energy at a later starting time we use the Bjorken scaling for the entropy \cite{Bjorken:1983qr}. We calculate the entropy density $s_0$ at $\tau_0$ from the energy density $\varepsilon_0$ and scale it by the following formula
\begin{equation}
s_0 \tau_0 = s_{\rm i} \tau_{\rm i} = \hbox{const.} \quad \to
\quad s_{\rm i} = s_0 \frac{\tau_0}{\tau_{\rm i}}.
\label{eqn:resrhic2_Borken_s}
\end{equation}
As a result the initial central temperature for the most central collision (\hbox{$c=0-5\%$}) is changed from $T_{\rm i} = 500 $ MeV down to 331 MeV and for the peripheral case (\hbox{$c=20-30\%$}) the temperature is decreased from 460 MeV to 305 MeV. This type of scaling works surprisingly well as the method of construction the initial conditions at $\tau_{\rm i}=1.0$ fm. If we compare the hypersurfaces from the previous section, Fig.~~\ref{fig:resrhic2nofs-hs}, with the new ones obtained with the free-streaming, Fig.~\ref{fig:resrhic2fs-hs}, then one can notice that they are almost identical. Both the shapes (starting from $\tau = 1.0$ fm) and values of the velocity on the hypersurface are very similar.
\par
\begin{figure}[!ht]
  \begin{center}
    \subfigure{\includegraphics[width=0.49\textwidth]{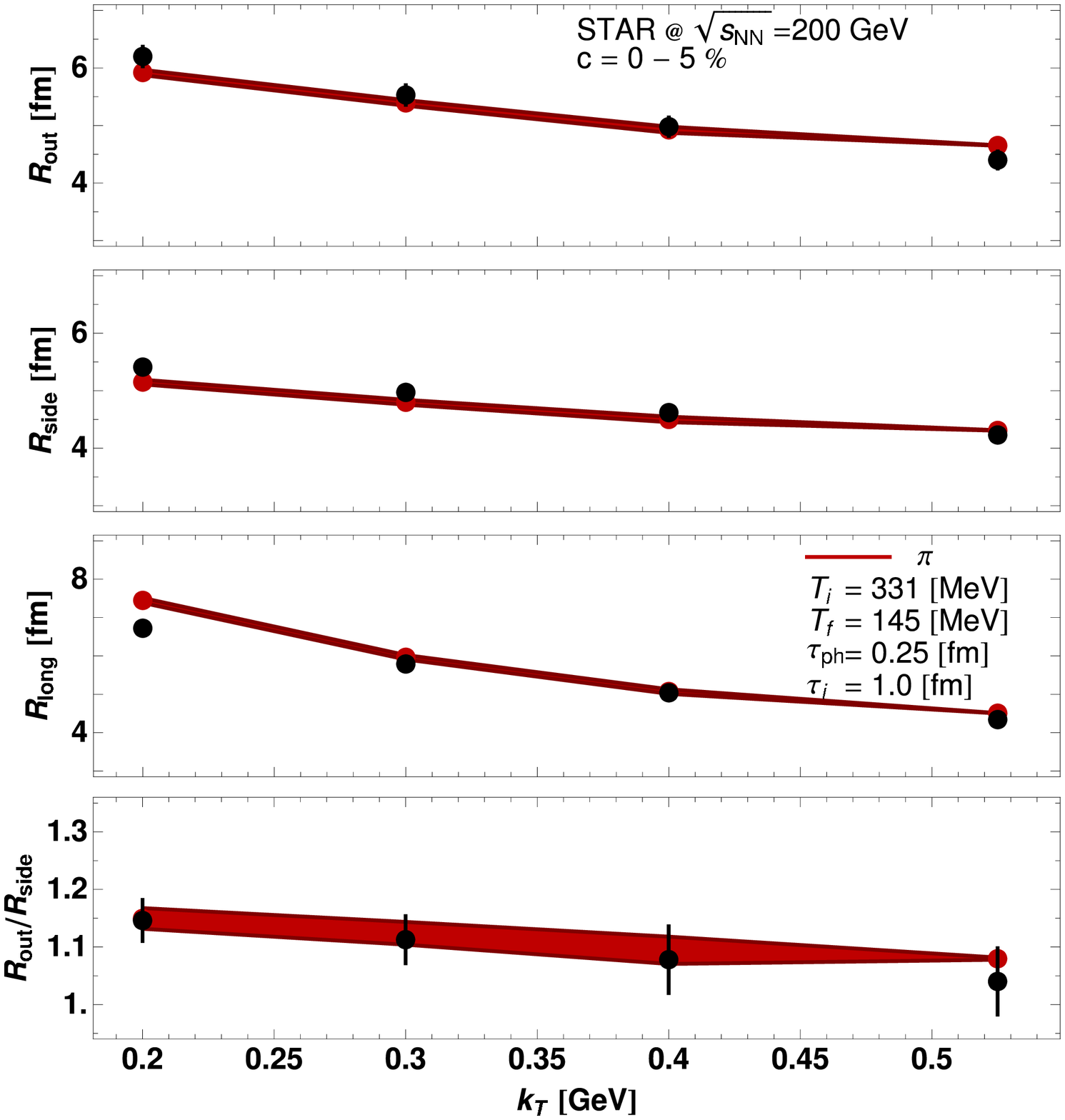}}
    \subfigure{\includegraphics[width=0.49\textwidth]{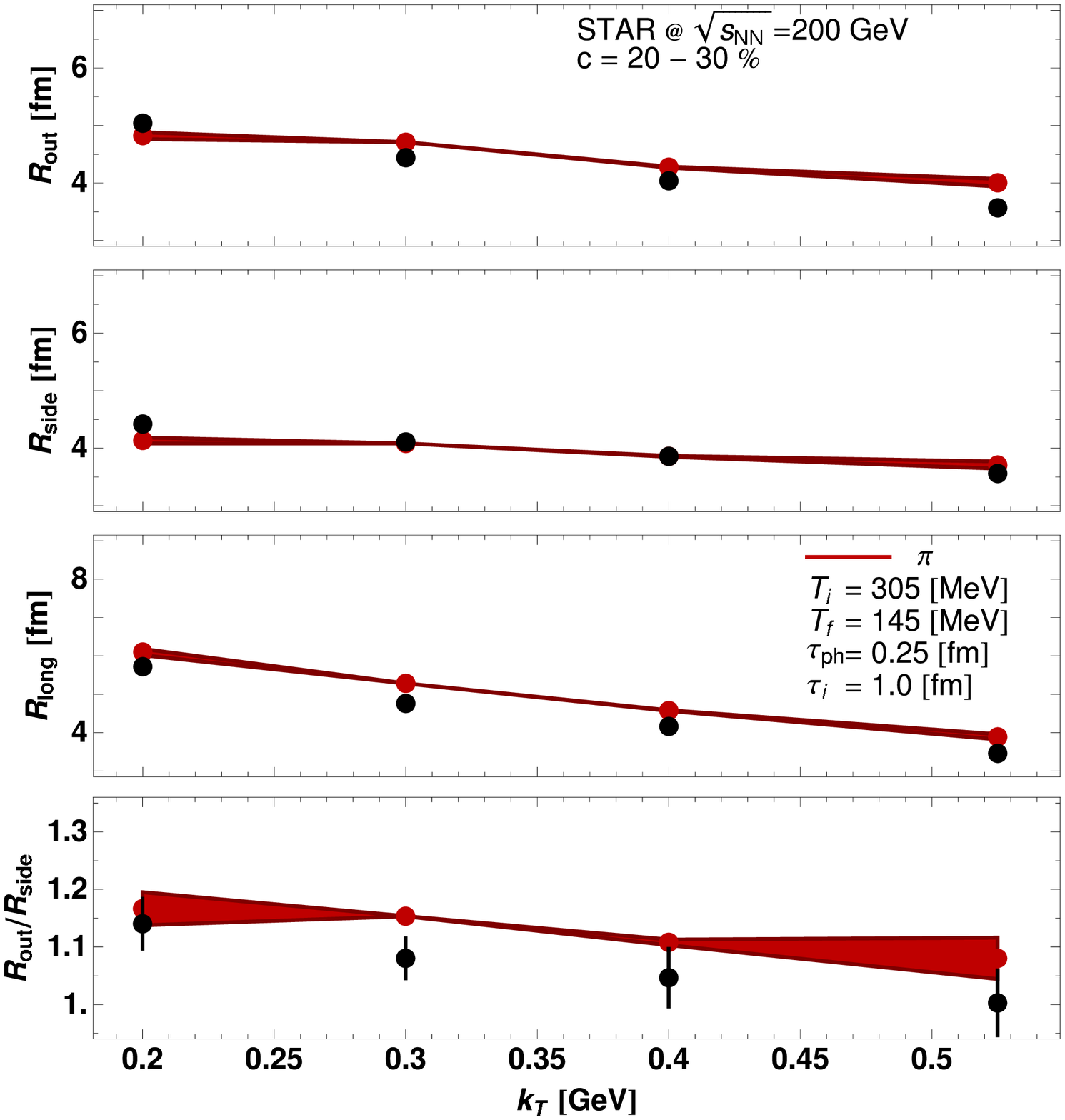}}\\
  \end{center}
  \caption{\small The pion HBT radii $R_{\rm side}$ , $R_{\rm out}$ , $R_{\rm long}$, and the ratio $R_{\rm out}/R_{\rm side}$ for central collisions, compared to the data from \cite{Adams:2004yc}. Model parameters same as in Fig. \ref{fig:resrhic2fs-hs}.}
  \label{fig:resrhic2fs-hbt}
\end{figure}

 The results of the calculations with a pre-hydrodynamics stage are shown in Figs.~\ref{fig:resrhic2fs-spv2pt} and \ref{fig:resrhic2fs-hbt}. As in Section \ref{section:resrhic2-early-start} we have started from the Gaussian profile (\ref{eqn:resrhic2_e0}). We notice very similar results to the ones obtained in the previous Section, not to mention the very good description of the data. Larger free-streaming times ($\tau-\tau_0 \sim 1.5$~fm) spoil this agreement, as the flow becomes too strong. Again we have achieved a uniform agreement for soft physics at RHIC. In particular, the transverse-momentum spectra, the elliptic-flow, and the HBT correlation radii, including the notorious ratio $R_{\rm out}/R_{\rm side}$, are all properly described.\\

\chapter{Summary}
\label{chaper:concl}

Our recently developed 2+1 (boost-invariant) hydrodynamic model has been presented and used to i) describe the soft hadronic data collected in the central region of the relativistic heavy-ion collisions at RHIC and ii) to make predictions for the heavy-ion collisions at the LHC energies. We have addressed both the one- and two-particle observables: the transverse momentum spectra, the elliptic flow coefficient $v_2$, and the pion HBT radii. The realistic equation of state for strongly interacting matter has been constructed that interpolates between the hadron gas model and the results of the QCD lattice simulations. The computational platform has been constructed, which combines the results of our hydrodynamic code with the statistical hadronization model {\tt THERMINATOR}. The satisfactory description of the soft hadronic RHIC data has been achieved with the standard initial conditions obtained from the optical limit of the Glauber model. Predictions for the future heavy-ion collisions at LHC have been formulated. The solution of the RHIC HBT puzzle has been proposed. It suggests the use of the modified Gaussian-type initial conditions for the energy density in the transverse plane. Finally, the processes of the free streaming of partons followed by the sudden equilibration have been incorporated in the model. The inclusion of the free-streaming stage allows for the delayed start of the hydrodynamic evolution, which is a desirable effect in the context of the early thermalization problem. 
\appendix
\chapter{Relativistic thermodynamics of perfect gases}
\label{chapter:Athermo}
%
\section{Grand canonical potential $\Omega$}
\label{section:Athermo_Omega}
%
The first law of thermodynamics for systems described by the grand canonical ensemble reads
\begin{equation}
  d E = T\,dS - P\,dV+\mu\,dN,
  \label{eqn:Athermo_GCE1stlaw}
\end{equation}
where $E$ is the energy, $S$ is the entropy, $P$ stands for pressure, and $N$ is the number of particles in the system (for relativistic systems the number of particles is not conserved and $N$ should be interpreted as the baryon number or any other conserved quantity, for example strangeness). Energy is an extensive function of three extensive variables, $E = E(S,V,N)$. By scaling the entropy, volume, and particle number by a factor $\lambda$, we obtain the energy rescaled by the same factor
\begin{equation}
E(\lambda S,\lambda V,\lambda N) = \lambda E(S,V,N).
\label{eqn:Athermo_GCEscaleE}
\end{equation}
By calculating the total derivative with respect to the scaling parameter $\lambda$ and using the relation (\ref{eqn:Athermo_GCE1stlaw}) we find
\begin{eqnarray}
  \left.\frac{\partial E}{\partial S}\right|_{V,N} S + \left.\frac{\partial E}{\partial V}\right|_{S,N} V + \left.\frac{\partial E}{\partial N}\right|_{S,V} N&=& E, \nonumber \\
  \label{eqn:Athermo_GCErelE}
\end{eqnarray}
which leads to
\begin{eqnarray}
  T\,S - P\,V + \mu\,N &=& E.
  \label{eqn:Athermo_GCErelE2}
\end{eqnarray}
The grand canonical potential $\Omega$ is defined as
\begin{equation}
  \Omega = E - T S - \mu N = - P V.
  \label{eqn:Athermo_omega_def}
\end{equation}
To calculate the change of the potential $\Omega$ we calculate the total derivative from Eq. (\ref{eqn:Athermo_omega_def}) and use (\ref{eqn:Athermo_GCE1stlaw}). In this way we find
\begin{equation}
  d\Omega = - S dT - P dV - N d\mu.
  \label{eqn:Athermo_Domega}
\end{equation}
%
\subsection{The grand canonical partition function}
\label{subsection:Athermo_Omega_Zfun}
%
It is useful to start our consideration with the introduction of the partition function for perfect gases. In this case the grand canonical partition function is defined in the following way
\begin{equation}
  Z = \prod_{i=1}^\infty z_i = \prod_{i=1}^\infty \sum_{n_i=0}^\infty e^\frac{n_i \left(\mu - \epsilon_i \right)}{T},
  \label{eqn:Athermo_def_partfun}
\end{equation}
where the index $i$ runs over all microstates of the system and $z_i$ is an individual partition function for that microstate. The index $n_i$ is the number of particles with the energy $\epsilon_i$ in the state $i$. In the cases where the product is taken over the states with equal energy rather then over all microscopic states, we introduce the degeneration factor $g_i$ giving the number of states with the same energy
\begin{equation}
  Z = \prod_{i=1}^\infty \left( z_i \right)^{g_i}.
    \label{eqn:Athermo_partfun_gi}
\end{equation}
For indistinguishable particles satisfying the Bose-Einstein or Fermi-Dirac statistics we are able to evaluate explicitly the sum in Eq. (\ref{eqn:Athermo_def_partfun}) 
\begin{eqnarray}
  z_i &=& \sum_{n_i=0}^\infty e^\frac{n_i \left(\mu - \epsilon_i \right)}{T} = \frac{1}{ 1 - e^\frac{\mu - \epsilon_i}{T}} \qquad \mbox{(Bose-Einstein)}, \label{eqn:Athermo_ziBE} \\
  z_i &=& \sum_{n_i=0}^1      e^\frac{n_i \left(\mu - \epsilon_i \right)}{T} = \left(    1 + e^\frac{\mu - \epsilon_i}{T} \right) \quad \mbox{(Fermi-Dirac)}. \label{eqn:Athermo_ziFD}
\end{eqnarray}
%
\subsection{Quantum statistics}
\label{subsection:Athermo_Omega_QS}
%
From the statistical physics we know the connection between $Z$ and $\Omega$,
\begin{equation}
  \Omega (T, V, \mu) = -T \ln Z,
  \label{eqn:Athermo_omega}
\end{equation}
which gives
\begin{equation}
  \Omega = -T \ln Z = -T \ln \prod_{i=1}^\infty \left( z_i \right)^{g_i} = -T \sum_{i=1}^\infty g_i \ln z_i.
  \label{eqn:Athermo_omega_zi}
\end{equation}
This formula can be further rewritten in the form including Bose-Einstein ($\epsilon = -1$) or Fermi-Dirac ($\epsilon = +1$) statistics, Eqs. (\ref{eqn:Athermo_ziBE}) and (\ref{eqn:Athermo_ziFD}),
\begin{equation}
  \Omega = - T\, \epsilon\, \sum_{i=1}^\infty g_i \ln \left( 1 + \epsilon\, e^\frac{\mu - \epsilon_i}{T} \right).
  \label{eqn:Athermo_omega_fin_sum}
\end{equation}
In the classical approach we can replace the sum in Eq. (\ref{eqn:Athermo_omega_fin_sum}) by the integration over the one-particle phase-space. This will lead us to the general form of the grand potential
\begin{equation}
  \Omega(T, V, \mu) = - T\, V\, g\, \epsilon\, \int \frac{d^3 p}{\left( 2 \pi \right)^3} \ln \left( 1 + \epsilon\, e^\frac{\mu - E_p}{T} \right).
  \label{eqn:Athermo_omega_fin_int}
\end{equation}
Here we assumed that all the degeneracies are equal $g_i = g$ -- this is the typical case where $g$ is related to the spin degeneracy where $g = 2 s + 1$. 
\par In the following we shall assume that the condition $\mu - E < 0$ is fulfilled. This condition is always true for bosons and it is also valid for fermions in the case $\mu < m$. In this case we use the series expansion of the logarithm,
\begin{equation}
\ln \left( 1 + x \right) = - \sum_{\kappa=1}^{\infty} \frac{(-1)^\kappa}{\kappa} x^\kappa,
\label{eqn:Athermo_expoflog}
\end{equation}
to rewrite  Eq. (\ref{eqn:Athermo_omega_fin_int}) in the form
\begin{equation}
  \Omega_{\rm QS}(T, V, \mu) = T\, V\, g\, \epsilon\, \int \frac{d^3 p}{\left( 2 \pi \right)^3} \sum_{\kappa=1}^\infty \frac{(-\epsilon)^\kappa}{\kappa} e^{\frac{\mu - E_p}{T}\kappa}.
  \label{eqn:Athermo_omegaQS}
\end{equation}
%
\subsection{Boltzmann classical limit}
\label{subsection:Athermo_Omega_BL}
%
For particles which have small average occupation numbers one can use the approximation of the classical Boltzmann gas. This assumption is realized by using only the leading term in the expansion (\ref{eqn:Athermo_omegaQS}), that corresponds to $\kappa = 1$,
\begin{equation}
  \Omega_{\rm BL}(T, V, \mu) = - T\, V\, g\, \int \frac{d^3 p}{\left( 2 \pi \right)^3} e^\frac{\mu - E_p}{T}.
  \label{eqn:Athermo_omegaBL}
\end{equation}
We note that this approximation is independent of quantum statistics since in Eq. (\ref{eqn:Athermo_omegaQS}) we have always the term $\epsilon^{\kappa+1}$ which equals 1 for $\kappa = 1$.
%
\subsection{Massive particles}
\label{subsection:Athermo_Omega_mass}
%
The relativistic relation between the particle energy $E_p$, momentum $p$, and the rest mass $m$ is $E_p = \sqrt{p^2 + m^2}$. Introducing this relation into  Eq. (\ref{eqn:Athermo_omegaQS}) enables us to calculate the potential $\Omega$ for massive particles
\begin{equation}
  \Omega(T, V, \mu) = T\, V\, g\, \epsilon\, \sum_{\kappa=1}^\infty \frac{(-\epsilon)^\kappa}{\kappa} \int \frac{d^3 p}{\left( 2 \pi \right)^3} e^{\frac{\mu - \sqrt{m^2+p^2}}{T}\kappa}.
  \label{eqn:Athermo_omegaMass1}
\end{equation}
Changing to the spherical coordinates and performing simple calculations we find
\begin{equation}
  \Omega(T, V, \mu) = \frac{1}{2\pi^2}\, T\, V\, g\, \epsilon\, \sum_{\kappa=1}^\infty \frac{(-\epsilon)^\kappa}{\kappa}\, e^{\frac{\mu}{T}\kappa}\, \int_0^\infty dp\, p^2\, e^{-\frac{\sqrt{m^2 + p^2}}{T}\kappa}.
  \label{eqn:Athermo_omegaMass2}
\end{equation}
Let us now consider only the integral on the right-hand-side of Eq. (\ref{eqn:Athermo_omegaMass2}). By changing the integration variables from $p$ to $x$ with the substitution $m^2 + p^2 = m^2\,x^2$, this integral takes the following form
\begin{equation}
  \int_0^\infty dp\, p^2\, e^{-\frac{\sqrt{m^2 + p^2}}{T}\kappa} = m^3 \int_1^\infty dx\, x\, \left(x^2-1\right)^\frac{1}{2} e^{-\frac{m\,\kappa}{T}x}.
  \label{eqn:Athermo_intBesselK_1}
\end{equation}
The formula above may be reduced to the integral definition of the modified Bessel function of the second kind $K_\nu(z)$, see Eq. (\ref{eqn:Zmath_BesselK_int}) with $\nu = 2$ and $z = m\, \kappa/T$,
\begin{equation}
  \frac{m^3}{3}\frac{m\, \kappa}{T} \int_1^\infty dx\, \left(x^2-1\right)^{2-\frac{1}{2}} e^{-\frac{m\, \kappa}{T}x} = m^2\, \, \frac{T}{\kappa} K_2\left( \frac{m}{T}\kappa\right).
  \label{eqn:Athermo_intBesselK_2}
\end{equation}
Thus we write the final form of the grand canonical potential for massive particles with finite chemical potential and quantum statistics as follows
\begin{equation}
  \Omega(T, V, \mu) = \frac{1}{2\pi^2}\, T^2\, V\, m^2\, g\, \epsilon\, \sum_{\kappa=1}^\infty \frac{(-\epsilon)^\kappa}{\kappa^2} e^{\frac{\mu}{T}\kappa}\, K_2\left( \frac{m}{T}\kappa\right) \quad (m > 0, \mu < m).
  \label{eqn:Athermo_omegaMass3}
\end{equation}
%
\subsection{Massless particles}
\label{subsection:Athermo_Omega_nomass}
%
\begin{table}[t]
  \caption{Grand canonical potential $\Omega$ for various choices of $m$. }
  \label{tab:Athermo_Omega}
\begin{center}
  {\footnotesize
  \begin{tabular}{c|cc} 
    \multicolumn{1}{c}{} & \multicolumn{2}{c}{\bf quantum statistics} \\
    {\tiny \vphantom{.}} & & \\
     \hline
      {\tiny $\vphantom{.}$} & & \\
      $ m \neq 0 	$ &
      $ \displaystyle   \frac{1}{2\pi^2}\, T^2\, V\, g\, m^2\, \epsilon\, \sum_{\kappa=1}^\infty \frac{(-\epsilon)^\kappa}{\kappa^2} e^{\frac{\mu}{T}\kappa}\, K_2\left( \frac{m}{T}\kappa\right)	$\\
    {\tiny \vphantom{.}} & & \\
      $ m = 0		$ &
      $ \displaystyle   \frac{1}{ \pi^2}\, T^4\, V\, g\,       \epsilon\, \hbox{Li}_4\left(-\epsilon\, e^\frac{\mu}{T} \right)	$ \\
    \multicolumn{3}{c}{} \\
    \multicolumn{1}{c}{} & \multicolumn{2}{c}{\bf classical limit} \\
    {\tiny \vphantom{.}} & & \\
    \hline 
    {\tiny \vphantom{.}} & & \\
      $ m \neq 0	$ &
      $ \displaystyle - \frac{1}{2\pi^2}\, T^2\, V\, g\, m^2\, e^\frac{\mu}{T}\, K_2\left( \frac{m}{T}\right)	$ \\
    {\tiny \vphantom{.}} & & \\
      $ m = 0 		$ &
      $ \displaystyle - \frac{1}{ \pi^2}\, T^4\, V\, g\, e^\frac{\mu}{T}	$ \\
  \end{tabular}
  }
\end{center}
\end{table}
Of the special interest is the case of massless particles. Such a limiting case is appropriate for gluons and also for quarks at sufficiently high temperatures \footnote{Typically the light quarks $u$ and $d$ may be regarded as massless particles. If the temperature is sufficiently high, in addition the strange quark may be regarded as a massless particles.}. The thermodynamical properties  of massless particles follow directly from the formula (\ref{eqn:Athermo_omega_fin_int}) where we set $m=0$ and $\mu=0$.  The alternative way is to consider the limiting case of Eq. (\ref{eqn:Athermo_omegaMass3}). For massless particles the modified Bessel function $K_\nu(z)$ in Eq. (\ref{eqn:Athermo_omegaMass3})  may be expanded into a power series in the vicinity of the point $m = 0$ according to Eqs. (\ref{eqn:Zmath_BesselK_powser}). This leads us to the expression
\begin{equation}
  \Omega(T, V, \mu) = \frac{1}{2\pi^2}\, T^2\, V\, m^2\, g\, \epsilon\, \sum_{\kappa=1}^\infty \frac{(-\epsilon)^\kappa}{\kappa^2} e^{\frac{\mu}{T}\kappa}\, \left( 
    \frac{2}{\left( \frac{m}{T}\kappa \right)^2} - \frac{1}{2} + O\left( 2 \right) 
  \right).
  \label{eqn:Athermo_omegaNomass1}
\end{equation}
Since only the first term of this expansion gives non zero contribution in the limit $m \to 0$, the grand potential for massless particles reads
\begin{equation}
  \Omega(T, V, \mu) = \frac{1}{\pi^2}\, T^4\, V\, g\, \epsilon\, \sum_{\kappa=1}^\infty 
\frac{(-\epsilon \, e^\frac{\mu}{T})^\kappa}{\kappa^4} = \frac{1}{\pi^2}\, T^4\, V\, g\, \epsilon\, 
\hbox{Li}_4 \left(- \epsilon\, e^\frac{\mu}{T} \right).
  \label{eqn:Athermo_omegaNomass2}
\end{equation}
Here we have introduced the  polylogarithm function defined by the series
\begin{equation}
\hbox{Li}_n(z) = \sum_{k=1}^\infty \frac{z^k}{k^n}.
\label{eqn:Athermo_Li}
\end{equation}
In the limit of vanishing chemical potential the values $\hbox{Li}_4(1)$ (required for bosons) and $\hbox{Li}_4(+1)$ (required for fermions) should be obtained from the analytic continuation which gives \mbox{$\hbox{Li}_4(1) = \pi^4/90$} and \mbox{$\hbox{Li}_4(-1) = - 7 \pi^4/720$}. In this way we obtain
\begin{equation}
  \Omega(T, V, \mu=0,m=0) = -\frac{\pi^2}{90}\,g V\, T^4  \qquad \mbox{(Bose-Einstein)}
  \label{eqn:Athermo_omegaNomass2b}
\end{equation}
and
\begin{equation}
  \Omega(T, V, \mu=0,m=0) = -\frac{7 \pi^2}{720}\,g V\, T^4 \qquad \mbox{(Fermi-Dirac)}. 
  \label{eqn:Athermo_omegaNomass2f}
\end{equation}
\begin{table}[t]
  \caption{Mean number of particles $N$ for various choices of $m$.}
  \label{tab:Athermo_N}
\begin{center}
  {\footnotesize
  \begin{tabular}{c|cc} 
    \multicolumn{1}{c}{} & \multicolumn{2}{c}{\bf quantum statistics} \\
    {\tiny \vphantom{.}} & & \\
    \hline
    {\tiny \vphantom{.}} & & \\
      $ m \neq 0 	$ &
      $ \displaystyle - \frac{1}{2\pi^2}\, T\,   V\, g\, m^2\, \epsilon\, \sum_{\kappa=1}^\infty \frac{(-\epsilon)^\kappa}{\kappa} e^{\frac{\mu}{T}\kappa}\, K_2\left( \frac{m}{T}\kappa\right)	$ \\
    {\tiny \vphantom{.}} & & \\
      $ m = 0		$ &
      $ \displaystyle - \frac{1}{ \pi^2}\, T^3\, V\, g\,       \epsilon\, \hbox{Li}_3\left(-\epsilon\, e^\frac{\mu}{T} \right)	$ \\
    \multicolumn{3}{c}{} \\
    \multicolumn{1}{c}{} & \multicolumn{2}{c}{\bf classical limit} \\
    {\tiny \vphantom{.}} & & \\
    \hline
    {\tiny \vphantom{.}} & & \\
      $ m \neq 0	$ &
      $ \displaystyle   \frac{1}{2\pi^2}\, T\,   V\, g\, m^2\, e^\frac{\mu}{T}\, K_2\left( \frac{m}{T}\right)	$ \\
    {\tiny \vphantom{.}} & & \\
      $ m = 0 		$ &
      $ \displaystyle   \frac{1}{ \pi^2}\, T^3\, V\, g\,       e^\frac{\mu}{T}	$ \\
  \end{tabular} 
  }
\end{center}
\end{table}
%
\subsection{Vanishing chemical potential}
\label{subsection:Athermo_Omega_nomu}
%
Sometimes one assumes that the chemical potential is very small and may be neglected. In this case the grand canonical potential, Eq. (\ref{eqn:Athermo_omegaQS}),  takes the form 
\begin{equation}
  \Omega(T, V, \mu=0) = - T\, V\, g\, \epsilon\, \int \frac{d^3 p}{\left( 2 \pi \right)^3} \sum_{\kappa=1}^\infty \frac{(-\epsilon)^\kappa}{\kappa} e^{-\frac{E_p}{T}\kappa},
  \label{eqn:Athermo_omegaNomu1}
\end{equation}
or in the case of classical massless gas with no chemical potential
\begin{equation}
  \Omega_{\rm BL}(T, V, \mu=0) = - \frac{1}{\pi^2}\, g V \, T^4.
  \label{eqn:Athermo_omegaNomu2}
\end{equation}
%
\section{Other thermodynamic variables}
\label{section:Athermo_N}
%
Given the formula for the differential of the grand potential (\ref{eqn:Athermo_Domega}) we can evaluate the mean number of particles using the following expression
\begin{equation}
N = - \left. \frac{\partial \Omega}{\partial \mu} \right\vert_{T,V}.
\label{eqn:Athermo_Ndef}
\end{equation}
With the most general definition of $\Omega$, valid for massive gas with quantum statistics, see Eq. (\ref{eqn:Athermo_omegaMass3}), we find
\begin{equation}
N(T, V, \mu) = - \frac{1}{2\pi^2}\, T\, V\, m^2\, g\, \epsilon\, \sum_{\kappa=1}^\infty \frac{(-\epsilon\,)^\kappa}{\kappa^2} e^{\frac{\mu}{T}\kappa}\, K_2\left( \frac{m}{T}\kappa\right).
  \label{eqn:Athermo_NMass}
\end{equation}
Neglecting the quantum statistics will lead to the Boltzmann classical approximation in the form
\begin{equation}
N_{\rm BL}(T, V, \mu) = \frac{1}{2\pi^2}\, T\, V\, e^\frac{\mu}{T}\, m^2\, g\, K_2\left( \frac{m}{T}\right).
\label{eqn:Athermo_NBL}
\end{equation}
In the further step we can do the massless-gas assumption and following the same formalism as in subsection \ref{subsection:Athermo_Omega_nomass} one gets
\begin{equation}
  N_{\rm BL}(T, V, \mu) = \frac{1}{\pi^2}\, T^3\,g V\, e^\frac{\mu}{T}.
  \label{eqn:Athermo_NBLNomass}
\end{equation}
The formula for the entropy follows from the equation
\begin{equation}
  S = -\left. \frac{\partial\Omega}{\partial T} \right\vert_{\mu,V},
  \label{eqn:Athermo_Sdef}
\end{equation}
which for massive particles obeying the quantum statistics gives
\begin{equation}
  S = - \frac{1}{2\pi^2}\, V\, m^2\, g\, \epsilon\, \sum_{\kappa=1}^\infty \frac{(-\epsilon\,)^\kappa}{\kappa^2} e^{\frac{\mu}{T}\kappa}\, \left[ \left( 4\, T -\mu \right) K_2\left( \frac{m}{T}\kappa\right) + m\, \kappa\, K_1\left( \frac{m}{T}\kappa\right)\right].
  \label{eqn:Athermo_SMass}
\end{equation}
Similarly, to calculate the pressure we use the formula
\begin{equation}
  P = -\left. \frac{\partial\Omega}{\partial V} \right\vert_{T,\mu}
  \label{eqn:Athermo_Pdef}
\end{equation}
and get
\begin{equation}
  P = - \frac{1}{2\pi^2}\, T^2\, m^2\, g\, \epsilon\, \sum_{\kappa=1}^\infty \frac{(-\epsilon)^\kappa}{\kappa^2} e^{\frac{\mu}{T}\kappa}\, K_2\left( \frac{m}{T}\kappa\right).
  \label{eqn:Athermo_PMass}
\end{equation}
\begin{table}[t]
  \caption{Entropy for various choices of $m$.}
  \label{tab:Athermo_S}
\begin{center}
  {\footnotesize 
  \begin{tabular}{c|cc} 
    \multicolumn{1}{c}{} & \multicolumn{2}{c}{\bf quantum statistics} \\
    {\tiny \vphantom{.}} & & \\
    \hline
    {\tiny \vphantom{.}} & & \\
      $ m \neq 0 	$ &
      $ \displaystyle - \frac{1}{2\pi^2}\,       V\, g\, m^2\, \epsilon\, \sum_{\kappa=1}^\infty \frac{(-\epsilon)^\kappa}{\kappa^2} e^{\frac{\mu}{T}\kappa}\, \left[ \left( 4\,T - \mu\,\kappa \right) K_2\left( \frac{m}{T}\kappa \right) + m\,\kappa\, K_1\left( \frac{m}{T}\kappa \right) \right] $ \\
    {\tiny \vphantom{.}} & & \\
      $ m = 0		$ &
      $ \displaystyle - \frac{1}{ \pi^2}\, T^2\, V\, g\,       \epsilon\, \left[ 4\, T\, \hbox{Li}_4\left(-\epsilon\, e^\frac{\mu}{T} \right) - \mu\, \hbox{Li}_3\left(-\epsilon\, e^\frac{\mu}{T} \right) \right]$ \\
    \multicolumn{3}{c}{} \\
    \multicolumn{1}{c}{} & \multicolumn{2}{c}{\bf classical limit} \\
    {\tiny \vphantom{.}} & & \\
    \hline 
    {\tiny \vphantom{.}} & & \\
      $ m \neq 0	$ &
      $ \displaystyle   \frac{1}{2\pi^2}\,       V\, g\, m^2\, e^\frac{\mu}{T}\, \left[ \left( 4\,T - \mu \right) K_2\left( \frac{m}{T} \right) + m\, K_1\left( \frac{m}{T} \right) \right]$ \\
    {\tiny \vphantom{.}} & & \\
      $ m = 0 		$ &
      $ \displaystyle   \frac{1}{ \pi^2}\, T^2\, V\, g\,       e^\frac{\mu}{T}\, \left( 4\,T - \mu \right)	$\\
  \end{tabular}
  }
\end{center}
\end{table}
\begin{table}[!ht]
  \caption{Pressure for various choices of $m$.}
  \label{tab:Athermo_P}
\begin{center}
  {\footnotesize
  \begin{tabular}{c|cc} 
    \multicolumn{1}{c}{} & \multicolumn{2}{c}{\bf quantum statistics} \\
    {\tiny \vphantom{.}} & & \\
    \hline
    {\tiny \vphantom{.}} & & \\
      $ m \neq 0 	$ &
      $ \displaystyle - \frac{1}{2\pi^2}\, T^2\,     g\, m^2\, \epsilon\, \sum_{\kappa=1}^\infty \frac{(-\epsilon)^\kappa}{\kappa^2} e^{\frac{\mu}{T}\kappa}\, K_2\left( \frac{m}{T}\kappa\right)	$ \\
    {\tiny \vphantom{.}} & & \\
      $ m = 0		$ &
      $ \displaystyle - \frac{1}{ \pi^2}\, T^4\,     g\,       \epsilon\, \hbox{Li}_4\left(-\epsilon\, e^\frac{\mu}{T} \right)	$  \\
    \multicolumn{3}{c}{} \\
    \multicolumn{1}{c}{} & \multicolumn{2}{c}{\bf classical limit} \\
    {\tiny \vphantom{.}} & & \\
    \hline 
    {\tiny \vphantom{.}} & & \\
      $ m \neq 0	$ &
      $ \displaystyle   \frac{1}{2\pi^2}\, T^2\,     g\, m^2\, e^\frac{\mu}{T}\, K_2\left( \frac{m}{T}\right)	$ \\
    {\tiny \vphantom{.}} & & \\
      $ m = 0 		$ &
      $ \displaystyle   \frac{1}{ \pi^2}\, T^4\,     g\,       e^\frac{\mu}{T}\,	$ \\
  \end{tabular} 
  }
\end{center}
\end{table}
Finally, the energy is obtained from the relation
\begin{equation}
  E = T\,S - P\,V +\mu\,N
  \label{eqn:Athermo_Edef}
\end{equation}
which gives
\begin{equation}
  E = - \frac{1}{2\pi^2}\, T\,   V\, g\, m^2\, \epsilon\, \sum_{\kappa=1}^\infty \frac{(-\epsilon)^\kappa}{\kappa^2} e^{\frac{\mu}{T}\kappa}\, \left[ 3\, T\, K_2\left( \frac{m}{T}\kappa \right) + m\, \kappa\, K_1\left( \frac{m}{T}\kappa \right) \right].
  \label{eqn:Athermo_EMass}
\end{equation}
\begin{table}[t]
  \caption{Energy for various choices of $m$.}
  \label{tab:Athermo_E}
\begin{center}
  { \footnotesize
  \begin{tabular}{c|cc} 
    \multicolumn{1}{c}{} & \multicolumn{2}{c}{\bf quantum statistics} \\
    {\tiny \vphantom{.}} & & \\
    \hline
    {\tiny \vphantom{.}} & & \\
      $ m \neq 0 	$ &
      $ \displaystyle - \frac{1}{2\pi^2}\, T\,   V\, g\, m^2\, \epsilon\, \sum_{\kappa=1}^\infty \frac{(-\epsilon)^\kappa}{\kappa^2} e^{\frac{\mu}{T}\kappa}\, \left[ 3\, T\, K_2\left( \frac{m}{T}\kappa \right) + m\, \kappa\, K_1\left( \frac{m}{T}\kappa \right) \right] $ \\
    {\tiny \vphantom{.}} & & \\
      $ m = 0		$ &
      $ \displaystyle - \frac{3}{ \pi^2}\, T^4\, V\, g\,       \epsilon\, \hbox{Li}_4 \left(-\epsilon\, e^\frac{\mu}{T}\right)	$ \\
    \multicolumn{3}{c}{} \\
    \multicolumn{1}{c}{} & \multicolumn{2}{c}{\bf classical limit} \\
    {\tiny \vphantom{.}} & & \\
    \hline 
    {\tiny \vphantom{.}} & & \\
      $ m \neq 0	$ &
      $ \displaystyle   \frac{1}{2\pi^2}\, T\,   V\, g\, m^2\, e^\frac{\mu}{T}\,\left[ 3\, T\, K_2\left( \frac{m}{T} \right) + m\, K_1\left( \frac{m}{T} \right) \right] $ \\
    {\tiny \vphantom{.}} & & \\
      $ m = 0 		$ &
      $ \displaystyle   \frac{3}{ \pi^2}\, T^4\, V\, g\,	e^\frac{\mu}{T}\,	$ \\
  \end{tabular} 
  }
\end{center}
\end{table}

\chapter[Properties of hydrodynamic equations]{Properties of relativistic \\ hydrodynamic equations for $\mu_B=0$}
\label{chapter:Bhydro}
%
This Appendix presents the details of transformations of the relativistic hydrodynamic equations which were applied to obtain the final form used in the numerical calculations. It also introduces the basic hydrodynamic definitions that are used in our work.
%
\section{Basic definitions}
\label{section:Bhydro_def}
%
The fluid four-velocity is defined as
\begin{equation}
  u^\mu = (u^0,u^i) = \gamma\, (1,{\bf v}\,) = \gamma\, (1, v_x, v_y, v_z),
  \label{eqn:Bhydro_def_umu}
\end{equation}
where $\gamma$ is the Lorentz factor defined as
\begin{equation}
  \gamma = u^0 = \frac{1}{\sqrt{1-v^2}}.
  \label{eqn:Bhydro_def_gamma}
\end{equation}
The derivative of $\gamma$ equals
\begin{equation}
  d\gamma = \gamma^3 v^i d v^i = \gamma^3 v dv,
  \label{eqn:Bhydro_noncov_Dgamma}
\end{equation}
where
\begin{equation}
v dv = \gamma^{-3} d\gamma,    \qquad  v = \sqrt{v_x^2 + v_y^2 + v_z^2} = \sqrt{v_\perp^2 + v_z^2}.
  \label{eqn:Bhydro_def_vel}
\end{equation}
Here $v_\perp$ is the magnitude of the transverse velocity ${\bf v}_\perp$. The total time derivative in the Cartesian coordinates reads
\begin{equation}
 \frac{d}{dt} = 
          \,\frac{\partial }{\partial t} 
    + v^j \,\frac{\partial }{\partial x^j}
    =     \,\frac{\partial }{\partial t} 
    + v_x \,\frac{\partial }{\partial x} 
    + v_y \,\frac{\partial }{\partial y} 
    + v_z \,\frac{\partial }{\partial z},
  \label{eqn:Bhydro_def_dtCar}
\end{equation}
while in the cylindrical coordinates it has the form
\begin{equation}
 \frac{d}{dt} = 
				\,\frac{\partial}{\partial t}
    + v_\perp \cos\alpha		\,\frac{\partial}{\partial r}
    + \frac{v_\perp \sin \alpha}{r} \,\frac{\partial}{\partial \phi}
    + v_z			\,\frac{\partial}{\partial z}.
  \label{eqn:Bhydro_def_dtCyl}
\end{equation}
Here $\alpha$ is the angle between the transverse velocity ${\bf v}_\perp$ and radial velocity ${\bf v}_r$, \mbox{$v_r = v_\perp \cos \alpha$}, \mbox{$v_\phi = v_\perp \sin \alpha$}.
%
\section{Covariant form}
\label{section:Bhydro_cov}
%
The conservation of the energy-momentum tensor of the perfect fluid in its most general form yields
\begin{equation}
  \partial_\mu\left[\left(\varepsilon + P \right) u^\mu u^\nu - P\, g^{\mu\nu} \right] = 0.
  \label{eqn:Bhydro_cov_0}
\end{equation}
We transform Eq. (\ref{eqn:Bhydro_cov_0}) using thermodynamic identities  $\varepsilon + P = T\,s$ and $dP = s\, dT$ to get
\begin{equation}
  T u^\nu\, \partial_\mu (s u^\mu) + s u^\mu\, \partial_\mu (T u^\nu) = s \partial^\nu T.
  \label{eqn:Bhydro_cov_1}
\end{equation}
Multiplying both sides of Eq. (\ref{eqn:Bhydro_cov_1}) by $u_\nu$ and using the normalization condition for the four-velocity, $u^\nu u_\nu=1$, we arrive at the following formula
\begin{equation}
  T \partial_\mu (s u^\mu) + s u^\mu T u_\nu \partial_\mu u^\nu + s u^\mu \partial_\mu T = s u_\nu \partial^\nu T. 
  \label{eqn:Bhydro_cov_s1}
\end{equation}
The term that includes $u_\nu \partial_\mu u^\nu$ is equal to zero because of the four-velocity normalization condition. Thus, we find the expression 
\begin{equation}
  T \partial_\mu (s u^\mu) + s u^\mu \partial_\mu T = s u_\nu \partial^\nu T.
  \label{eqn:Bhydro_cov_s2}
\end{equation}
This equation directly implies that
\begin{equation}
  \partial_\mu (s u^\mu) = 0.
  \label{eqn:Bhydro_cov_s3}
\end{equation}
Eq. (\ref{eqn:Bhydro_cov_s3}) states that the entropy is conserved in the system -- the hydrodynamic expansion is adiabatic.

In the next step we return to Eq. (\ref{eqn:Bhydro_cov_1}) and use Eq. (\ref{eqn:Bhydro_cov_s3}) to rewrite it in the following form
\begin{equation}
  u^\mu \partial_\mu (T u^\nu) = \partial^\nu T.
  \label{eqn:Bhydro_cov_T1}
\end{equation}
Eq. (\ref{eqn:Bhydro_cov_T1}) is the acceleration equation -- a relativistic generalization of the Euler equation known from the classical fluid dynamics. Equation (\ref{eqn:Bhydro_cov_T1}) includes only tree non-trivial equations. It is easy to check this property if we project this equation on the four-velocity $u_\nu$. The final form of the hydrodynamic equations in the covariant form consists of a set of four equations
\begin{eqnarray}
  u^\mu \partial_\mu (T u^\nu)	&=& \partial^\nu T, 	\nonumber \\ 
  \partial_\mu (s u^\mu) 	&=& 0,			\label{eqn:Bhydro_cov_2}
\end{eqnarray}
which should be supplemented by the equation of state.
%
\section{Non-covariant form}
\label{section:Bhydro_noncov}
%
In this Section we show how to rewrite relativistic hydrodynamic equations in the non-covariant form. We start with the acceleration equation in the covariant form, Eq. (\ref{eqn:Bhydro_cov_T1}), and show that the corresponding non-covariant form is 
\begin{equation}
  \frac{\partial}{\partial t}(T\gamma {\bf v}) + {\bf \nabla}(T\gamma)={\bf v} \times \left[{\bf \nabla}
 \times (T\gamma {\bf v}) \right].
\label{eqn:Bhydro_noncov_Tfin}
\end{equation}
In order to prove this property we write the $i$th (spatial)  component of Eq. (\ref{eqn:Bhydro_noncov_Tfin}) as
\begin{equation}
  \frac{\partial}{\partial t} (T \gamma v^i) + \nabla^i(T \gamma) = \epsilon^{ijk} v^j \epsilon^{klm} \nabla^l (T \gamma v^m).
  \label{eqn:Bhydro_noncov_T1}
\end{equation}
The next step is to contract the $\epsilon$ tensors with the help of the formula $$\epsilon^{ijk} \epsilon^{klm} = \delta^{il} \delta^{jm} - \delta^{im} \delta^{jl}$$ and to carry out the sums with the Kronecker delta. In this way we find
\begin{equation}
  \frac{\partial}{\partial t} (T \gamma v^i) + \nabla^i (T \gamma) = v^j \nabla^i (T \gamma v^j) - v^j \nabla^j(T \gamma v^i).
  \label{eqn:Bhydro_noncov_T2}
\end{equation}
We move the terms in Eq. (\ref{eqn:Bhydro_noncov_T2}) that include $T\gamma v^i$ to the left-hand-side and multiply both sides of this equation by the Lorentz $\gamma$ factor. This procedure yields the equation whose sides are given by the expressions
\begin{eqnarray}
  \hbox{LHS} &=& \gamma\, \frac{\partial}{\partial t} (T \gamma v^i) + \gamma\, v^j \nabla^j(T \gamma v^i) = u^0\partial_0(T u^i) + u^j\partial_j(T u^i) = u^\mu \partial_\mu (T u^i),
  \nonumber \\
  \hbox{RHS} &=& \gamma\, v^j \nabla^i (T \gamma v^j) - \gamma\, \nabla^i (T \gamma) = \gamma^2 (v^2 - 1) \nabla^i T + T \left[ \gamma (v^2 - 1) + \frac{1}{\gamma}\right] \nabla^i \gamma \nonumber \\ 
  &=& -\nabla^i T = \partial^i T.
  \nonumber
\end{eqnarray}
One observes that both sides of Eq. (\ref{eqn:Bhydro_cov_T1}) are reproduced after those manipulations. 
The non-covariant form of the entropy conservation law, Eq. (\ref{eqn:Bhydro_cov_s3}),  may be written in the straightforward way as
\begin{equation}
  \frac{\partial}{\partial t}(s \gamma) + {\bf \nabla} (s \gamma {\bf v}) = 0.
  \label{eqn:Bhydro_noncov_s}
\end{equation}
%
\section{Temperature equation}
\label{section:Bhydro_Teq}
%
In this section we study in more detail the temperature equation (\ref{eqn:Bhydro_noncov_Tfin}). We start with the $i$th component of Eq. (\ref{eqn:Bhydro_noncov_Tfin}), which in the tensor notation has the form 
\begin{equation}
  \frac{\partial}{\partial t} (T \gamma v^i) + \nabla^i (T \gamma) = 
v^j \nabla^i (T \gamma v^j) - v^j \nabla^j(T \gamma v^i).
\label{eqn:Bhydro_Teq_TiCar}
\end{equation}
We expand separately the left- and the right-hand-side of Eq. (\ref{eqn:Bhydro_Teq_TiCar}) to find
\begin{eqnarray}
  \hbox{LHS} &=& \gamma v^i \frac{\partial T}{\partial t} + T v^i \frac{\partial \gamma}{\partial t} + T \gamma \frac{\partial v^i}{\partial t} + \gamma \frac{\partial T}{\partial x^i} + T \frac{\partial \gamma}{\partial x^i} \nonumber \\
  &=& \gamma v^i \frac{\partial T}{\partial t} + T v^i \gamma^3 v^k \frac{\partial v^k}{\partial t} + T \gamma \frac{\partial v^i}{\partial t} + \gamma \frac{\partial T}{\partial x^i} + T \gamma^3 v^k \frac{\partial v^k}{\partial x^i} \nonumber \\
  &=& T \gamma \left[ v^i \frac{\partial \ln T}{\partial t} + \gamma^2 v^i \,v^k \frac{\partial v^k}{\partial t} + \frac{\partial v^i}{\partial t} + \frac{\partial \ln T}{\partial x^i} + \gamma^2 \,v^k \frac{\partial v^k}{\partial x^i} \right], \nonumber \\ 
  \end{eqnarray}
\begin{eqnarray}
& &  \hbox{RHS} = \nonumber \\
& &  \nonumber \\
&=& v^j \gamma v^j \frac{\partial T}{\partial x^i} + v^j T v^j \frac{\partial \gamma}{\partial x^i} + v^j T \gamma \frac{\partial v^j}{\partial x^i} - v^j \gamma v^i \frac{\partial T}{\partial x^j} - v^j T v^i \frac{\partial \gamma}{\partial x^j} - v^j T \gamma \frac{\partial v^i}{\partial x^j} \nonumber \\
  &=& \gamma v^2 \frac{\partial T}{\partial x^i} + T \gamma^3 v^2\, v^k \frac{\partial v^k}{\partial x^i} + T \gamma v^j \frac{\partial v^j}{\partial x^i} - \gamma v^j v^i \frac{\partial T}{\partial x^j} - T \gamma^3 v^j v^i\, v^k \frac{\partial v^k}{\partial x^j} - T \gamma v^j  \frac{\partial v^i}{\partial x^j} \nonumber \\
  &=& T \gamma \left[ v^2 \frac{\partial \ln T}{\partial x^i} + \gamma^2 v^2 \,v^k \frac{\partial v^k}{\partial x^i} + v^k \frac{\partial v^k}{\partial x^i} - v^i v^j \frac{\partial \ln T}{\partial x^j} - \gamma^2 v^i v^j\, v^k \frac{\partial v^k}{\partial x^j} - v^j \frac{\partial v^i}{\partial x^j}\right]. \nonumber
\end{eqnarray}
By comparing LHS with RHS we get
\begin{eqnarray}
\hspace{-1cm}  &&(1-v^2) \frac{\partial \ln T}{\partial x^i} + v^i \left[ \frac{\partial \ln T}{\partial t} + v^j \frac{\partial \ln T}{\partial x^j}\right]
  + \frac{\partial v^i}{\partial t} + v^j \frac{\partial v^i}{\partial x^j} + \gamma^2 v^i \,v^k \left[ \frac{\partial v^k}{\partial t} + v^j \frac{\partial v^k}{\partial x^j}\right] = 0. \nonumber
\end{eqnarray}
The final form of the temperature equations in the Cartesian coordinates is
\begin{eqnarray}
 (1-v^2) \frac{\partial \ln T}{\partial x^i} + v^i \frac{d \ln T}{dt} + \frac{d v^i}{d t} + \gamma^2 v^i\, v^k \frac{d v^k}{dt} = 0,
 \label{eqn:Bhydro_TiGen_Car}
\end{eqnarray}
or 
\begin{eqnarray}
  (1-v^2) \frac{\partial \ln T}{\partial x} + v_x \frac{d \ln T}{dt} + \frac{d v_x}{d t} + \gamma^2 v_x\, v \frac{d v}{dt} &=& 0, \label{eqn:Bhydro_Tx_Car}\\
  (1-v^2) \frac{\partial \ln T}{\partial y} + v_y \frac{d \ln T}{dt} + \frac{d v_y}{d t} + \gamma^2 v_y\, v \frac{d v}{dt} &=& 0, \label{eqn:Bhydro_Ty_Car}\\
  (1-v^2) \frac{\partial \ln T}{\partial z} + v_z \frac{d \ln T}{dt} + \frac{d v_z}{d t} + \gamma^2 v_z\, v \frac{d v}{dt} &=& 0. \label{eqn:Bhydro_Tz_Car}
\end{eqnarray}
For further use it is convenient to consider linear combinations of Eqs. (\ref{eqn:Bhydro_Tx_Car}) - (\ref{eqn:Bhydro_Tz_Car}). At first we multiply Eqs. (\ref{eqn:Bhydro_Tx_Car}) - (\ref{eqn:Bhydro_Tz_Car}) by the appropriate components of the velocity (for example,  Eq. (\ref{eqn:Bhydro_Tx_Car}) is multiplied by $v_x$, etc.) and add them together. The second combination is obtained if we multiply Eq. (\ref{eqn:Bhydro_Tx_Car}) by $v_y$ and subtract Eq. (\ref{eqn:Bhydro_Ty_Car}) multiplied by $v_x$. As the third independent equation we take the unchanged formula (\ref{eqn:Bhydro_Tz_Car}). In this way one finds
\begin{eqnarray}
  (v^2 - 1) \frac{\partial \ln T}{\partial t} + \frac{d \ln T}{dt} + \frac{1}{1 - v^2} v \frac{d v}{dt} &=& 0,
  \label{eqn:Bhydro_T1_Car} \\
  (1 - v^2) \left( v_y \frac{\partial \ln T}{\partial x} - v_x \frac{\partial \ln T}{\partial y} \right) + v_y \frac{d v_x}{dt} - v_x \frac{d v_y}{dt} &=& 0,
  \label{eqn:Bhydro_T2_Car} \\
   (1-v^2) \frac{\partial \ln T}{\partial z} + v_z \frac{d \ln T}{dt} + \frac{d v_z}{d t} + \frac{v_z}{1-v^2}\, v \frac{d v}{dt} &=& 0. \label{eqn:Bhydro_T3_Car}
\end{eqnarray}
These equations can be also rewritten  in the cylindrical coordinates in the form
\begin{eqnarray}
  (v^2 - 1) \,\frac{\partial \ln T}{\partial t} + \frac{d \ln T}{dt} + \frac{1}{1-v^2}\, v \frac{d v}{dt} &=& 0, \nonumber \\
  \label{eqn:Bhydro_T1_Cyl}\\
  (1 - v^2) \left( v_r \sin \alpha \,\frac{\partial \ln T}{\partial r} - \frac{v_r \cos \alpha}{r} \,\frac{\partial \ln T}{\partial \phi}\right) - v_r^2 \left( \frac{d \alpha}{dt} + \frac{v_r \sin \alpha}{r} \right) &=& 0, \nonumber \\
  \label{eqn:Bhydro_T2_Cyl} \\
  (1 - v^2) \frac{\partial \ln T}{\partial z} + v_z \frac{d \ln T}{dt} + \frac{d v_z}{d t} + \frac{v_z}{1-v^2}\, v \frac{d v}{dt} &=& 0. \nonumber \\
  \label{eqn:Bhydro_T3_Cyl}
\end{eqnarray}
We note that the total derivative in the above equations has the form of Eq. (\ref{eqn:Bhydro_def_dtCyl}).
%
\subsection{Boost-invariance}
\label{subsection:Bhydro_Teq_BI}
%
For the boost invariant systems the longitudinal velocity has the form $v_z = z/t$. We may insert this form into the hydrodynamic equations, calculate all the necessary derivatives and then set $z=0$. In this way we obtain the hydrodynamic equations in the plane $z=0$.
\begin{eqnarray}
  (v_r^2 - 1) \frac{\partial \ln T}{\partial t} + \frac{d \ln T}{dt} + \frac{1}{1 - v_r^2} v_r \frac{d v_r}{dt} &=& 0,
  \label{eqn:Bhydro_T1_Car_BI} \\
  (1 - v_r^2) \left( v_y \frac{\partial \ln T}{\partial x} - v_x \frac{\partial \ln T}{\partial y} \right) + v_y \frac{d v_x}{dt} - v_x \frac{d v_y}{dt} &=& 0.
  \label{eqn:Bhydro_T2_Car+BI}
\end{eqnarray}
Again we can make use of the cylindrical coordinates to get
\begin{eqnarray}
  (v_r^2 - 1) \,\frac{\partial \ln T}{\partial t} + \frac{d \ln T}{dt} + \frac{1}{1-v^2}\, v \frac{d v}{dt} = 0, \nonumber \\
  \label{eqn:Bhydro_T1_Cyl_BI} \\
  (1 - v_r^2) \left( v_r \sin \alpha \,\frac{\partial \ln T}{\partial r} - \frac{v_r \cos \alpha}{r} \,\frac{\partial \ln T}{\partial \phi}\right) - v_r^2 \left( \frac{d \alpha}{dt} + \frac{v_r \sin \alpha}{r} \right) = 0. \nonumber \\
  \label{eqn:Bhydro_T2_Cyl_BI}
\end{eqnarray}
The total derivatives for the boost-invariant systems are reduced at $z=0$ to the expressions
\begin{eqnarray}
  \frac{d}{dt} &=& \frac{\partial}{\partial t} + v_x \frac{\partial}{\partial x} + v_y \frac{\partial}{\partial y} 
  \qquad \qquad \qquad \,\,\,\, \hbox{(Cartesian)},
  \label{eqn:Bhydro_dt_Car_BI} \\
  \frac{d}{dt} &=& \frac{\partial}{\partial t} + v_r \cos \alpha \frac{\partial}{\partial r} + \frac{v_r \sin \alpha}{r} \frac{\partial}{\partial \phi}
  \qquad \hbox{(cylindrical)}.
  \label{eqn:Bhydro_dt_Cyl_BI}
\end{eqnarray}
%
\subsection{Cylindrical symmetry}
\label{subsection:Bhydro_Teq_CS}
%
In certain physical situations the system described by the relativistic hydrodynamics has cylindrical symmetry. In this case
the hydrodynamic equations are reduced to the form
\begin{eqnarray}
  (v^2 - 1) \frac{\partial \ln T}{\partial t} + \frac{d \ln T}{dt} + \frac{1}{1 - v^2} v \frac{d v}{dt} &=& 0,
  \label{eqn:Bhydro_T1_CS} \\
  (1-v^2) \frac{\partial \ln T}{\partial z} + v_z \frac{d \ln T}{dt} + \frac{d v_z}{d t} + \frac{v_z}{1-v^2}\, v \frac{d v}{dt} &=& 0. \label{eqn:Bhydro_T2_CS}
\end{eqnarray}
These equations have the same form in cylindrical coordinates.
Combining both the boost-invariance and cylindrical symmetry gives us only one independent equation 
\begin{eqnarray}
  v_r \frac{\partial \ln T}{\partial t} + \frac{\partial \ln T}{\partial r} + \frac{1}{1 - v_r^2} \frac{\partial v_r}{\partial t} +\frac{v_r}{1 - v_r^2} \frac{\partial v_r}{\partial r} = 0.
  \label{eqn:Bhydro_T1_BICS}
\end{eqnarray}
%
\section{Entropy equation}
\label{section:Bhydro_Seq}
%
In this Section we discuss the entropy conservation law, Eq. (\ref{eqn:Bhydro_noncov_s}). Following the same steps as in the previous Section we obtain
\begin{eqnarray}
  \gamma \frac{\partial s}{\partial t} + s \frac{\partial \gamma}{\partial t} + \gamma v^i \frac{\partial s}{\partial x^i} + s v^i \frac{\partial \gamma}{\partial x^i} + s \gamma \frac{\partial v^i}{\partial x^i} &=& 0,
  \nonumber \\
  \gamma s \left[ \frac{\partial \ln s}{\partial t} + v^i \frac{\partial \ln s}{\partial x^i} + \gamma^2 \, v^k \left( \frac{\partial v^k}{\partial t} + v^i \frac{\partial v^k}{\partial x^i} \right) + \frac{\partial v^i}{\partial x^i}\right] &=& 0.\nonumber
\end{eqnarray}
The final form of the entropy equation has the form
\begin{equation}
  \frac{d \ln s}{dt} + \frac{v}{1-v^2} \frac{d v}{dt} + \frac{\partial v^i}{\partial x^i} = 0,
  \label{eqn:Bhydro_sGen}
\end{equation}
which in the Cartesian and cylindrical coordinates has the form
\begin{eqnarray}
  \frac{d \ln s}{dt} + \frac{v}{1-v^2} \frac{d v}{dt} + \frac{\partial v_x}{\partial x} + \frac{\partial v_y}{\partial y} + \frac{\partial v_z}{\partial z} = 0,
  \label{eqn:Bhydro_s_Car}
\end{eqnarray}
and 
\begin{eqnarray}
  &&\frac{d \ln s}{dt} + \frac{v}{1-v^2} \frac{d v}{dt} + \cos \alpha \frac{\partial v_r}{\partial r} + \frac{\sin \alpha}{r} \frac{\partial v_r}{\partial \phi} \nonumber \\
  &-& v_r \sin \alpha \frac{\partial \alpha}{\partial r} + \frac{v_r \cos \alpha}{r} \left( \frac{\partial \alpha}{\partial \phi} + 1 \right) + \frac{\partial v_z}{\partial z}= 0.
  \label{eqn:Bhydro_s_Cyl}
\end{eqnarray}
%
\subsection{Boost-invariance}
\label{subsection:Bhydro_seq_BI}
%
The $z$ component of velocity equals $v_z = z/t$, hence its derivative with the respect to that coordinate gives a term $\frac{1}{t}$. With this remark in mind we perform the rest of the calculations at $z=0$ and find
\begin{eqnarray}
  \frac{d \ln s}{dt} +\frac{v_r}{1 - v_r^2} \frac{d v_r}{dt} + \frac{\partial v_x}{\partial x} + \frac{\partial v_y}{\partial y} + \frac{1}{t} = 0,
  \label{eqn:Bhydro_s_Car_BI}
\end{eqnarray}
or
\begin{eqnarray}
   v_r \frac{d \ln s}{dt} + \frac{1}{1-v_r^2} \frac{d v_r}{dt} - \frac{\partial v_r}{\partial t} - v_r^2 \sin \alpha \frac{\partial \alpha}{\partial r} + \frac{v_r^2 \cos \alpha}{r} \left( \frac{\partial \alpha}{\partial \phi} + 1 \right) + \frac{v_r}{t}= 0. \nonumber \\
  \label{eqn:Bhydro_s_Cyl_BI}
\end{eqnarray}
%
\subsection{Cylindrical symmetry}
\label{subsection:Bhydro_seq_CS}
%
For cylindrically symmetric systems entropy density and transverse velocities are independent of direction in transverse plane. The entropy equation with cylindrical symmetry has the form
\begin{eqnarray}
   &&\frac{d \ln s}{dt} + \frac{v_r}{1-v_r^2} \frac{\partial v_r}{\partial t} + \frac{1}{1-v_r^2} \frac{\partial v_r}{\partial r} + \frac{v_r}{r} + \frac{\partial v_z}{\partial z} = 0.
  \label{eqn:Bhydro_s_Cyl_CS}
\end{eqnarray}
where we have used the definition for the transverse velocity divergence
\begin{equation}
  \frac{\partial v_x}{\partial x} + \frac{\partial v_y}{\partial y} = \frac{\partial v_r}{\partial r} + \frac{v_r}{r}.
  \label{eqn:Bhydro_DIVvr_CS}
\end{equation}
When both the boost-invariance and cylindrical symmetry is applied to the entropy equation we get
\begin{eqnarray}
   &&\frac{\partial \ln s}{\partial t} + v_r \frac{\partial \ln s}{\partial r} + \frac{v_r}{1-v_r^2} \frac{\partial v_r}{\partial t} + \frac{1}{1-v_r^2} \frac{\partial v_r}{\partial r} + \frac{v_r}{r} + \frac{1}{t} = 0.
  \label{eqn:Bhydro_s_Cyl_BICS}
\end{eqnarray}

\chapter{Notation}
\label{chapter:Ysymbol}
%
In this Appendix we collect the symbols used in the Thesis.
\begin{table}[!ht]
{\footnotesize
  \begin{tabular}{cl}
    \hline
    \multicolumn{2}{c}{}\\
    \multicolumn{2}{c}{\bf Thermodynamics}\\
    \multicolumn{2}{c}{}\\
    \hline\\
    \hphantom{$\alpha = \tan^{-1}(v_y/v_x) - \phi $} & \hphantom{ distance from the point $(\tau=\tau_{\rm i},x=0,y=0)$ to the hypersurface}\\
    $\Omega$		& grand canonical potential,\\
    $Z$ or $z$		& grand canonical partition function,\\
    $T$			& temperature,\\
    $V$			& volume,\\
    $E$			& energy,\\
    $\displaystyle \varepsilon = \frac{E}{V}$	& energy density,\\
    $S$			& entropy,\\
    $\displaystyle s = \frac{S}{V}$	& entropy density,\\
    $P$			& pressure,\\
    $N$			& particle number,\\
    $\mu$		& chemical potential,\\
    $\mu_{\rm B}$	& baryon chemical potential,\\
    $\mu_{\rm S}$	& strange chemical potential,\\
    $\mu_{\rm I_3}$	& isospin chemical potential,\\
    $\epsilon$		& statistics identifier (+1 for FD, -1 for BE),\\
    $g$			& degeneration factor,\\
    $c_{\rm s}$		& sound velocity,\\
    $T_{\rm c}$		& critical temperature,\\
    $m_\pi$		& pion mass.\\
  \end{tabular}
}
\end{table}
\begin{table}[!ht]
{\footnotesize
  \begin{tabular}{cl}
    \hline
    \multicolumn{2}{c}{}\\
    \multicolumn{2}{c}{\bf Hydrodynamics}\\
    \multicolumn{2}{c}{}\\
    \hline\\
    $T^{\mu\nu}$	& energy momentum tensor,\\
    $u^\mu = \gamma\, (1,{\bf v})$	& four-velocity,\\
    ${\bf v} = (v_x,v_y,v_z)$	& velocity,\\
    ${\bf v}_r$		& radial velocity,\\
    ${\bf v}_\perp$	& transverse velocity,\\
    $\alpha = \tan^{-1}(v_y/v_x) - \phi $	& angle between transverse velocity ${\bf v}_\perp$ and the radial velocity ${\bf v}_r$,\\
    $\eta_\perp = \tanh^{-1} v_\perp$	& transverse fluid rapidity,\\ 
    $\gamma = (1-v^2)^{-\frac{1}{2}}$	& Lorentz factor,\\
    $j^\mu_{\rm B} $	& baryon four-current,\\
    $\tau $		& proper time,\\
    $\Phi $		& thermodynamic potential used in the Baym formalism,\\
    $A_+, A_-, A$	& auxiliary functions used in the Baym formalism.\\
    \multicolumn{2}{c}{}\\
    \hline\\
    \multicolumn{2}{c}{\bf Initial conditions for hydrodynamics}\\
    \multicolumn{2}{c}{}\\
    \hline\\
    $\rho$		& nuclear profile,\\
    $\rho_0, r_0, a$	& parameters for Woods-Saxon nuclear profile,\\
    $\rho_{\rm WN}$	& wounded nucleon profile,\\
    $\rho_{\rm BC}$	& binary collisions profile,\\
    $T_A$		& thickness function,\\
    $\sigma_{\rm in}$	& total inelastic $p p$ cross-section,\\
    $\sigma_{\rm in}^{\rm Au Au}$	& total inelastic cross-section in Au Au,\\
    $\kappa$		& mixing factor between WN and BC nuclear profiles,\\
    ${\bf b}$		& impact vector,\\
    $\tau_{\rm i}$	& initial proper time,\\
    $s_{\rm i}$		& initial entropy density at $r=0$,\\
    $\varepsilon_{\rm i}$	& initial energy density at $r=0$,\\
    $T_{\rm i}$		& initial temperature at $r=0$,\\
    $H_0$		& initial Hubble flow.\\
    \multicolumn{2}{c}{}\\
    \hline\\
    \multicolumn{2}{c}{\bf Freeze-out}\\
    \multicolumn{2}{c}{}\\
    \hline\\
    $\phi$		& spatial azimuthal angle,\\
    $\zeta$		& angle in the $\tau - \rho $ plane,\\
    $\displaystyle \eta_\parallel = \frac{1}{2} \ln \frac{t+z}{t-z} $& spacetime rapidity, \\
    $d(\phi,\zeta)$	& distance from the point $(\tau=\tau_{\rm i},x=0,y=0)$ to the hypersurface\\
                        & point with coordinates $(\phi,\zeta)$,\\
    ${\bf v}_\perp$	& transverse velocity on the hypersurface,\\
    $\alpha$		& angle between $v_\perp$ and $v_r$ on the hypersurface,\\
    $p_T$ or $p_\bot$	& particle's transverse momentum,\\
    $\phi_p$		& particle's azimuthal transverse momentum angle,\\
    $y$			& particle's rapidity,\\
    $m$			& particle's mass,\\
    $d\Sigma$		& three-dimensional element of the hypersurface.\\
  \end{tabular}
}
\end{table}

\chapter{Mathematical supplement}
\label{chapter:Zmath}
%
\section{Modified Bessel functions}
\label{section:Zmath_BesselK}
%
The second-order ordinary differential equation
\begin{equation}
  z^2\frac{d^2 w(z)}{d\,z^2}+z\frac{d w(z)}{dz}-\left(z^2+\nu^2\right)w(z)=0
  \label{eqn:Zmath_BesselK_diffeqn}
\end{equation}
has a solution in form of the modified Bessel function of the first kind $I_\nu(z)$ and second kind $K_\nu(z)$
\begin{equation}
  w(z) = c_1 I_\nu(z) + c_2 K_\nu(z).
  \label{eqn:Zmath_BesselK_soldiffeqn}
\end{equation}
Modified Bessel function $K_\nu(z)$ has a simple integral representations through the exponential function and power functions in the integrand
\begin{equation}
  K_n(z) = \frac{\sqrt{\pi}\,z^\nu}{2^\nu\,\Gamma\left(\nu+\frac{1}{2}\right)} \int_1^\infty e^{-z t} \left(t^2-1\right)^{\nu-\frac{1}{2}} dt; \quad {\rm Re}(\nu) >-\frac{1}{2}\,\wedge\, {\rm Re}(z) >0.
  \label{eqn:Zmath_BesselK_int}
\end{equation}
The Bessel function $K_\nu(z)$ satisfies the following recurrence identities
\begin{equation}
  \begin{array}{l}
    \displaystyle K_\nu(z) = K_{\nu+2}(z) - \frac{2(\nu+1)}{z} K_{\nu+1}(z), \\
    \\
    \displaystyle K_\nu(z) = K_{\nu-2}(z) + \frac{2(\nu-1)}{z} K_{\nu-1}(z).
  \end{array}
  \label{eqn:Zmath_BesselK_recur}
\end{equation}
The derivative of the Bessel function $K_\nu(z)$ has a rather simple and symmetrical representation that can be expressed through other Bessel $K_\nu(z)$ functions with different indices
\begin{equation}
  \frac{\partial K_\nu(z)}{\partial z} = -\frac{1}{2} \left(K_{\nu-1}(z) + K_{\nu+1}(z)\right),
  \label{eqn:Zmath_BesselK_deriv1}
\end{equation}
this derivative can also be represented in other form
\begin{equation}
  \frac{\partial K_\nu(z)}{\partial z} = - K_{\nu-1}(z) - \frac{\nu}{z} K_\nu(z) = \frac{\nu}{z} K_\nu(z) - K_{\nu+1}(z).
  \label{eqn:Zmath_BesselK_deriv2}
\end{equation}
Series representations of the modified Bessel function $K_\nu(z)$ about the point $z=0$ is very complex. We will limit ourselves to those series expansions useful in our calculations
\begin{equation}
  \begin{array}{l}
    \displaystyle K_1(z) = \frac{ 1}{z  }                                                + O(z^1), \\
    \\
    \displaystyle K_2(z) = \frac{ 2}{z^2} - \frac{1}{2  }                                + O(z^2), \\
    \\
    \displaystyle K_3(z) = \frac{ 8}{z^3} - \frac{1}{z  } + \frac{z}{8}                  + O(z^3), \\
    \\
    \displaystyle K_4(z) = \frac{48}{z^4} - \frac{4}{z^2} + \frac{1}{4} - \frac{z^2}{48} + O(z^4)
  \end{array}
  \label{eqn:Zmath_BesselK_powser}
\end{equation}
The asymptotic expansion of the modified Bessel function $K_\nu(z)$ about $z \rightarrow \infty$ has the form
\begin{equation}
  K_\nu(z) = e^{-z} \left( \sqrt{\frac{\pi}{2}}\, z^{-\frac{1}{2}} + O\left( z^{-\frac{3}{2}} \right) \right)
  \label{eqn:Zmath_BesselK_powser2}
\end{equation}

\providecommand{\href}[2]{#2}\begingroup\raggedright

\endgroup

\end{document}